\documentclass[prb,twocolumn,showpacs,amsmath,amssymb]{revtex4}

\usepackage{graphicx}
\usepackage{bm}
\usepackage{epsf}
\def\be{\begin{equation}}
\def\ee{\end{equation}}
\def\bea{\begin{eqnarray}}
\def\eea{\end{eqnarray}}

\renewcommand{\theequation}{\arabic{section}.\arabic{equation}}

\begin{document}
\title{Coulomb drag between ballistic quantum wires}
\author{A. P. Dmitriev$^{1,2}$}
\author{I. V. Gornyi$^{2,1}$}
\author{D. G. Polyakov$^{2}$}
\affiliation{
$^1$A.F.Ioffe Physico-Technical Institute, 194021 St.Petersburg, Russia
\\
$^2$Institut f\"ur Nanotechnologie, Karlsruhe Institute of Technology, 76021 Karlsruhe, Germany
}

\begin{abstract}
We develop a kinetic equation description of Coulomb drag between ballistic one-dimensional electron systems, which enables us to demonstrate that equilibration processes between right- and left-moving electrons are crucially important for establishing dc drag. In one-dimensional geometry, this type of equilibration requires either backscattering near the Fermi level or scattering with small momentum transfer near the bottom of the electron spectrum. Importantly, pairwise forward scattering in the vicinity of the Fermi surface alone is not sufficient to produce a nonzero dc drag resistivity $\rho_{\rm D}$, in contrast to a number of works that have studied Coulomb drag due to this mechanism of scattering before. We show that slow equilibration between two subsystems of electrons of opposite chirality, ``bottlenecked" by inelastic collisions involving cold electrons near the bottom of the conduction band, leads to a strong suppression of Coulomb drag, which results in an activation dependence of $\rho_{\rm D}$ on temperature---instead of the conventional power law. We demonstrate the emergence of a drag regime in which $\rho_{\rm D}$ does not depend on the strength of interwire interactions, while depending strongly on the strength of interactions inside the wires.
\end{abstract}

\maketitle

\section{Introduction}
\label{s1}
\setcounter{equation}{0}

A remarkable property of a system of two conductors placed in proximity to each other is the occurrence of the phenomenon of Coulomb drag. This consists of inducing an electric field or current in one of the conductors by sending a current through the other---with the friction force being due to electron-electron interactions---in the absence of transfer of electrons between the two subsystems. As such, Coulomb drag is a sensitive probe of electron-electron correlations and, specifically, of inelastic electron-electron scattering.

The key quantity describing friction is the drag resistivity $\rho_{\rm D}$ conventionally defined for two homogeneous conductors parallel to each other as
\be
\rho_{\rm D}=-E_2/j_1~,
\ee
where $j_1$ is the electric current density in (``active") conductor 1 and $E_2$ is the electric field applied to (``passive") conductor 2 to compensate for the friction force under the condition that no current flows in the passive conductor. Since its prediction \cite{pogrebinskii77,price83} a third of a century ago, for two-dimensional geometry of two parallel conducting sheets, Coulomb drag has been extensively studied experimentally in double-layer semiconductor structures, \cite{gramila91,sivan92,kellogg02,pillarisetty02,pillarisetty05,price07,seamons09} also in a transverse magnetic field. \cite{hill96,rubel97,lilly98,feng98,lok01,kellogg02a,kellogg03,muraki04,tutuc09} Recent experimental work has addressed a similar phenomenon in double-layer graphene. \cite{kim11,kim12,gorbachev12} In one-dimensional geometry, a number of experiments have explored Coulomb drag between quantum wires. \cite{debray00,debray01,debray02,yamamoto06,laroche10} Drag experiments have also been done on electron systems of other geometry: between two- and essentially three-dimensional electron systems \cite{solomon89} or between quantum-point contacts. \cite{khrapai07}

\subsection{``Orthodox theory"}
\label{s1.1}

A great deal of understanding of the mechanism of Coulomb drag has been achieved by calculating the friction force perturbatively, at second order, in the dynamically screened interaction $V_{12}(\omega,q)$ between two two-dimensional electron systems (``orthodox theory"). \cite{gramila91,jauho93,zheng93,kamenev95,flensberg95} Within this framework, Coulomb drag is represented as rectification of nonequilibrium current fluctuations induced in the passive layer and, consequently, the linear-response resistivity $\rho_{\rm D}$ is related to dynamical correlations in thermal fluctuations of the electron densities in different layers at equilibrium. Equivalently, $\rho_{\rm D}$ within the orthodox theory is proportional to the rate of momentum transfer between the layers at order $V_{12}^2$. One important result of the orthodox theory is that $\rho_{\rm D}$ at order $V_{12}^2$ scales with temperature $T$ in the limit of small $T$ as $T^2$ (for ballistic electron systems, \cite{gramila91,jauho93,zheng93,kamenev95} or as $T^2\ln T$ in the diffusive limit \cite{zheng93,kamenev95}). The power-law vanishing of $\rho_{\rm D}$ as $T$ decreases is associated with the constraints on the phase space available for inelastic electron-electron scattering.

To the best of our knowledge, in all works where Coulomb drag in two-dimensional systems was studied within the framework of the orthodox theory, $\rho_{\rm D}$ was derived under the {\it tacit} assumption that the intralayer relaxation processes (determined, e.g., by disorder) are faster than the processes of momentum transfer between the layers. Within the kinetic equation approach, which we employ in this paper, this condition implies that an iterative solution \cite{jauho93} of the kinetic equation (equivalent, in the diagrammatic language, to the evaluation of the Aslamazov-Larkin-type diagrams \cite{kamenev95,flensberg95}) is justified. A delicate point here is that the resulting drag resistivity in the presence of disorder does not necessarily depend on the strength of disorder, which might seem to imply that the thus obtained $\rho_{\rm D}$ describes the clean limit as well. However, in the absence of relaxation processes induced by disorder or inelastic intralayer interactions, the nonequilibrium part of the electron distribution function is governed by interactions between the layers, so that the lowest-order expansion in the interlayer collision integral, assumed in the orthodox theory, is generally not sufficient. Therefore, the orthodox theory should not be expected to be generically valid in the clean limit---even for an arbitrarily weak interaction between the conductors. One particular example that demonstrates a dramatic departure from the orthodox theory in the clean case is Coulomb drag between ballistic quantum wires, addressed in this paper.

The orthodox theory also explicitly points to the important role of electron-hole asymmetry in a degenerate Fermi gas, in the absence of which the electron and hole contributions to $\rho_{\rm D}$ at order $V_{12}^2$ cancel each other. The cancellation \cite{disclaimer} has the consequence that, in the case of particle-hole asymmetry produced by a finite curvature $1/m$ of the electron dispersion relation, where $m$ is the electron mass, $\rho_{\rm D}$ is small in the parameter $(T/\epsilon_F)^2$ with $\epsilon_F$ being the Fermi energy. In the diffusive case, $\rho_{\rm D}$ for sufficiently small transferred momenta can be directly related at order $V_{12}^2$ to the dependence of the local conductivity on the local electron density, \cite{narozhny01,vonoppen01} absent in the particle-hole symmetric case.

Apart from a nonzero curvature $1/m$, particle-hole asymmetry can also result from the energy dependence of the electron density of states in the vicinity of $\epsilon_F$. The latter contribution to Coulomb drag is important in the presence of a transverse magnetic field \cite{vonoppen01,gornyi04} because of the modification of the density of states by Landau quantization. It is also important in two-dimensional electron systems with a linear dispersion relation $(1/m=0)$, in which the density of states varies linearly with energy; in particular, in graphene. \cite{narozhny12} Particle-hole symmetry in two dimensions is realized in graphene at the charge-neutrality point, where the orthodox theory gives zero $\rho_{\rm D}$. Therefore, possible deviations from the orthodox theory in the vicinity of this point in graphene become particularly important. \cite{schuett12} The prevailing notion that Coulomb drag is entirely due to particle-hole asymmetry is justified in the case of a disordered two-dimensional electron system only at order $V_{12}^2$. Beyond the golden-rule level, already at third order in $V_{12}$, rectification of interaction-induced current fluctuations in a diffusive double-layer system yields nonzero $\rho_{\rm D}$ even in a particle-hole symmetric system. \cite{levchenko08}

\subsection{Coulomb drag in one dimension: Backward scattering}
\label{s1.2}

In one-dimensional geometry, the connection between Coulomb drag and particle-hole asymmetry is subtler. Processes of electron scattering due to interwire interaction separate into two classes: backscattering, in which an electron changes its chirality, and forward scattering, in which it does not. In a ballistic system with no disorder, the contribution of interwire forward-scattering processes to Coulomb drag vanishes if the electron dispersion relation is linearized (Luttinger-liquid model \cite{giamarchi04}); however, the contribution of interwire backscattering processes to $\rho_{\rm D}$ remains nonzero even in the particle-hole symmetric limit. \cite{flensberg98,nazarov98,klesse00,ponomarenko00,komnik01,trauzettel02,fuchs05,fiete06,peguiron07} Much of the prior work on Coulomb drag between quantum wires has therefore focused on the backscattering processes. At the golden-rule level, $\rho_{\rm D}$ induced by backscattering between identical wires is linearly proportional \cite{hu96,remark1} to $T$ with
\begin{equation}
\rho_{\rm D}\sim {2\pi\over e^2}\,\beta_{\rm b}^2\,{T\over v_F}~,
\label{1.0}
\end{equation}
where $\beta_{\rm b}$ is the dimensionless coupling constant describing interwire backscattering at the Fermi level with momentum transfer $2k_F$, $v_F$ is the Fermi velocity, and $2\pi/e^2$ is the resistance quantum (here and below $\hbar=1$). At higher orders in the strength of interaction, both intra- and interwire, a power-law renormalization of the backscattering amplitude develops as $T$---or the drive current in the active wire in the nonlinear response regime---decreases \cite{flensberg98,nazarov98,klesse00,ponomarenko00,komnik01,schlottmann04,fuchs05,peguiron07} (a similar renormalization of $\rho_{\rm D}$ in the strongly-interacting limit of a ``spin-incoherent" Luttinger liquid has been discussed in Ref.~\onlinecite{fiete06}).

Below a characteristic energy scale (at which the renormalized amplitude $g_1$ is of the order of unity), electrons in two wires form a ``zig-zag ordered" charge-density wave and the power-law behavior crosses over into an exponential growth of $\rho_{\rm D}$ with lowering $T$ or, in finite-size systems at sufficiently low $T$, into an exponential growth of the drag resistance with increasing system size. \cite{nazarov98,klesse00,ponomarenko00,fuchs05} In the limit $T\to 0$, the resistivity $\rho_{\rm D}$ (defined as the linear resistance per unit length under the condition that the size of the system is made infinite before any other limit is taken, in particular, that of zero $T$) is infinitely large (``absolute current drag" in the terminology of Ref.~\onlinecite{nazarov98}). By contrast, the linear drag resistance between finite-size wires vanishes to zero as $T^2$ in the limit of low $T$, independently of the strength of intrawire interaction and on whether the wires are long enough to form the zig-zag order or not. \cite{ponomarenko00,remark3} In the former case, however, there exists a parametrically wide range of $T$ in which the drive and drag currents are almost equal to each other (almost absolute current drag) up to an exponentially small unbalance due to transport of solitons in the charge-density wave. In the nonlinear response regime, the drag resistance is finite at zero $T$ and shows oscillations \cite{nazarov98,peguiron07} as the drive current varies, related to the interference of plasmon waves reflected from the boundaries of the wires, which are suppressed \cite{peguiron07} as $T$ is increased. If the length of the region in which interwire backscattering occurs is much smaller than the total length of the wires, Coulomb drag in the limit of low $T$ and small bias voltages can be described in terms of backscattering at effectively pointlike contact. \cite{flensberg98,komnik01,trauzettel02,remark2} In this model, the drag conductance induced by electron-electron backscattering is expressible \cite{flensberg98,komnik01,remark4} in a particularly simple form in terms of the conductance of a single wire with a single static backscattering impurity.

One of the conclusions that follow from the above results for backscattering-induced Coulomb drag is that---even if the bare (before the renormalization) backscattering amplitude is small---the drag effects can be strong in the infrared limit, which for the linear drag resistivity $\rho_{\rm D}$ means sufficiently low $T$. However, the backscattering amplitude falls off with increasing distance $a$ between the wires as $\exp (-2k_Fa)$, where $k_F$ is the Fermi wavelength (assuming the electron densities in the wires to be equal to each other). As a result, for $k_Fa\gg 1$ the drag effects that originate from backscattering are strongly suppressed unless $T$ is exponentially small and electrons are zig-zag ordered. Moreover, the effect of backscattering is also suppressed in the case of nonidentical wires. \cite{fuchs05}

\subsection{Coulomb drag in one dimension: Forward scattering}
\label{s1.3}

An alternative mechanism of drag is interwire scattering with small momentum transfer. \cite{pustilnik03,aristov07,rozhkov08} Despite relying on electron-hole asymmetry (e.g., a nonlinear dispersion relation for electrons) or the presence of disorder, \cite{remark6} this mechanism of drag is expected to be more effective than backscattering if quantum wires are sufficiently separated from each other, and is further favored by an imbalance in the electron densities.

Against this background, it is desirable to explore the possibility of Coulomb drag due to interwire forward scattering in the absence of any backward scattering. An important advance in this direction was made in Ref.~\onlinecite{pustilnik03} which extended the orthodox theory \cite{jauho93,zheng93,kamenev95,flensberg95} for electron systems with a parabolic dispersion relation to one dimension (see Ref.~\onlinecite{imambekov12} for a review of one-dimensional non-Luttinger liquid models)---under the assumption that electrons are ballistic (no disorder) and can only exchange momenta much smaller than $k_F$. Specifically, the calculation in Ref.~\onlinecite{pustilnik03} suggests \cite{remark34} that $\rho_{\rm D}$ induced by forward scattering in the vicinity of the Fermi level for $T\ll v_F/a$ reads
\be
\rho_{\rm D}\stackrel{?}{\sim} {2\pi\over e^2}\,\beta_{\rm f}^2k_F\!\left({T\over\epsilon_F}\right)^2~,
\label{1.1}
\ee
where $\beta_{\rm f}$ is the dimensionless coupling constant describing interwire forward scattering. The meaning of the question mark will become clear in the next paragraph. At higher $T$, in the interval $v_F/a\ll T\ll\epsilon_F$ (which exists for $k_Fa\gg 1$), $\rho_{\rm D}$ shows a plateau in the dependence on $T$ according to Ref.~\onlinecite{pustilnik03}. Other work \cite{aristov07,rozhkov08} has reached similar conclusions by employing a bosonic description of the one-dimensional electron liquid with a nonlinear dispersion relation for electrons; in particular, reproduced Eq.~(\ref{1.1}). According to Ref.~\onlinecite{aristov07}, however, the $T^2$ scaling of $\rho_{\rm D}$ for the case of identical wires is only valid for $\beta_{\rm f}\epsilon_F\ll T\ll v_F/a$ (provided that $v_F/a$ is larger than $\beta_{\rm f}\epsilon_F$ and such an interval of $T$ exists). The energy scale $\beta_{\rm f}\epsilon_F$ describes splitting between symmetric and antisymmetric plasmon modes in the double-wire system. In the low-$T$ limit, for $T\ll\beta_{\rm f}\epsilon_F$,  the drag resistivity between identical wires is predicted \cite{aristov07,remark7} to vanish with decreasing $T$ as $\rho_{\rm D}\sim\beta_{\rm f}^{-1}(2\pi/e^2)k_F(T/\epsilon_F)^5$. Importantly, all the prior work \cite{pustilnik03,aristov07,rozhkov08,pereira10} on Coulomb drag due to interactions with small momentum transfer obtained a {\it nonzero} drag resistivity (unless $T=0$) from forward scattering in the vicinity of the Fermi level.

One of the purposes of this paper is to demonstrate that in fact
\be
\rho_{\rm D}(T)\equiv 0
\label{1.2}
\ee
in the absence of scattering that changes the chirality of electrons. That is, forward scattering itself cannot produce a nonzero dc drag resistivity. \cite{remark24} As will be shown, the orthodox theory, \cite{jauho93,zheng93,kamenev95,flensberg95} with the use of which (or of its equivalent in the bosonized formulation of the problem) the nonzero result was obtained in the earlier works, \cite{pustilnik03,aristov07,rozhkov08,pereira10} fails entirely---at the conceptual level---in one dimension. The basic question behind this problem is under what conditions the second-order expansion \cite{jauho93,zheng93,kamenev95,flensberg95} of $\rho_{\rm D}$ in powers of $V_{12}$, which constitutes the essence of the orthodox theory, is justified. Clearly, this is correct if drag is sufficiently weak. The question is about {\it how} weak. The answer, as we will demonstrate in the paper, contains a delicate but crucially important point which does not appear to have been generally appreciated in the literature.

\subsection{Kinetic-equation approach vs the orthodox theory}
\label{s1.4}

The Kubo-type formula for the {\it bulk resistivity} \cite{remark24} $\rho_{\rm D}$ derived within the orthodox theory in one dimension reads:
\bea
\rho_{\rm D}&=&{1\over 2e^2n_1n_2T}\int\!{d\omega\over 2\pi} \!\int\!{dq\over 2\pi}\, q^2 |V_{12}(\omega,q)|^2\nonumber\\
&\times&{{\rm Im}\Pi_1(\omega,q)\,{\rm Im}\Pi_2(\omega,q)\over \sinh^2(\omega/2T)}~,
\label{1.3}
\eea
where $\Pi_{1,2} (\omega,q)$ and $n_{1,2}$ are the (retarded) polarization operators and the electron densities, respectively, in wires 1 and 2. The product of the imaginary parts of $\Pi_{1,2} (\omega,q)$ results from the application of the fluctuation-dissipation theorem to the equilibrium dynamical structure factors for density fluctuations $S_1(\omega,q)$ and $S_2(-\omega,-q)$. The legitimacy of the use of the lowest (second) order in $V_{12}$ for $\rho_{\rm D}$ is based on the assumption that the density fluctuations in the active conductor (wire 1) are {\it equilibrium} in the frame moving with the drift velocity $v_d=-j_1/en_1$ (throughout the paper the electron charge is defined as $-e$, i.e., $e>0$). Indeed, as demonstrated in Ref.~\onlinecite{pustilnik03}, the expansion of $S_1(\omega-qv_d,q)$ to first order in $j_1$ in the cross-correlation function of the electric forces in two conductors gives \cite{remark8a} the linear-response dc electric field $E_2=-\rho_{\rm D}j_1$ in wire 2 (for $j_2$ maintained at zero) with $\rho_{\rm D}$ from Eq.~(\ref{1.3}).

However, the assumption about the density fluctuation being equilibrium in the moving frame (``drift ansatz") is by no means innocent: actually, in one dimension, it strongly limits the applicability of Eq.~(\ref{1.3}). If one assumes, for definiteness, that the wires are identical (with the electron density $n$) and represents $\rho_{\rm D}$ as
\begin{equation}
\rho_{\rm D}=m/e^2n\tau_{\rm D}
\label{1.3a}
\end{equation}
by introducing the ``drag rate" $1/\tau_{\rm D}$ which describes momentum exchange between two conductors in the dc limit, the drift ansatz is only legitimate---as will be shown below---if $1/\tau_{\rm D}$ is much smaller than the thermalization rate. \cite{remark8} In quantum wires, thermalization means not only energy relaxation within the same chirality branch but also ``right-left" relaxation. The latter, however, can only occur if backscattering is allowed, so that in the model of Refs.~\onlinecite{pustilnik03,aristov07,rozhkov08,pereira10}, in which there is no backscattering ``by construction", the use of Eq.~(\ref{1.3}) is not legitimate. At this point, one might think that a deviation of the exact shape of the distribution function of electrons in the active wire from equilibrium in the moving frame does not change the result (\ref{1.1}) qualitatively, i.e., only the numerical coefficient in Eq.~(\ref{1.1}) depends on the shape but remains of the order of unity. This is, however, not the case; on the contrary, the shape is exactly such that in the absence of backscattering $\rho_{\rm D}$ is zero [Eq.~(\ref{1.2})].

\subsection{Drift ansatz and the contact drag resistance}
\label{s1.5}

Naively, one might expect that the results of Refs.~\onlinecite{pustilnik03,aristov07,rozhkov08} for drag induced by forward scattering are valid for sufficiently long ballistic wires, namely for wires whose length $L$ is much larger than the characteristic scale of the right-left relaxation. We emphasize, however, that the equilibrium state in the moving frame in the double-wire system {\it cannot} be reached by increasing $L$ beyond this scale (as would be the case \cite{rech09,micklitz10} in a single wire). Nonzero friction prevents this from happening by constantly exciting electron-hole pairs---even in the limit of an infinitely large system size, as follows from our calculation of $\rho_{\rm D}$.

Although we focus in this paper on the calculation of the bulk drag resistivity \cite{remark24} $\rho_{\rm D}$, there is one more point worth noting here: the drag resistance $R_{\rm D}(L)$ of finite-$L$ wires depends on the setup of the contacts between the wires and the leads---and thus is not, generally, expressible solely in terms of $\rho_{\rm D}$. Similar to the resistance of a single wire, one can introduce the bulk drag resistance and the contact drag resistance. The drag resistivity $\rho_{\rm D}$ is then understood as the drag resistance per unit length in infinitely long wires. As such, $\rho_{\rm D}$ describes the bulk properties of the wires, not affected by the contacts, and the emergence of a {\it homogeneous} response to current flow.

The largest spatial scale that determines the characteristic size of the ``contact regions" (inside the wires), within which the distribution function is generically different from that in the bulk, is the right-left thermalization length $l_{\rm b}$. As shown in this paper, $l_{\rm b}$ is, in effect, the elementary scale for the drag problem in one dimension. This means, in particular, that the drag resistance measured in the limit $L\gg l_{\rm b}$ between two points (potential probes) in the bulk, separated by a distance $L_p$, scales linearly with $L_p$ as $R_{\rm D}(L_p)=\rho_{\rm D}L_p$. The total drag resistance $R_{\rm D}(L)$, however, is affected by the relaxation processes that provide matching between our bulk solution and the distribution functions in the leads, thus depending on $L$ in a nonuniversal way.

In fact, the $T^2$ behavior of the drag resistance [Eq.~(\ref{1.1})] is obtainable, perhaps counterintuitively, in the limit of small $L$, where $R_{\rm D}(L)$ is given by the contact resistance. It is important here that Coulomb drag crucially depends, as demonstrated in this paper, on the relative strength of two types of relaxation processes that differ in whether they lead to thermal equilibration of the {\it difference} of the distribution functions in two wires in the moving or stationary frame. The former tend to establish much stronger drag. As will be shown below, interwire pair collisions tend to equilibrate the difference of the distribution functions in the {\it stationary} frame, in contrast to the drift ansatz that leads to Eq.~(\ref{1.1}). \cite{remark27}  In the limit of small $L$, however, one can---in principle---create the distribution function in the active wire in the form of the drift ansatz by ``preparing" it in this form in the leads, where interwire interactions are absent. \cite{remark25}

If $L$ is so small that the distribution function in the active wire is only slightly modified by interwire interactions, the friction force can be calculated perturbatively (similar to Ref.~\onlinecite{remark1} for the case of backscattering). For the drift-ansatz distribution function ``incident" on the wire from the leads, the drag resistance is then given by $R_{\rm D}(L)\sim (2\pi/e^2)L/l_{\rm f}$, where $l_{\rm f}\sim (1/\beta^2_{\rm f}k_F)(\epsilon_F/T)^2$ (for $T\ll v_F/a$, and independent of $T$ for higher $T$), in accordance with Eq.~(\ref{1.1}). \cite{remark25,remark30} The spatial scale $l_{\rm f}$ characterizes interwire momentum exchange between particles of the same chirality in the drift-ansatz solution in the limit of small momentum transfer. \cite{remark29} Note, however, that if the incident distribution function is equilibrium in the stationary frame (Landauer-B\"uttiker setup with ``Fermi leads"), the perturbative drag resistance in the limit of small $L$ is exponentially suppressed (in the parameter $\epsilon_F/T$). This follows directly from Eqs.~(\ref{b5}) and (\ref{b6}) or, equivalently, from a golden-rule calculation \cite{remark32,remark35} of the interwire momentum transfer rate expressed in terms of the dynamical structure factors. The comparison of the perturbative results in the above two setups emphasizes the nonuniversality of the contact drag resistance.

Thus, forward scattering {\it can} contribute to the contact drag resistance if there is a mismatch between the distribution function incident from the leads and the distribution function that describes bulk drag in the limit of large $L$. The mismatch in the case of pair collisions is minimized if parallel wires are directly connected to the Fermi leads. The full description of the contact drag resistance (also including triple collisions \cite{remark27}) as a function of $L$, depending on the setup, is out of the scope of this paper.

\subsection{Outline of the results}
\label{s1.6}

Our main results can be described as follows. We demonstrate that $\rho_{\rm D}$ in one-dimensional geometry vanishes in the case of electrons not changing their chirality in scattering processes. A key consequence of this is that the drag resistivity is necessarily suppressed compared to the conventional theories [epitomized by Eq.~(\ref{1.3})] if the right-left equilibration is not fast enough. In the case of energy relaxation being mainly due to processes with momentum transfer much smaller than $k_F$ (the exact condition depends on $T$), the right-left equilibration is ``bottlenecked" by inelastic collisions that involve cold electrons near the bottom of the electron spectrum. Hence $\rho_{\rm D}$ acquires the activation factor $\exp (-\epsilon_F/T)$ in the low-$T$ limit. The temperature dependence of $\rho_{\rm D}$ is shown in Fig.~\ref{f2} in Sec.~\ref{s3b}. Remarkably, the drag resistivity in the activation regime does not depend on the distance between the wires.

The power-law behavior of $\rho_{\rm D}$ with $T$ that follows from the conventional approaches \cite{pustilnik03,aristov07,rozhkov08} is only recovered if the drag rate $1/\tau_{\rm D}$ [Eq.~(\ref{1.3a})] resulting from the drift ansatz in these approaches is smaller than the equilibration rate, proportional to $\exp (-\epsilon_F/T)$ in the case of soft collisions. At low $T\ll\epsilon_F$, this can only occur if the distance $a$ between the wires is exponentially large in the parameter $\epsilon_F/T$. This answers the question formulated below Eq.~(\ref{1.2}): the orthodox theory for the drag resistivity is only justified when drag is exponentially weak in $\epsilon_F/T$. Conversely, for fixed $a$, the range of applicability of the orthodox theory is limited to temperatures which are only ``logarithmically smaller" than $\epsilon_F$.

On a more detailed note, our results show a nontrivial interplay between triple and pair collisions. The activation behavior $\rho_{\rm D}\propto\exp (-\epsilon_F/T)$ is determined by triple collisions with one electron scattered near the bottom of the conduction band and two electrons scattered near the Fermi level. If the {\it intrawire} triple collisions are less effective in the right-left equilibration than {\it interwire} pair collisions between two cold electrons, there exists a range of $T$ in which $\rho_{\rm D}$ acquires one more activation factor and behaves as $\exp (-2\epsilon_F/T)$, crossing over into the regime dominated by three-electron scattering as $T$ decreases. However, in any case, $\rho_{\rm D}$ is exponentially suppressed at low $T$. One more conclusion that comes from the comparison of the effect that pair and triple collisions have on $\rho_{\rm D}$ is that the orthodox theory of the drag resistivity for the case of forward scattering is totally unjustifiable if only pair collisions are present and hinges entirely on the triple-collision rate being sufficiently high. Schematically, the dependence of $\rho_{\rm D}$ on the rate of three-particle collisions is illustrated in Fig.~\ref{f1} in Sec.~\ref{s3b}.

Our theory of Coulomb drag is built on the quasiclassical kinetic equation approach. Although we will focus most of our attention on scattering with momentum transfer much smaller than $k_F$, this approach allows us to easily incorporate backscattering near the Fermi level as well. Throughout the paper, however, we assume that $T$ is still higher than the crossover temperature below which interlocked charge-density waves \cite{nazarov98,klesse00,ponomarenko00,fuchs05} induced by backscattering are formed. In this paper, we specialize to the case of ballistic quantum wires (no disorder) and spinless electrons.

The paper is organized as follows. Section~\ref{s2} is devoted to Coulomb drag due to pair collisions. In Sec.~\ref{s2a}, we introduce the kinetic equation for a double wire and obtain the high-frequency drag resistivity. In Sec.~\ref{s2b}, we formulate and solve a model which contains interwire forward scattering but explicitly forbids backscattering---to show that there is no dc drag resistivity in one dimension without backscattering. In Sec.~\ref{s2c}, we obtain the dc drag resistivity induced by pair collisions with small momentum transfer and demonstrate its activation behavior for low temperatures. In Sec.~\ref{s2d}, we discuss drag resulting from direct backscattering at the Fermi level. Section~\ref{s3} deals with Coulomb drag in the presence of both pair and triple collisions and emphasizes the important role of the latter. In Sec.~\ref{s3a}, we write down the kinetic equation that describes triple collisions in a double wire. In Sec.~\ref{s3a1}, we discuss singularities that arise in the calculation of the three-particle scattering probabilities. In Sec.~\ref{s3b1}, we describe soft triple collisions within the Fokker-Planck approach. In Sec.~\ref{s3b2}, we compare various channels of three-particle scattering in the double-wire system and identify those that are most important for Coulomb drag. In Sec.~\ref{s3b}, we consider the effect of triple collisions on Coulomb drag induced by soft pair collisions and show that three-particle scattering dramatically enhances drag at low temperature. Our results are summarized in Sec.~\ref{s4}. Some of the technical details are placed in the appendices.

\section{Coulomb drag in one dimension: Pair collisions}
\label{s2}

\setcounter{equation}{0}

\subsection{Kinetic equation for pair collisions}
\label{s2a}

Our point of departure is the kinetic equation for {\it pair} collisions in a system of two spatially homogeneous quantum wires. In one dimension and for the quadratic dispersion relation, this type of scattering does not affect the distribution function if both electrons are in the same wire---but does lead to a relaxation of the distribution function if electrons are in different wires. Throughout the paper we neglect tunneling between wires, so that the exchange processes for electrons from different wires are absent. We thus have:
\begin{equation}
\partial_tf_\sigma(k_1)-eE_\sigma\partial_{k_1}f_\sigma(k_1)={\rm St}_\sigma\{f\}~,
\label{1}
\end{equation}
where the symbol $\sigma=1,2$ distinguishes wires 1 and 2, $E_\sigma$ is the electric field in wire $\sigma$, and the collision integral ${\rm St}_\sigma\{f\}$ for the case of pair collisions is given [at the lowest (second) order in interaction] for $\sigma=1$ by
\begin{widetext}
\begin{eqnarray}
{\rm St}^{(2)}_1\{f\}&=&(2\pi)^2\!\int\!{dk_{1'}\over 2\pi}\!\int\! {dk_2\over 2\pi}\!\int\! {dk_{2'}\over
2\pi}\,|V(k_{1'}-k_1)|^2\delta(k_1+k_2-k_{1'}-k_{2'})\,\delta(\epsilon_1+\epsilon_2-\epsilon_{1'}-\epsilon_{2'})\nonumber\\
&\times&\left\{f_1(k_{1'})f_2(k_{2'})[1-f_1(k_1)][1-f_2(k_2)]-f_1(k_1)f_2(k_2)[1-f_1(k_{1'})][1-f_2(k_{2'})]\right\}~,
\label{2}
\end{eqnarray}
where $\epsilon_1=k_1^2/2m$, etc., and $V(q)$ is the Fourier component of the interaction potential of electrons in different wires with the momentum transfer $q$ [given, e.g., by $V_{12}(q)$ from Appendix~\ref{aA}: throughout Sec.~\ref{s2} we omit the subscript of $V_{12}$]. The superscript $M=2$ in ${\rm St}^{(M)}_1\{f\}$ in Eq.~(\ref{2}) is meant to indicate that this is a contribution to the collision integral of two-particle collisions ($M$-particle scattering with $M>2$ will be discussed in Sec.~\ref{s3}). For ${\rm St}_2\{f\}$ in Eq.~(\ref{1}), the wire indices of $f_\sigma(k)$ in Eq.~(\ref{2}) should be transposed (momenta $1\leftrightarrow 2,1'\leftrightarrow 2'$). The product of the delta-functions in Eq.~(\ref{2}) reduces in the case of quadratic dispersion to
\begin{equation}
\delta(k_1+k_2-k_{1'}-k_{2'})\,\delta(\epsilon_1+\epsilon_2-\epsilon_{1'}-\epsilon_{2'})={m\over |k_{1'}-k_1|}\,\delta(k_1-k_{2'})\delta(k_2-k_{1'})~, \label{3}
\end{equation}
which gives
\begin{equation}
{\rm St}^{(2)}_1\{f\}=m\int\!{dk_{1'}\over 2\pi}\,{|V(k_{1'}-k_1)|^2\over |k_{1'}-k_1|}\left\{f_1(k_{1'})f_2(k_1)[1-f_1(k_1)][1-f_2(k_{1'})]-f_1(k_1)f_2(k_{1'})[1-f_1(k_{1'})][1-f_2(k_1)]\right\}~.
\label{4}
\end{equation}
\end{widetext}
Below, we will focus on the linear response under the assumption that the wires are identical; in particular, that their chemical potentials and temperature are the same. It is then convenient to represent the solution of Eq.~(\ref{1}) in terms of the functions $g_\sigma(k)$ as
\begin{equation}
f_\sigma(k)=f_T+g_\sigma(k)T\partial_\epsilon f_T~,
\label{5}
\end{equation}
where $f_T=[1+e^{(\epsilon-\epsilon_F)/T}]^{-1}$ is the thermal distribution function. Linearizing Eq.~(\ref{1}) in $g_\sigma$, we then obtain (in the $\omega$ representation):
\begin{eqnarray}
&&-i\omega g_1(k)-{eE_1k\over mT}={\rm st}^{(2)}_1\{g\}~,\nonumber\\ &&-i\omega g_2(k)-{eE_2k\over mT}=-{\rm st}^{(2)}_1\{g\}~,
\label{6}
\end{eqnarray}
where
\begin{eqnarray}
{\rm st}^{(2)}_1\{g\}&=&{m\over 4}\int\!{dk'\over 2\pi}\,\zeta^2(k'){|V(k'-k)|^2\over |k'-k|}\nonumber\\
&\times&[\,g_1(k')+g_2(k)-g_1(k)-g_2(k')\,]
\label{7}
\end{eqnarray}
and
\begin{equation}
\zeta(k)={1\over \cosh[(\epsilon-\epsilon_F)/2T]}~.
\label{8}
\end{equation}
By introducing the functions $g_\pm(k)=[\,g_1(k)\pm g_2(k)\,]/2$, we thus have:
\begin{eqnarray}
&&g_+(k)={e(E_1+E_2)k\over 2mT}{1\over -i\omega+0}~,
\label{9}\\
&&-i\omega g_-(k)-{e(E_1-E_2)k\over 2mT}={\rm st}^{(2)}_-\{g\}~,
\label{10}
\end{eqnarray}
where
\begin{equation}
{\rm st}^{(2)}_-\{g\}={m\over 2}\int\!{dk'\over 2\pi}\,\zeta^2(k'){|V(k'-k)|^2\over |k'-k|}[\,g_-(k')-g_-(k)\,]~.
\label{11}
\end{equation}
The electric current in wire 1 [sign $+$ in Eq.~(\ref{i12})] and wire 2 ($-$) is given in terms of the functions $g_\pm(k)$ by
\begin{equation}
j_{1,2}={e\over 4m}\int_{-\infty}^{\infty}\!\frac{dk}{2\pi}\,\zeta^2(k)\,k\,[\,g_+(k)\pm g_-(k)\,]~.
\label{i12}
\end{equation}

One simple result that follows immediately from Eq.~(\ref{10}) gives the drag conductivity $\sigma_{21}$ (defined as $\sigma_{21}=j_2/E_1$ under the condition that $E_2=0$) in the high-frequency limit. Iterating Eq.~(\ref{10}) in the limit of large $\omega$ once yields ${\rm Re}\,\sigma_{21}\simeq -e^2n/m\omega^2\tau_{\rm D}^\infty$ for $\omega\tau_{\rm D}^\infty\gg 1$, where
\begin{equation}
n={1\over 4mT}\int\!{dk\over 2\pi}\,\zeta^2k^2
\label{19}
\end{equation}
is the electron density in one wire and
\begin{equation}
{1\over\tau_{\rm D}^\infty}={1\over 32nT}\int\!{dk\over 2\pi}\,\zeta^2(k)\!\int\!{dk'\over 2\pi}\,\zeta^2(k')|V(k'-k)|^2|k'-k|~.
\label{11a}
\end{equation}
The sign $\infty$ is intended to emphasize that the scattering rate (\ref{11a}) describes high-frequency drag. \cite{remark17} If the main contribution to $1/\tau^\infty_{\rm D}$ comes from momentum transfers with $|k'-k|\sim T/v_F$, then $1/\tau^\infty_{\rm D}\propto T^2$ [cf.\ Eq.~(\ref{1.1})], while if it comes from backscattering with $|k'-k|\simeq 2k_F$, then $1/\tau^\infty_{\rm D}\propto T$ [cf.\ Eq.~(\ref{1.0})]. Naively, one might think---in the spirit of the Drude theory or, for that matter, the memory-function formalism with the memory function expanded to second order in interaction---that $1/\tau^\infty_{\rm D}$ determines drag also at $\omega\to 0$, with the dc drag resistivity $\rho_{\rm D}\propto1/\tau_{\rm D}^\infty$. As will be seen below, this assumption is correct for drag induced by backscattering in the close vicinity of the Fermi level; however, it is totally wrong for the case of forward scattering.

\subsection{Absence of friction from forward scattering}
\label{s2b}

As mentioned in Sec.~\ref{s1}, dc drag resistivity vanishes [Eq.~(\ref{1.2})] in the absence of interwire backscattering. To see this, consider a model in which the interaction matrix element does not connect electron states with opposite chirality. In this model, backscattering processes both near the Fermi level (momentum transfer about $2k_F$) and near the bottom of the spectrum (momentum transfer much smaller than $k_F$) are forbidden by construction. It is important that the model excludes the latter possibility as well, because the backscattering processes with small-momentum transfer, while being exponentially suppressed for $T\ll\epsilon_F$, still can lead to a ``leakage of current" between the  subsystems of right- and left-moving electrons. For definiteness, let us substitute for $V_{k'-k}$ in Eq.~(\ref{11}) a function of $k$ and $k'$ that is proportional to the $\theta$-function of the product $kk'$, which explicitly forbids backscattering: \cite{remark9}
\begin{equation}
V(k'-k)\to{\cal V}(k'-k)\,\theta(kk')~.
\label{Vtheta}
\end{equation}
The model (\ref{Vtheta}) is compatible with those used for studying Coulomb drag due to forward scattering in Refs.~\onlinecite{pustilnik03,aristov07,rozhkov08}. In the Luttinger-liquid formalism, generalized to the finite-curvature case, it corresponds to retaining only the $g_{4\perp}$ type of interaction. \cite{giamarchi04}

Within the model (\ref{Vtheta}), the equation for the distribution function $g_-(k\!>\!0)$ of right-moving electrons can be written in a {\it closed} form:
\begin{eqnarray}
&&\phantom{a}\hspace{-1.6cm}-i\omega g_-(k)-{e(E_1-E_2)k\over 2mT}\nonumber\\
&&\phantom{a}\hspace{-1.6cm}={m\over 2}\!\int_0^\infty\!\!{dk'\over 2\pi}\,\zeta^2(k'){|{\cal V}(k'-k)|^2\over |k'-k|}[\,g_-(k')-g_-(k)\,]~,
\label{king-}
\end{eqnarray}
while the distribution function of left-moving electrons $g_-(k\!<\!0)$ is related to $g_-(k\!>\!0)$ by the condition $g_-(-k)=-g(k)$ which follows from the fact that the source term in Eq.~(\ref{10}) is odd in $k$. Importantly, the collision integral in Eq.~(\ref{king-}) is nullified if $g_-(k>0)$ does not depend on $k$ [i.e., $g_-(k)={\rm const}(k){\rm sgn}(k)$]. The solution of Eq.~(\ref{king-}) can therefore be represented as a sum of two terms,
\begin{equation}
g_-(k)=h_0+h(k)~,
\label{i3}
\end{equation}
where the zero-mode term $h_0$ does not depend on $k$ and has a pole at $\omega=0$,
\begin{equation}
h_0=\frac{A(\omega)}{-i\omega+0}{e(E_1-E_2)v_F\over 2T}~,
\label{i4}
\end{equation}
with a residue proportional to a yet unknown constant $A(0)$. The equation for $h(k)$ then reads
\begin{eqnarray}
&&\phantom{a}\hspace{-1.5cm}-i\omega h(k)+\left[\,A(\omega)-{k\over k_F}\,\right]{e(E_1-E_2)v_F\over 2T}\nonumber\\
&&\phantom{a}\hspace{-1.5cm}={m\over 2}\int_0^\infty\!{dk'\over 2\pi}\,\zeta^2(k'){|{\cal V}(k'-k)|^2\over |k'-k|}[\,h(k')-h(k)\,]~.
\label{kinh}
\end{eqnarray}
Multiplying Eq.~(\ref{kinh}) by $\zeta^2(k)$ and integrating both sides over $k$, we eliminate the collision integral to obtain the connection between $A(\omega)$ and $h(k)$ in a form that does not contain the collision kernel explicitly:
\begin{equation}
A(\omega)={1\over\pi v_F}{\partial\mu\over\partial n}\left[\,1+{i\pi\omega\over e(E_1-E_2)}\int_0^\infty\!{dk\over 2\pi}\,\zeta^2h\,\right]~,
\label{i5}
\end{equation}
where $\partial\mu/\partial n=2T/\int_0^\infty\!(dk/2\pi)\zeta^2$ is the inverse compressibility. For $T\ll\epsilon_F$,
\begin{equation}
{1\over\pi v_F}\,{\partial\mu\over\partial n}\simeq 1-{\pi^2\over 8}\left({T\over\epsilon_F}\right)^2~.
\label{compress}
\end{equation}
The difference of the compressibility from $1/\pi v_F$ at finite $T$ will be important for the calculation of the singular (at $\omega\to 0$) part of $\sigma_{21}$.

From Eq.~(\ref{i5}), the closed equation for $h$ is written as
\begin{eqnarray}
&&\phantom{a}\hspace{-1.5cm}-i\omega [\,h(k)-\bar{h}\,]-{e(E_1-E_2)\over 2mT}(k-\bar{k})\nonumber\\
&&\phantom{a}\hspace{-1.5cm}={m\over 2}\int_0^\infty\!{dk'\over 2\pi}\,\zeta^2(k'){|{\cal V}(k'-k)|^2\over |k'-k|}[\,h(k')-h(k)\,]~,
\label{kinh1}
\end{eqnarray}
where
\begin{equation}
\bar{h}={1\over 2T}{\partial\mu\over\partial n}\int_0^\infty\!{dk\over 2\pi}\,\zeta^2h~,\quad\bar{k}={m\over\pi}{\partial\mu\over\partial n}~.
\end{equation}
The solution of Eq.~(\ref{kinh1}) does not contain, by construction, a zero-mode part and is regular at $\omega\to 0$. It follows from Eq.~(\ref{i5}), then, that $A(\omega\to 0)$ is finite (neither diverging nor vanishing) and determined by the first term in the square brackets in Eq.~(\ref{i5}), namely $A(0)=(1/\pi v_F)\partial\mu/\partial n$.
The solution of Eq.~(\ref{king-}) can thus be represented as a sum of the singular (at $\omega=0$) term and the regular term as follows:
\begin{equation}
g_-(k)=\left[\,{A(\omega)\over -i\omega+0}+{B(\omega,k)\over -i\omega+M(\omega,k)}\,\right]{e(E_1-E_2)v_F\over 2T}~,
\label{i9}
\end{equation}
where $A(\omega)$, the ``source renormalization" factor $B(\omega,k)$, and the ``memory function" $M(\omega,k)$ are all regular at $\omega\to 0$, and $M(0,k)>0$. \cite{remark10} This form of $g_-$, together with Eq.~(\ref{9}) for $g_+$, dictates a very special type of behavior of the conductivity and resistivity tensors (in the space of the wire indices) as $\omega\to 0$, as is seen below.

Using the relation (\ref{i5}) between the singular $(h_0)$ and regular $(h)$ parts of $g_-$, the current [Eq.~(\ref{i12})] can be expressed in terms of only the regular part as
\begin{widetext}
\begin{equation}
j_{1,2}=\frac{e^2}{4m^2 T}{1\over -i\omega +0}\int_0^\infty\!{dk\over 2\pi}\,\zeta^2k\left[\,(E_1+E_2)k\pm (E_1-E_2)\bar{k}\,\right]\pm\frac{e}{2m}\int_0^\infty\!{dk\over 2\pi}\,\zeta^2k\,(h-\bar{h})~.
\label{i13}
\end{equation}
The conductivity matrix resulting from Eq.~(\ref{i13}) reads
\begin{equation}
\hat\sigma(\omega)=
\frac{e^2 v_F}{\pi}\left[\,\frac{1}{-i\omega+0}\left(\begin{array}{cc}
\lambda_1+\lambda_2  & \lambda_1-\lambda_2 \\
\lambda_1-\lambda_2 & \lambda_1+\lambda_2 \\
\end{array}\right)+
C\left(\omega \right) \left(\begin{array}{rr}
1 & -1 \\
-1 & 1 \\
\end{array}\right)\,\right]~,
\label{i14a}
\end{equation}
where
\begin{equation}
\lambda_1={\pi n\over 2k_F}~,\qquad\lambda_2={1\over 2\pi v_F}{\partial\mu\over\partial n}~,
\label{i14b}
\end{equation}
and
\begin{equation}
C(\omega)={\pi\over 16mT^2}\,{\partial\mu\over\partial n}\int_0^\infty\!{dk\over 2\pi}\,\zeta^2(k)\!\int_0^\infty\!{dk'\over 2\pi}\,\zeta^2(k')\,(k-k')
\left[\,\frac{B(\omega,k)}{-i \omega +M(\omega,k)}-\frac{B(\omega,k')}{-i \omega + M(\omega,k')}\,\right]~.
\label{i15}
\end{equation}
\end{widetext}
For $T\ll\epsilon_F$, the coefficients $\lambda_{1,2}$ are given by [cf.\ Eq.~(\ref{compress})]
\begin{equation}
\lambda_1\simeq\frac{1}{2}-{\pi^2\over 48}\left({T\over\epsilon_F}\right)^2~,\quad
\lambda_2\simeq\frac{1}{2}-{\pi^2\over 16}\left({T\over\epsilon_F}\right)^2~,
\label{i14c}
\end{equation}
i.e., the diagonal elements of the first matrix in Eq.~(\ref{i14a}) are close to unity in the limit of small $T$, whereas the nondiagonal ones vanish as $T^2$. Note that the singular behavior of the nondiagonal elements is determined by the $T$ dependent corrections to the coefficients $\lambda_{1,2}$ in Eq.~(\ref{i14b}). The function $C(\omega)$ in front of the second matrix is also proportional to $T^2$ at $T\to 0$. Indeed, the integrals over $k$ and $k'$ in Eq.~(\ref{i15}) are determined [because of the factors $\zeta(k)$ and $\zeta(k')$] by the close vicinity of $k=k'=k_F$, while the integrand contains a product of two factors each of which is zero at $k=k'$. The vanishing of $C(\omega)$ at $T\to 0$ can also be seen from the sum rule for the conductivity (see, e.g., Ref.~\onlinecite{aristov07}), according to which
\begin{equation}
\int_{-\infty}^{\infty}\! {d\omega\over 2\pi}\,{\rm Re}\,\hat\sigma(\omega)={e^2n\over 2m}\left(\begin{array}{cc}
1 & 0 \\
0 & 1 \\
\end{array}\right)
\label{sum}
\end{equation}
independently of the strength of interaction. Equations (\ref{i14a}) and (\ref{sum}), combined together, tell us that
\begin{equation}
\int_{-\infty}^{\infty}\!{d\omega\over 2\pi}\,{\rm Re}\, C(\omega)={1\over 2}(\lambda_1-\lambda_2)~,
\label{sumC}
\end{equation}
which, in view of Eq.~(\ref{i14c}), means the $T^2$ behavior also for the integral characteristic of the regular part of $\hat\sigma (\omega)$.

It is instructive to represent the conductivity matrix for $T\ll\epsilon_F$ as
\begin{eqnarray}
\hat\sigma(\omega)&\simeq& \frac{e^2 v_F}{\pi}\,\left[\,{1-\eta\over 2}\frac{\hat\Sigma_1}{-i\omega+\delta_1}\right.\nonumber\\&+&\left.{1-3\eta\over 2}\frac{\hat\Sigma_2}{-i\omega+\delta_2}+C(\omega)\hat\Sigma_2~\,\right]~,
\label{i16}
\end{eqnarray}
where
\begin{equation}
\eta={\pi^2\over 24}\left({T\over\epsilon_F}\right)^2~,
\end{equation}
the matrices $\hat\Sigma_{1,2}$ are given by
\begin{equation}
\hat\Sigma_1=\left(\begin{array}{cc}
1 & 1 \\
1 & 1 \\
\end{array}\right)~,
\quad\hat\Sigma_2=\left(\begin{array}{rr}
1 & -1 \\
-1 & 1 \\
\end{array}\right)~,
\label{i17}
\end{equation}
and the infinitesimally small frequency shifts $i\delta_{1}$ and $i\delta_2$ in the singular terms proportional to $\hat\Sigma_1$ and $\hat\Sigma_2$, respectively, are denoted differently to emphasize the different origin of possible damping in the two terms. Specifically, the singular term proportional to $\Sigma_1$ comes from the symmetric (in the wire indices) function $g_+$ whose singularity is associated with total-momentum conservation. Hence $\delta_1=0$ in homogeneous wires, independently of the type and strength of electron-electron interaction. In contrast, the singular term proportional to $\Sigma_2$ stems from the zero-mode function $h_0$ whose singularity reflects particle number conservation within each chirality. That is, $\delta_2=0$ in the model of only forward electron-electron scattering. The last (nonsingular) term in Eq.~(\ref{i16}) is the contribution of both $h$ and the regular part of $h_0$ [the last term in Eq.~(\ref{i13})]: its damping is related to the equilibration between electrons of the same chirality in different wires.

The matrix structure of Eq.~(\ref{i16}) with $\delta_1=\delta_2=0$ differs in an essential way from that proposed for the same case of forward electron-electron scattering in Ref.~\onlinecite{aristov07}. The crucial difference is that the prefactor of $\hat\Sigma_2$ in Eq.~(\ref{i16}) is singular at $\omega=0$, i.e., behaves in the limit of small $\omega$ as $1/(-i\omega + \delta_2)$ with $\delta_2=0$, whereas in Ref.~\onlinecite{aristov07} it is proportional to $1/(-i\omega + 2/\tau_{\rm D}^\infty)$, where the scattering rate $1/\tau_{\rm D}^\infty$, describing high-frequency drag, is given by Eq.~(\ref{11a}). Inversion of the conductivity matrix in Ref.~\onlinecite{aristov07} yielded a nonzero dc drag resistivity $\rho_{\rm D}=\pi/e^2v_F\tau_{\rm D}^\infty$, i.e., $1/\tau_{\rm D}$ [Eq.~(\ref{1.3a})] equal to $1/\tau_{\rm D}^\infty$, which also agrees with the result of Refs.~\onlinecite{pustilnik03,rozhkov08}. In contrast, the inverse of the matrix $\hat\sigma (\omega)$ from Eq.~(\ref{i16}) is
\begin{eqnarray}
&&\phantom{a}\hspace{-1.0cm}{\hat\rho}(\omega)=\frac{\pi}{e^2 v_F} \frac{i \omega}{(1-\eta)(1-3\eta -2 i\omega C)}\nonumber\\
&&\phantom{a}\hspace{-1.0cm}\times\left(
\begin{array}{cc}
-1+2\eta+i \omega C & \eta+i\omega C \\
\eta+i\omega C & -1+2\eta+i \omega C \\
\end{array}\right)~,
\label{i18}
\end{eqnarray}
which at $\omega\to 0$ gives
\begin{equation}
{\rm Re}\,\rho_{21}\propto\omega^2~.
\label{i19}
\end{equation}
That is, in the model of only forward scattering the dc drag resistivity
\begin{equation}
\rho_{\rm D}=-\rho_{21}(\omega=0)
\label{19a}
\end{equation}
is strictly zero [Eq.~(\ref{1.2})]. Note that the diagonal dissipative resistivity ${\rm Re}\,\rho_{11}$ also vanishes with decreasing $\omega$ as $\omega^2$, similar to ${\rm Re}\,\rho_{21}$. The coefficient in front of $\omega^2$ is, in both cases, proportional to $C(0)$ which scales as $T^2$ in the low-$T$ limit.

It is worth mentioning that nonzero, in contrast to the solution of the kinetic equation, drag in the model of forward scattering was obtained in Refs.~\onlinecite{pustilnik03,aristov07,rozhkov08} in two ways. In Ref.~\onlinecite{aristov07}, $\rho_{\rm D}\neq 0$ was found as a direct consequence of the conjectured Lorentzian shape of the $\omega$ dependence of $\sigma_{21}$. On the other hand, in Refs.~\onlinecite{pustilnik03,rozhkov08}, the same expression for the drag resistivity followed from the one-dimensional version of the orthodox theory \cite{jauho93,zheng93,kamenev95,flensberg95} at $\omega=0$. In particular, in Ref.~\onlinecite{pustilnik03} the orthodox theory was cast in the form of the drift ansatz. The relation between the two approaches and the step in the solution of the kinetic equation at which the drift ansatz fails are further discussed in Appendix \ref{aB}.

Coulomb drag in the dc limit would only occur if $\delta_2\neq 0$ in Eq.~(\ref{i16}), namely
\begin{equation}
\hat\rho(\omega=0)={\pi\over 4e^2v_F}\,{\delta_2\over C(0)\delta_2+\lambda_2}\,\hat\Sigma_2~.
\label{i20}
\end{equation}
We thus see that the scattering processes that change the chirality of electrons---recall that it is these processes that yield $\delta_2\neq 0$---are the only source of nonzero dc Coulomb drag. As already discussed in Sec.~\ref{s1}, one can distinguish two main types of backscattering: in the vicinity of the Fermi level and at the bottom of the spectrum. The contribution of the former to $\rho_{\rm D}$ is exponentially suppressed, as $\exp (-4k_Fa)$, if the distance $a$ between the wires is much larger than the Fermi wavelength. The contribution of the latter is also exponentially suppressed, as $\exp (-\epsilon_F/T)$ [or $\exp (-2\epsilon_F/T)$, depending on the parameters], if $T$ is much smaller than the Fermi energy. It follows that the important parameter that controls the relative weight of these two types of backscattering in $\rho_{\rm D}$ is the ratio of $a$ and the ``thermal length" $v_F/T$. For $a\gg v_F/T$, backscattering with momentum transfer much smaller than $k_F$ is expected to provide the main contribution to $\rho_{\rm D}$. This type of backscattering is discussed in Sec.~\ref{s2c} below.

\vspace{5mm}
\subsection{Coulomb drag due to soft pair collisions: Fokker-Planck description}
\label{s2c}

Let us consider the limit in which the characteristic momentum transfer in Eq.~(\ref{11}) is much smaller than $T/v_F$. For concreteness, we can think of the interaction potential given by Eq.~(\ref{a1}) and simplify Eq.~(\ref{11}) in the limit $T/v_F\gg |k'-k|\sim 1/a$ [see Eq.~(\ref{a3})]. In this limit, the collision integral (\ref{11}) can be written in the differential form:
\begin{eqnarray}
{\rm st}^{(2)}_-\{g\}&\simeq& {mc\over 2}\left({\partial \zeta^2\over \partial k}\,{\partial g_-\over \partial k}+{1\over 2}\zeta^2\,{\partial^2g_-\over \partial
k^2}\right)\nonumber\\
&=&{mc\over 4}\,{1\over\zeta^2}\,{\partial\over\partial k}\left(\zeta^4\,{\partial g_-\over\partial k}\right)~,
\label{12}
\end{eqnarray}
where
\begin{equation}
c=\int\!{dq\over 2\pi}\,|q||V(q)|^2~. \label{13}
\end{equation}
In the limit $Ta/v_F\gg 1$, the scattering rate $1/\tau_{\rm D}^\infty$ [Eq.~(\ref{11a})], which describes high-frequency drag, and $c$ are related to each other as follows:
\begin{equation}
c=24{\epsilon_F\over\tau_{\rm D}^\infty}~.
\label{13a}
\end{equation}
Being rewritten in terms of the function
\begin{equation}
f_-=(f_1-f_2)/2=g_-T\partial_\epsilon f_T=-g_-\zeta^2/4
\label{13b}
\end{equation}
(i.e., going back from the ``$g$-functions" to the distribution functions $f_\sigma$), Eq.~(\ref{12}) can be cast in the form of the Fokker-Planck equation:
\begin{equation}
-i\omega f_-+{e(E_1-E_2)\zeta^2k\over 8mT}=-{\partial J^{(2)}\over \partial k}~,
\label{14}
\end{equation}
where the current in momentum space $J^{(2)}(k)$ [related to $f_-(k)$ and $\partial f_-(k)/\partial k$ locally---at one point $k$] is given by
\begin{equation}
J^{(2)}=-D{\partial f_-\over \partial k}+f_-{\partial D\over \partial k}
\label{15}
\end{equation}
with the $k$-dependent diffusion coefficient in momentum space \cite{remark15}
\begin{equation}
D(k)={mc\over 4}\,\zeta^2(k)~. \label{16}
\end{equation}

The solution of Eq.~(\ref{14}) in the dc limit can be found exactly. At $\omega=0$, $J^{(2)}(k)$ is obtained by integrating Eq.~(\ref{14}) [assuming that $\lim_{\omega\to 0}(\omega f_-)=0$, which will be confirmed by the solution]:
\begin{widetext}
\begin{equation}
J^{(2)}(k)=-{e(E_1-E_2)\over 8mT}\int_{-\infty}^k\!\!dp\,p\zeta^2(p)={1\over 4}e(E_1-E_2)\left[\,1-\tanh\left({k^2-k_F^2\over 4mT}\right)\,\right]~.
\label{20}
\end{equation}
The boundary condition used in Eq.~(\ref{20}) is
$J^{(2)}(k\to\pm\infty)=0$. Substituting Eq.~(\ref{20}) in Eq.~(\ref{15}) yields a first-order equation for $f_-(k)$:
\begin{equation}
{\partial f_-\over\partial k}+{k\over mT}\tanh\left({k^2-k_F^2\over 4mT}\right)f_-=-{e(E_1-E_2)\over mc}\,{1\over \zeta^2}\left[\,1-\tanh \left({k^2-k_F^2\over4mT}\right)\,\right]~,
\label{21}
\end{equation}
which should be solved for the boundary condition $f_-(k=0)=0$. The solution reads:
\begin{eqnarray}
f_-&=&-{e(E_1-E_2)\over mc}\,\zeta^2\int_0^k\!dp\left[\,1-\tanh \left({p^2-k_F^2\over 4mT}\right)\,\right]\,{1\over \zeta^4(p)} \label{22}\\ &=&-{e(E_1-E_2)\over mc}\,\zeta^2\int_0^k\!dp\,\exp \left(-{p^2-k_F^2\over 4mT}\right){1\over \zeta^3(p)}~.
\label{23}
\end{eqnarray}
\end{widetext}

Using the parameter $T/\epsilon_F\ll 1$, Eq.~(\ref{23}) can be simplified to
\begin{equation}
f_-\simeq -{e(E_1-E_2)\over 16mc}\left(\pi mT\right)^{1/2}e^{2\epsilon_F/T}\,\zeta^2 \,\Phi\left({k\over \sqrt{mT}}\right)~,
\label{24}
\end{equation}
where $\Phi(x)=(2/\sqrt{\pi})\int_0^x\exp (-t^2)dt$ is the error function. Equation (\ref{24}) is the asymptotically exact expression valid for not too large energies
$\epsilon<\epsilon^*$ (more accurately, for $\epsilon^*-\epsilon\gg T$), where
\begin{equation}
\epsilon^*=3\epsilon_F+{T\over 2}\ln{\epsilon_F\over T}~.
\label{25}
\end{equation}
For larger energies $\epsilon-\epsilon^*\gg T$, it follows from Eq.~(\ref{23}) that $f_-$ falls off as a power law: \cite{remark11}
\begin{equation}
f_-\simeq -{e(E_1-E_2)\over 2c}{T\over k}~.
\label{26}
\end{equation}
Specifically, for all energies $\epsilon-\epsilon_F\gg T$, $f_-$ is given by the sum of two contributions to the integral (\ref{23}) coming from $|q|$ of order $(mT)^{1/2}$ and
from $|q|\simeq (2m\epsilon)^{1/2}$:
\begin{eqnarray}
f_-&\simeq&-{e(E_1-E_2)\over 4}\left({\pi T\over mc^2}\right)^{1/2}\nonumber\\
&\times&\left[\,\exp\left({3k_F^2-k^2\over 2mT}\right){\rm sgn}(k)+{2(mT/\pi)^{1/2}\over k}\,\right]~.\nonumber\\
\label{27}
\end{eqnarray}
The ranges of applicability of Eqs.~(\ref{24}) and (\ref{27}) overlap. For all energies $T\ll\epsilon <\epsilon^*$ (which includes momenta around the peaks of $f_-$ at $k=\pm k_F$), the shape of $f_-$ as a function of $k$ is given \cite{remark12} simply by $\zeta^2$:
\begin{equation}
f_-\simeq -{e(E_1-E_2)\over 16mc}\left(\pi mT\right)^{1/2}e^{2\epsilon_F/T}\zeta^2\,{\rm sgn}(k)~.
\label{28}
\end{equation}

The electric current
\begin{equation}
j_-={1\over 2}(j_1-j_2)=-{e\over m}\int\!{dk\over 2\pi}\,kf_-~,
\label{17}
\end{equation}
calculated by integrating Eq.~(\ref{28}), reads
\begin{equation}
j_-\simeq {e^2(E_1-E_2)n\over 2m\gamma}~,
\label{31}
\end{equation}
where
\begin{equation}
\gamma=2c\left({2\epsilon_F\over \pi T^3}\right)^{1/2}e^{-2\epsilon_F/T}~,
\label{29}
\end{equation}
i.e., the difference of the dc conductivities $\sigma_{11}-\sigma_{21}=2j_-/(E_1-E_2)\simeq e^2v_F/\pi\gamma$. Taking into account that the sum $\sigma_{11}+\sigma_{21}=e^2v_F/\pi(-i\omega+0)$ [as it follows from Eq.~(\ref{9})] and inverting the conductivity matrix, we obtain the dc resistivity matrix in the form
\begin{equation}
\hat\rho(\omega=0)=\rho_{\rm D}\hat\Sigma_2~,
\label{29a}
\end{equation}
where the matrix $\hat\Sigma_2$ is given by Eq.~(\ref{i17}), i.e.,
\begin{equation}
\rho_{11}=-\rho_{21}
\label{29c}
\end{equation}
in the dc limit. Equation~(\ref{29a}) yields the following expression \cite{remark33} for the dc drag resistivity $\rho_{\rm D}$ [Eq.~(\ref{19a})]:
\begin{equation}
\rho_{\rm D}={E_1-E_2\over 4j_-}~.
\label{29b}
\end{equation}
From Eq.~(\ref{31}) we thus have
\begin{equation}
\rho_{\rm D}\simeq {\pi\gamma\over 2e^2v_F}={\pi c\over e^2v_F}\left({2\epsilon_F\over \pi T^3}\right)^{1/2}e^{-2\epsilon_F/T}~.
\label{32}
\end{equation}
\begin{figure}
\centerline{\includegraphics[width=0.95\columnwidth]{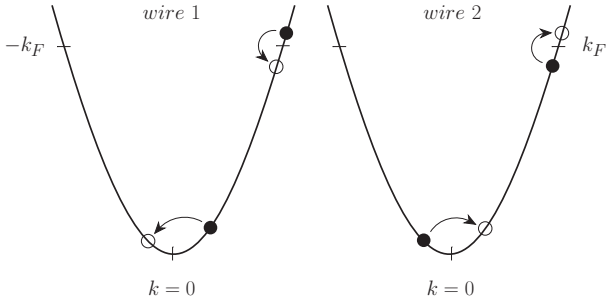}}
\caption{Electron-electron scattering with small momentum transfer, much smaller than the Fermi momentum $k_F$, in a double-wire system. Electrons and holes are shown on the parabolic dispersion curves as full and empty circles, respectively. Scattering in the vicinity of the Fermi level does not contribute to the bulk drag resistivity $\rho_{\rm D}$ in the dc limit---only scattering that changes the chirality of electrons does. In the limit of soft scattering, $\rho_{\rm D}$ is determined by scattering processes involving cold electrons near the bottom of the spectrum at $k=0$.
}
\label{forback}
\end{figure}
Equation (\ref{32}) is in agreement with the conclusion of Sec.~\ref{s2b} that backscattering (Fig.~\ref{forback}) is the only source of dc Coulomb drag (and should be contrasted with the result of the orthodox theory \cite{pustilnik03,aristov07,rozhkov08} that yields nonzero $\rho_{\rm D}$ from forward scattering in the absence of any backscattering). The scattering rate $1/\tau_{\rm D}$ [Eq.~(\ref{1.3a})], which describes drag in the dc limit, is seen to be given by $\gamma/2$. This means that $1/\tau_{\rm D}$ is much smaller, for $T\ll\epsilon_F$, than the scattering rate $1/\tau_{\rm D}^\infty$ [Eqs.~(\ref{11a}),(\ref{13a})] describing high-frequency drag:
\begin{equation}
{\tau_{\rm D}^\infty\over\tau_{\rm D}}\simeq 24\left({2\over\pi}\right)^{1/2}\left({\epsilon_F\over T}\right)^{3/2}e^{-2\epsilon_F/T}~.
\label{32a}
\end{equation}
Importantly, the ratio (\ref{32a}) does not depend on the strength of interaction, with both scattering rates being quadratic in $V_{12}$.

\begin{figure}
\centerline{\includegraphics[width=0.8\columnwidth]{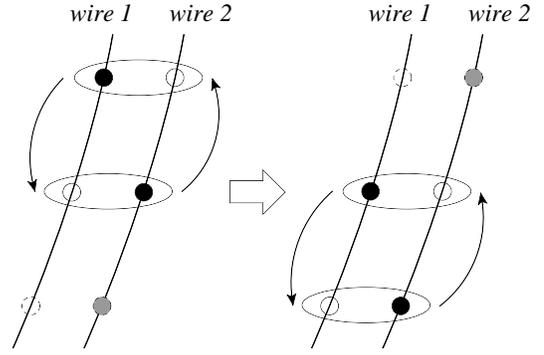}}
\caption{Diffusion in energy space due to two-particle scattering in a double-wire system. An electron (full circle) in one wire and a hole (empty circle) in the other, both having the same momentum, move as a whole along the dispersion curve (the same for wire 1 and wire 2, offset horizontally for clarity). Two consecutive steps in the diffusion process are shown, resulting in the cooling of the electron in wire 1 due to the heating of electrons in wire 2.
The electron and hole states that have not yet (left) or have already (right) participated in scattering are shown as dashed circles. Diffusion of the electron-hole pair along the dispersion curve between two Fermi points through the bottom of the spectrum (Fig.~\ref{forback}) amounts to effective backscattering at the Fermi level.
}
\label{twopart}
\end{figure}

It may be instructive to discuss the origin of the $T$ dependence in Eq.~(\ref{32}) in more detail. The factor $\exp (-2\epsilon_F/T)$ means that the relaxation of $f_-$ at the Fermi level in the dc limit is only due to the diffusion in energy space which leads to the cooling of an electron in wire 1---starting from the Fermi surface down to the very bottom at $k=0$---due to the heating of electrons in wire 2 (Fig.~\ref{twopart}), followed by backscattering of the electron at the bottom and its acceleration in the opposite direction, accompanied by the cooling of electrons in wire 2. This diffusion cycle, which amounts to effective backscattering at the Fermi level, is bottlenecked by electron-electron scattering at $k=0$ (requiring two holes, one in each of the wires, near the bottom)---hence the factor $\exp (-2\epsilon_F/T)$.

The above picture is also substantiated by the obtained $k$ dependence of the distribution function. Counting the exponential factors in Eq.~(\ref{24}), we observe that $\partial f_-/\partial k$ at $k=0$ is proportional to $\exp (\epsilon_F/T)$ and $f_-$ grows with increasing $|k|$ until it reaches maximum at $|k|\simeq k_F$, at which point it is proportional to $\exp (2\epsilon_F/T)$. This behavior of $f_-$ means that the characteristic relaxation rate for $f_-$ at given $k$ is small as $\exp (-2\epsilon_F/T)$ for {\it all} momenta both near the Fermi level and below it down to the bottom of the spectrum, i.e., the exponential factor in the relaxation rate does not depend on $k$. Moreover, Eq.~(\ref{28}) says that the relaxation rate at $\omega=0$, including the pre-exponential factor, is accurately approximated in the vicinity of the Fermi level for $|k-k_F|\ll k_F$ by a $k$-independent constant. This constant is precisely $\gamma$ given by Eq.~(\ref{29}). For $|k-k_F|\ll k_F$, the r.h.s.\ of the kinetic equation (\ref{14}) at $\omega=0$ can thus be written as $-\gamma f_-$, with the relaxation rate being independent of $k$ and determined by the slowest scattering processes at the very bottom.

It is worth noting that the condition $T\gg v_F/a$, which is necessary for the justification of the Fokker-Planck description [Eq.~(\ref{12})] in the {\it whole} range of momenta $|k|\alt k_F$, is not necessary for scattering within the same chiral branch to preserve its diffusive character in energy space for cold particles with $|k|\ll mTa$. If $T\ll v_F/a$, forward scattering with $|k|\ll mTa$ is still described by Eq.~(\ref{12}). In contrast, in the range $mTa\ll |k|<k_F$ it is modified in an essential way by strong asymmetry between hopping up and hopping down along the energy axis. Specifically, for a particle in this range of $k$, the characteristic probability of gaining energy in an elementary hop is much larger than the probability of losing energy. The asymmetry factor depends on how fast $|V^2(q)|$ falls off with increasing $|q|$ for $|q|a\gg 1$ compared to the growth of the thermal factor $\exp [(k+q)^2-k^2]/2mT$. For the case of $V(q)$ from Eq.~(\ref{a3}), the asymmetry factor for $Ta/v_F\ll 1$ is mainly given \cite{remark23} by $\exp [(k_F^2-k^2)/2mT]\gg 1$. The right-left relaxation via multiple scattering with small momentum transfer is hindered by the asymmetric hopping. However, for $T\ll v_F/a$, direct backscattering with momentum transfer $2k_F$ becomes important, as will be discussed in Sec.~\ref{s2d}.

\subsection{Coulomb drag due to backscattering at the Fermi level}
\label{s2d}

In Sec.~\ref{s2c}, we have calculated the contribution to $\rho_{\rm D}$ [Eq.~(\ref{32})] that comes for $T\gg v_F/a$ from pair collisions with momentum transfer much smaller than $k_F$. Friction from this type of scattering has been shown to be mediated by {\it effective} backscattering at the Fermi level, where ``effective" means that backscattering is a result of the diffusion in energy space through the bottom of the spectrum. Let us now turn to the contribution to $\rho_{\rm D}$ from backscattering at the Fermi level with the momentum $2k_F$ transferred in one transition. For $T\ll v_F/a$, this can be calculated from Eq.~(\ref{10}) straightforwardly by removing the ratio $|V(k'-k)|^2/|k'-k|$ from under the integral sign in Eq.~(\ref{11}) and substituting $|V(2k_F)|^2/2k_F$ for it. The collision integral (\ref{11}) reduces then to the out-scattering term (``relaxation time approximation"), which gives
\begin{equation}
g_-={e(E_1-E_2)k\over 2mT(-i\omega+\gamma_{\rm b})}
\label{32b}
\end{equation}
with
\begin{equation}
\gamma_{\rm b}=8\pi\beta_{\rm b}^2T~,
\label{32c}
\end{equation}
where $\beta_{\rm b}=V(2k_F)/2\pi v_F$ is the dimensionless amplitude of backscattering. The simple Lorentzian for the $\omega$ dependence of $g_-$ in Eq.~(\ref{32b}) means that the high-frequency and dc drag rates for the case of backscattering in the vicinity of the Fermi level coincide: $1/\tau_{\rm D}=1/\tau_{\rm D}^\infty=\gamma_{\rm b}/2$, in stark contrast to drag induced by scattering with small momentum transfer [Eq.~(\ref{32a})]. The result for $\rho_{\rm D}$ reads \cite{hu96}
\begin{equation}
\rho_{\rm D}\simeq {\pi\gamma_{\rm b}\over 2e^2v_F}={4\pi^2\over e^2}\beta_{\rm b}^2{T\over v_F}~.
\label{32d}
\end{equation}

By comparing the contributions to the dc drag resistivity from backscattering at the Fermi level [Eq.~(\ref{32d})] and effective backscattering due to soft collisions [Eq.~(\ref{32})],
\begin{equation}
{\rho_{\rm D}\,\,{\rm [Eq.\,(\ref{32d})]}\over \rho_{\rm D}\,\,{\rm [Eq.\,(\ref{32})]}}\sim \left[{V(2k_F)\over V(1/a)}\right]^2\left({Ta\over v_F}\right)^2\left({T\over\epsilon_F}\right)^{1/2}e^{2\epsilon_F/T}~,
\label{32e}
\end{equation}
one can see that the latter mechanism of backscattering provides more friction that the former for
\cite{remark13}
$T\gg v_F/a$. That is, despite the contribution of soft collisions being strongly suppressed compared to Eq.~(\ref{1.1}), for $k_Fa\gg 1$ there is still a wide range of temperature, $v_F/a\ll T\ll\epsilon_F$, in which soft collisions in a degenerate electron gas are more effective than direct backscattering. It is worth noting that the main contributions to $\rho_{\rm D}$ only come from backscattering at the Fermi level and from backscattering at the very bottom of the spectrum, while backscattering at intermediate energies plays no role. Indeed, the exponential factor $\exp [\,-4|k|a-2(\epsilon_F-\epsilon)/T\,]$ that describes (for $\epsilon<\epsilon_F$) direct backscattering with momentum transfer $2|k|$ is maximized either at $\epsilon=\epsilon_F$ for $T<v_F/4a$ or at $\epsilon=0$ for larger $T$ with a sharp (for $k_Fa\gg 1$) crossover of width in $T$ of the order of $1/ma^2$. If $T\ll 1/ma^2$, the main contribution to $\rho_{\rm D}$ comes from direct backscattering at the Fermi level independently of the parameter $k_Fa$.

\section{Coulomb drag in one dimension: Triple collisions}
\label{s3}
\setcounter{equation}{0}

In Secs.~\ref{s2b} and \ref{s2c}, we have shown that the processes of thermal equilibration between electrons with different (right-left) chirality are absolutely necessary for the bulk drag effect. Further, in Sec.~\ref{s2c}, we have demonstrated that {\it interwire} pair collisions are capable of establishing the right-left equilibration and that the resulting drag resistivity is proportional to $\exp (-2\epsilon_F/T)$. On the other hand, in a single wire, where pair collisions in the ballistic case do not change the electron distribution function, energy relaxation has been known to be associated with triple collisions, see Refs.~\onlinecite{sirenko94} and \onlinecite{lunde07,karzig10,micklitz10,micklitz11,levchenko11,levchenko11a} for the cases of a nondegenerate and degenerate electron gas, respectively. In the degenerate case, energy-relaxation processes that do not require changing the number of electrons with the same chirality \cite{karzig10,micklitz11,levchenko11} are much faster than those that do. \cite{lunde07,micklitz10,levchenko11a} Specifically, while the former are characterized by scattering rates that are power-law functions of the characteristic energy scales, the right-left equilibration rate due to triple collisions is proportional \cite{lunde07,micklitz10,levchenko11a} to $\exp (-\epsilon_F/T)$.

Despite being exponentially suppressed for $T\ll\epsilon_F$, the right-left relaxation is qualitatively important because triple collisions can change electric current (in a finite-size ballistic wire connected to the leads) only if they change the difference between the numbers of right- and left-moving electrons. \cite{lunde07,rech09,micklitz10,levchenko11a} In particular, it is the right-left equilibration that determines interaction-induced corrections to the conductance and to the thermopower: in short wires, whose length is smaller than the right-left equilibration length, the corrections \cite{lunde07,micklitz10,levchenko11a} are proportional to $\exp (-\epsilon_F/T)$ (for a similar consideration in the bosonic formulation, see Refs.~\onlinecite{matveev10,matveev11}). The right-left relaxation plays also a role in the transport properties of inhomogeneous wires. \cite{levchenko10}

Below, we study the contribution of triple collisions to the drag resistivity. As already mentioned in Sec.~\ref{s1}, triple collisions can strongly enhance drag at low $T$---this is precisely because of the right-left equilibration rate due to triple collisions being proportional to $\exp (-\epsilon_F/T)$, in contrast to $\exp (-2\epsilon_F/T)$ in the case of pair collisions in the double wire.

\subsection{Kinetic equation for triple collisions}
\label{s3a}

The contribution to the collision integral ${\rm St}_\sigma$ for the distribution function $f_\sigma(k_1)$ in Eq.~(\ref{1}) that comes from triple collisions reads (for $\sigma=1$ in ${\rm St}_\sigma$):
\begin{widetext}
\begin{eqnarray}
{\rm St}^{(3)}_1\{f\}&=&\sum\nolimits_{\sigma\sigma'}{1+\delta_{\sigma\sigma'}\over 2}\widetilde{\sum\nolimits^\prime}_{\!\!231'2'3'}w_{\sigma\sigma'}(1',2',3'|1,2,3)
\delta(\epsilon_1+\epsilon_2+\epsilon_3-\epsilon_{1'}-\epsilon_{2'}-\epsilon_{3'})\nonumber\\
&\times&\left\{f_1(k_{1'})f_\sigma(k_{2'})f_{\sigma'}(k_{3'})[1-f_1(k_1)][1-f_\sigma(k_2)][1-f_{\sigma'}(k_3)]\right.\nonumber\\
&-&\left.f_1(k_1)f_\sigma(k_2)f_{\sigma'}(k_3)[1-f_1(k_{1'})][1-f_\sigma(k_{2'})][1-f_{\sigma'}(k_{3'})]\right\}~.
\label{33}
\end{eqnarray}
The sign $^\prime$ in $\sum\nolimits^\prime$ means that the summation over momenta goes over {\it distinguishable} initial and final states [for ease of presentation, the integration in Eq.~(\ref{2}) is changed in Eq.~(\ref{33}) to the summation over quantized momenta: below, $L$ is the size of either wire 1 or wire 2]. With the three-particle state $|1,2,3\rangle$ (anti)symmetrized over permutations of electrons from the same wire and normalized to unity, this prevents the double counting of the partial scattering probabilities. The tilde over the summation sign in Eq.~(\ref{33}) denotes one more constraint on the momentum summation---this will be discussed in Sec.~\ref{s3a1}.

For spinless electrons, the three-particle state $|1,2,3\rangle$ is written as
\begin{eqnarray}
|1,2,3\rangle_a&=&{\rm D}_a(k_1,k_2,k_3)~, \label{34a}\\ |1,2,3\rangle_b&=&{1\over L^{1/2}}e^{ik_1x_1}{\rm D}_b(k_2,k_3)~, \label{34b}\\ |1,2,3\rangle_c&=&{1\over
L^{1/2}}e^{ik_3x_3}{\rm D}_c(k_1,k_2)~,
\label{34c}
\end{eqnarray}
where we distinguish three cases:
\begin{eqnarray}
&&(a): \quad {\rm all\,\,electrons\,\, are\,\,in\,\, wire\,\, 1}~,\,\,[{\rm Eq.}~(\ref{34a})]~,\nonumber\\ &&(b): \quad {\rm electron\,\, 1\,\, is\,\, in\,\, wire\,\,1,\,\, electrons\,\, 2\,\, and\,\, 3\,\,are\,\, in\,\, wire\,\, 2}~,\,\,[{\rm Eq.}~(\ref{34b})]~,\nonumber\\ &&(c): \quad {\rm electrons\,\, 1\,\, and\,\,2\,\,are\,\,in\,\, wire\,\, 1,\,\, electron\,\, 3\,\,is\,\, in\,\, wire\,\, 2}~,\,\,[{\rm Eq.}~(\ref{34c})]~,\nonumber
\end{eqnarray}
with the normalized three- and two-particle Slater determinants  (${\rm D}_a$ and ${\rm D}_{b,c}$, respectively) given by Eq.~(\ref{e1}) in Appendix \ref{aE}. The indices $\sigma,\sigma'$ in Eq.~(\ref{33}) are then grouped as follows: $\sigma=\sigma'=1$ in case (a) and $\sigma=\sigma'=2$ in case (b). In case (c), identically equal to each other contributions to ${\rm St_1}$ come from $\sigma=1,\sigma'=2$ and $\sigma=2,\sigma'=1$ [the two contributions are weighted with a factor of 1/2 each, as is accounted for by the factor $(1+\delta_{\sigma\sigma'})/2$ in Eq.~(\ref{33})]. The kernel $w_{\sigma\sigma'}$ in Eq.~(\ref{33}) is given in cases $(a),(b),(c)$ by
\begin{equation}
w_{a,b,c}(1',2',3'|1,2,3)=2\pi |A^{\rm irr}_{a,b,c}(1',2',3'|1,2,3)|^2~,
\label{36}
\end{equation}
where $A^{\rm irr}_{a,b,c}(1',2',3'|1,2,3)$ is the irreducible (not factorizable into independent blocks not connected by interaction) part of the three-particle amplitude $A_{a,b,c}(1',2',3'|1,2,3)$. At the lowest (second) order in interaction, the amplitude $A_{a,b,c}(1',2',3'|1,2,3)$ is written as
\begin{equation}
A_{a,b,c}(1',2',3'|1,2,3)=\sum\nolimits_{456}^\prime{A^{(1)}_{a,b,c}(1',2',3'|4,5,6)A^{(1)}_{a,b,c}(4,5,6|1,2,3)\over
(\epsilon_1+\epsilon_2+\epsilon_3)-(\epsilon_4+\epsilon_5+\epsilon_6)+i0}~,
\label{37}
\end{equation}
where
\begin{equation}
A^{(1)}_{a,b,c}(1,2,3|4,5,6)=\langle 1,2,3|{\rm v}(x_1-x_2)+{\rm v}(x_2-x_3)+{\rm v}(x_1-x_3)|4,5,6\rangle_{a,b,c}~.
\label{38}
\end{equation}
The potential ${\rm v}(x)$ in Eq.~(\ref{38}) is either ${\rm v}_{11}(x)$ or ${\rm v}_{12}(x)$, depending on whether it couples electrons in the same wire or in different wires.

The sign $^\prime$ in $\sum\nolimits^\prime$ in Eq.~(\ref{37}) has the same meaning as in Eq.~(\ref{33}). One can remove the constraint on the allowed momenta in Eq.~(\ref{37}) by introducing the following factors:
\begin{equation}
A_a={1\over 6}\sum\nolimits_{456}\{\ldots\}~,\quad A_b={1\over 2}\sum\nolimits_{456}\{\ldots\}~,\quad A_c={1\over 2}\sum\nolimits_{456}\{\ldots\}~, \label{39}
\end{equation}
where $\{\ldots\}$ is the same fraction as in Eq.~(\ref{37}). The factor 1/6=1/3! in $A_a$ comes from the permutations over all intermediate states (4,5,6). The factors 1/2 in $A_b$ and $A_c$ come from the permutations over states (5,6) and over states (4,5), respectively. Similarly, the restriction on the summation over momenta in Eq.~(\ref{33}) can be lifted after introducing additional factors in Eq.~(\ref{33}). Let us denote ${\rm St}_1^{(3a),(3b),(3c)}$ the contributions to ${\rm St}^{(3)}_1$ in Eq.~(\ref{33}) from processes (a),(b),(c). Then,
\begin{eqnarray}
&&{\rm St}_1^{(3a)}=2\pi\left({1\over 2}\times{1\over 6}\right)\widetilde{\sum}_{231'2'3'}|A_a^{\rm irr}|^2\delta(\ldots)\{\ldots\}~, \label{40}\\ &&{\rm St}_1^{(3b)}=2\pi\left({1\over 2}\times{1\over 2}\right)\widetilde{\sum}_{231'2'3'}|A_b^{\rm irr}|^2\delta(\ldots)\{\ldots\}~, \label{41}\\ &&{\rm St}_1^{(3c)}=2\pi\,{1\over 2}\,\widetilde{\sum}_{231'2'3'}|A_c^{\rm irr}|^2\delta(\ldots)\{\ldots\}~,
\label{42}
\end{eqnarray}
where $\delta(\ldots)$ denotes the delta-function from Eq.~(\ref{33}) and $\{\ldots\}$ is the sum of the products of the distribution functions as given by the expression in the curly brackets in Eq.~(\ref{33}). The factors 1/2 and 1/6 in Eq.~(\ref{40}) come from the summation over states (2,3) and ($1^\prime,2^\prime,3^\prime$), respectively. Two factors 1/2 in Eq.~(\ref{41}) come from the summation over states (2,3) and ($2^\prime,3^\prime$), respectively. The factor 1/2 in Eq.~(\ref{42}) comes from the summation over states ($1^\prime,2^\prime$). Note that the combinatorial factors in the collision integrals in Eqs.~(\ref{40})-(\ref{42}) are absent in the formalism of Ref.~\onlinecite{lunde07}, where the three-particle scattering rate (in a single wire) was plugged into the collision integral without any restriction on the summation over the initial and final states [cf.\ the sign $^\prime$ in Eq.~(\ref{33})]. The formalism from Ref.~\onlinecite{lunde07} was also used in writing the collision integral in Refs.~\onlinecite{karzig10,micklitz10,micklitz11,levchenko11,levchenko11a}. The irreducible three-particle amplitudes $A_{a,b,c}^{\rm irr}$ are given by Eqs.~(\ref{43a})-(\ref{43c}) in Appendix \ref{aE}.

Similar to Sec.~\ref{s2a}, it is convenient to rewrite the linearized kinetic equation that includes triple collisions [Eq.~(\ref{33})] in terms of the functions $g_\sigma(k)$ [Eq.~(\ref{5})]. By explicitly separating the zero-mode solution $g_+(k)$ [given \cite{remark31} by Eq.~(\ref{9}) independently of the number of colliding particles in the collision integral], the linearized kinetic equation is represented in a closed form for the function $g_-(k)$:
\begin{equation}
-i\omega g_-(k_1)-{e(E_1-E_2)k_1\over 2mT}={\rm st}^{(2)}_-\{g\}+{\rm st}^{(3)}_-\{g\}~,
\label{45}
\end{equation}
where the three-particle contribution ${\rm st}^{(3)}_-\{g\}$ to the collision integral depends, similar to ${\rm st}^{(2)}_-\{g\}$, only on $g_-(k)$. Specifically, ${\rm st}^{(3)}_-\{g\}$ is written as a sum of three terms associated, respectively, with processes (a),(b),(c):
\begin{eqnarray}
\phantom{a}\hspace{-14mm}&&{\rm st}^{(3a)}_-\{g\}={1\over 12}\,\widetilde{\sum}_{231'2'3'}{W_a(1',2',3'|1,2,3)\over\zeta^2(k_1)}\delta (\ldots)\,[\,g_-(k_{1'})+g_-(k_{2'})+g_-(k_{3'})-g_-(k_1)-g_-(k_2)-g_-(k_3)\,]~,\label{46}\\
\phantom{a}\hspace{-14mm}&&{\rm st}^{(3b)}_-\{g\}={1\over 4}\,\widetilde{\sum}_{231'2'3'}{W_b(1',2',3'|1,2,3)\over\zeta^2(k_1)}\delta (\ldots)\,[\,g_-(k_{1'})-g_-(k_{2'})-g_-(k_{3'})-g_-(k_1)+g_-(k_2)+g_-(k_3)\,]~,\label{47}\\
\phantom{a}\hspace{-14mm}&&{\rm st}^{(3c)}_-\{g\}={1\over 2}\,\widetilde{\sum}_{231'2'3'}{W_c(1',2',3'|1,2,3)\over\zeta^2(k_1)}\delta (\ldots)\,[\,g_-(k_{1'})+g_-(k_{2'})-g_-(k_{3'})-g_-(k_1)-g_-(k_2)+g_-(k_3)\,]~,\label{48}
\end{eqnarray}
with $\delta(\ldots)$ having the same meaning as in Eqs.~(\ref{40})-(\ref{42}) and
\begin{equation}
W_{a,b,c}(1',2',3'|1,2,3)=2\pi |A_{a,b,c}^{\rm irr}(1',2',3'|1,2,3)|^2\,{\zeta(k_1)\zeta(k_2)\zeta(k_3)\zeta(k_{1'})\zeta(k_{2'})\zeta_(k_{3'})\over 16}~.
\label{49}
\end{equation}
Note that the contributions to ${\rm st}^{(3a)}_-\{g\}$ of the differences $g(k_{2'})-g(k_2)$ and $g(k_{3'})-g(k_3)$ are equal in view of the symmetry of the kernel $W_a(1',2',3'|1,2,3)=W_a(1',3',2'|1,3,2)$, and the same is true for channel (b). In contrast, trading momenta $(2,2')\leftrightarrow (3,3')$ for given $(1,1')$ in channel (c) changes the kernel if $V_{11}(q)\neq V_{12}(q)$. If one neglects the difference between $V_{11}(q)$ and $V_{12}(q)$, the contributions to ${\rm st}^{(3c)}_-\{g\}$ of $g(k_{2'})-g(k_2)$ and $g(k_{3'})-g(k_3)$ cancel each other.

\end{widetext}

\subsection{Divergencies in the three-particle scattering rate}
\label{s3a1}

We now turn to the meaning of the tilde over the summation sign in Eqs.~(\ref{33}),(\ref{40})-(\ref{42}),(\ref{46})-(\ref{48}). Notice the energy denominators in the amplitudes $A_{a,b,c}^{\rm irr}$ [Eqs.~(\ref{43a})-(\ref{43c})]: being squared in the collision integral [Eq.~(\ref{36}) or (\ref{49})], they yield a singularity in the kernel of the collision integral of the type $1/\Delta^2$, where $\Delta$, defined for various scattering processes according to Eq.~(\ref{44}), is the energy transferred in a virtual transition to the intermediate state. The singularity is in general not integrable in the sense that the numerator does not vanish at $\Delta=0$. More specifically, it is not integrable if electrons possess spin (or pseudospin, as in the case of different wires)---for more details, see Appendix \ref{aC}. The tilde in Eqs.~(\ref{33}),(\ref{40})-(\ref{42}),(\ref{46})-(\ref{48}) is related to the proper handling of the $1/\Delta^2$ singularity in the kinetic theory, as explained below.

Let us first recall the relevant aspects of the many-particle scattering problem in the vacuum as we know them from quantum mechanical scattering theory. There are two conceptually important differences between the two-particle scattering problem and the $M$-particle scattering problem with $M>2$. One is related to the definition of the scattering matrix for $M>2$. In the former case, one can unambiguously define the (exact to arbitrary order in the interaction potential) two-particle scattering operator whose matrix elements modulus squared, taken on the mass-shell in the basis of {\it free} (with respect to the interparticle interaction potential) states, determine the scattering cross-section. In the case of $M>2$, the scattering states may not be definable as asymptotically free---this happens if particles can form bound states in the process of scattering. \cite{faddeev61,baz69} The $M$-particle $T$-matrix should then include the scattering states in which (a part of) interaction does not disappear at infinity and remains important for arbitrarily long times after the collision event (or, conversely, the bound states may exist before the collision event and be excited in the process of it). A general formalism that accounts for the proper boundary conditions in the $M$-particle scattering problem with arbitrary scattering channels is based on the Faddeev equations. \cite{faddeev61,baz69}

For the case of a repulsive interaction (assumed in this paper), there are no bound states. However, independently of the sign of interaction, there is another essential difference between the scattering problems with $M=2$ and $M>2$---which resembles the one mentioned above in that it is also related to scattering processes in which interaction remains relevant for arbitrarily long times. Specifically, $M$-particle collisions with $M>2$ occur not necessarily in a compact region in space and time even for the case of a short-range interaction potential. For example, three-particle scattering (contributing to the irreducible part of the scattering amplitude) occurs when all three particles are simultaneously within the range of the interaction, but it also includes processes which consist of two consecutive scattering events in one of which only two particles interact with each other and the other event in which one of those particles interacts with the third particle. \cite{resibois65,bezzerides68,newton76} The time separating the two events may be arbitrarily long. The $1/\Delta^2$ singularity \cite{resibois65,newton76} is associated with this type of scattering, for which the scattering probability increases not linearly but quadratically in time.

The $1/\Delta^2$ growth of the differential three-particle cross-section (defined diagrammatically as a sum squared of all connected diagrams for the three-particle scattering amplitude at given momenta of the incident and outgoing waves at infinity) as $\Delta\to 0$ is a no-nonsense singularity which requires, however, a proper regularization at the point $\Delta=0$. Clearly, there arises a question about the meaning of the cross-section integrated around $\Delta=0$. In Ref.~\onlinecite{resibois65}, a finite density of particles was introduced to regularize the divergency of the three-particle $T$-matrix in the collision integral. In effect, a similar regularization was used in Ref.~\onlinecite{bezzerides68}, where the quantum kinetic equation for triple collisions was derived in terms of scattering amplitudes in a ``medium" (the gas of interacting particles). In a different approach, \cite{newton76,merkur'ev71} it was pointed out that the limit $\Delta\to 0$ and the limit of the distance sent to infinity (in the definition of the $T$-matrix) do not commute. That is, the infinitesimal neighborhood of the point $\Delta=0$ in the differential cross-section requires delicate handling, depending on what quantity is calculated. In particular, the implications for the intensity of outgoing waves in a three-beam experiment in three dimensions were discussed in Ref.~\onlinecite{newton76}. Most importantly, by taking the limit of an infinitely large distance {\it after} the limit $\Delta\to 0$, the integral of the differential cross-section around the singularity was demonstrated to be finite. \cite{resibois65,bezzerides68,newton76} As follows from the results of Refs.~\onlinecite{resibois65,bezzerides68}, it is the latter order of taking the limits that determines the collision integral in the kinetic formulation. Specifically, $1/\Delta^2$ in the kernel of the collision integral should be regularized at $\Delta=0$ as the real part of a double pole: \cite{resibois65,bezzerides68,newton76}
\begin{equation}
{1\over |\Delta|^2}\to {\Delta^2-\varepsilon^2\over (\Delta^2+\varepsilon^2)^2}~,\quad\varepsilon\to 0~,
\label{50}
\end{equation}
which yields a finite result for the integral of the differential cross-section over a region that includes $\Delta=0$, and not as the modulus squared $1/|\Delta+i\varepsilon|^2$, which would give a divergent integral. \cite{kogan75} The tilde in Eqs.~(\ref{33}),(\ref{40})-(\ref{42}),(\ref{46})-(\ref{48}) denotes the regularization rule (\ref{50}).

The way the singularity at $\Delta=0$ is regularized in Eq.~(\ref{50}) has important ramifications for the structure of the collision integral expanded in a series in the number $M\geq 2$ of colliding particles, ${\rm St}_\sigma=\sum_M{\rm St}^{(M)}_\sigma$. As follows from Eq.~(\ref{50}), a naive extension of the $M=2$ result, assuming that the kernel of ${\rm St}^{(M)}_\sigma$ for given momenta is proportional to the modulus squared of the corresponding matrix element of the $M$-particle $T$-matrix, is incorrect. The thus defined ${\rm St}^{(M)}_\sigma$ would be divergent for $M>2$. As pointed out in Refs.~\onlinecite{resibois65,bezzerides68} for the case of $M=3$, the expansion over $M$ contains additionally counterterms. Specifically, for $M=3$:
\begin{equation}
{\rm St}^{(3)}_1=\widetilde{\sum}\{\ldots\}=\sum\{\ldots\}-I^{(3)}_1~,
\label{51}
\end{equation}
where $\{\ldots\}$ is the contribution to ${\rm St}^{(3)}_1$ of a given set of momenta as shown in Eq.~(\ref{33}), $\sum\{\ldots\}$ contains  $1/|\Delta+i\varepsilon|^2$ (modulus squared of the amplitude) and is thus diverging, and $I^{(3)}_1$ is a counterterm that cancels the divergent contribution to $\sum\{\ldots\}$ according to the rule (\ref{50}). The term $-I^{(3)}_1$ can be considered as removing from ${\rm St}^{(3)}_1$ the contribution of two consecutive two-particle collisions separated by an infinite time, so that in between the three-particle system returns to the mass-shell (the independent two-particle collisions are already accounted for by the term ${\rm St}^{(2)}_1$: subtracting the counterterm thus prevents double counting). \cite{remark19} The meaning of the tilde in Eq.~(\ref{51}) is thus that one should not include such ``real" states in the summation over virtual states in the three-particle scattering amplitudes. Ideologically, the subtraction of the counterterm in Eq.~(\ref{51}) bears similarity to the treatment of triple collisions in classical kinetic theory ($\S 17$ in Ref.~\onlinecite{pitaevskii81}).

The necessity to use the regularization (\ref{50}) in the collision integral has not been part of the discussion in the recent wave of interest in three-particle scattering in one dimension. \cite{lunde07,karzig10,micklitz10,micklitz11,levchenko11,levchenko11a} In all these works, the kernel of the collision integral is written simply as the modulus squared of the $T$-matrix element. This omission is, in fact, only justifiable in the case of spinless electrons [in our problem, this corresponds to three-particle scattering in channel (a), in which all electrons are (pseudo)spin-polarized]: the collision integral in this case does not diverge because of a cancellation between the contributions of direct and exchange scattering. In terms of the double counting discussed above, the absence of divergencies in the spinless case can be understood as a consequence of the fact that two-particle collisions do not affect the distribution function in a single wire. However, in the case of electrons with spin (or pseudospin, as is the case for two wires in the drag problem), it is absolutely necessary to specify in what way the singularity at $\Delta=0$ should be treated [Eq.~(\ref{50})], because the naive representation of the kernel as the modulus squared of the three-particle $T$-matrix element leads to divergency. \cite{remark22} The technical details of how the three-particle scattering rate behaves in the vicinity of the point $\Delta=0$ in the drag problem are further discussed, in the diagrammatic language, in Appendix \ref{aC}, where we calculate, as an example, the total scattering rate for a (pseudo)spinful particle (for a similar calculation in the spinless case, see Ref.~\onlinecite{khodas07}).

\subsection{Fokker-Planck description of soft triple collisions}
\label{s3b1}

As already noted at the beginning of Sec.~\ref{s3}, three-particle scattering may substantially enhance drag for the case of soft collisions, when the right-left equilibration is controlled by a slow diffusion in energy space. We therefore turn now to a description of three-particle scattering with small momentum transfer in terms of a Fokker-Planck equation, similar to Sec.~\ref{s2c} for two-particle scattering. Just as in the case of pair collisions, the Fokker-Planck approach is justified if $T$ is much larger than the characteristic energy transfer.

The current in momentum space $J^{(3)}$, induced by triple collisions, is related to ${\rm st}_-^{(3)}\{g\}$ in Eq.~(\ref{45}) by
\begin{equation}
{\rm st}_-^{(3)}\{g\}={4\over\zeta^2(k_1)}{\partial J^{(3)}(k_1)\over\partial k_1}
\label{52}
\end{equation}
[cf.\ Eq.~(\ref{14}) for the case of pair collisions]. For the linearized collision integral [Eqs.~(\ref{46})-(\ref{48})], the contributions $J^{(3a),(3b),(3c)}$ to $J^{(3)}$ of scattering processes (a),(b),(c) can be exactly rewritten as
\begin{widetext}
\begin{eqnarray}
&&J^{(3a)}(k)=-\int_{-\infty}^\infty\!{dq\over 2\pi}\,\int_{k-q}^k\!dk'\left[\,P_a(k',k'+q)g_-(k')+\bar{P}_a(k',k'+q)\,\right]~,\label{53}\\
&&J^{(3b)}(k)=-\int_{-\infty}^\infty\!{dq\over 2\pi}\,\int_{k-q}^k\!dk'\left[\,P_b(k',k'+q)g_-(k')-\bar{P}_b(k',k'+q)\,\right]~,\label{54}\\
&&J^{(3c)}(k)=-\int_{-\infty}^\infty\!{dq\over 2\pi}\,\int_{k-q}^k\!dk'\left[\,P_c(k',k'+q)g_-(k')+\bar{P}_c(k',k'+q)\,\right]~,\label{55}
\end{eqnarray}
where
\begin{equation}
P_a(k_1,k_{1'})={1\over 12}\times {1\over 4}\,L\,\widetilde{\sum}_{232'3'}W_a(1',2',3'|1,2,3)\delta (\ldots)~,
\label{56}
\end{equation}
$P_{b,c}(k_1,k_{1'})$ are defined similarly, with the numerical coefficient 1/12 in Eq.~(\ref{56}) being changed to 1/4 and 1/2 in cases (b) and (c), respectively, and
\begin{eqnarray}
&&\bar{P}_a(k_1,k_{1'})={1\over 12}\times {1\over 4}\,L\,\widetilde{\sum}_{232'3'}W_a(1',2',3'|1,2,3)\delta (\ldots)[\,g_-(k_2)-g_-(k_{2'})\,]~,\label{57}\\
&&\bar{P}_b(k_1,k_{1'})={1\over 4}\times {1\over 4}\,L\,\widetilde{\sum}_{232'3'}W_b(1',2',3'|1,2,3)\delta (\ldots)[\,g_-(k_2)-g_-(k_{2'})\,]~,\label{58}\\
&&\bar{P}_c(k_1,k_{1'})={1\over 2}\times {1\over 4}\,L\,\widetilde{\sum}_{232'3'}W_c(1',2',3'|1,2,3)\delta (\ldots){g_-(k_2)-g_-(k_3)-g_-(k_{2'})+g_-(k_{3'})\over 2}~.\label{59}
\end{eqnarray}
The integral over $k'$ in each of Eqs.~(\ref{53})-(\ref{55}) is taken over an interval whose width is the transferred momentum $q$. At this point, it is important to realize that it would be incorrect to simply expand in $q$ in the integrands of Eqs.~(\ref{53}) and (\ref{55}) [channels (a) and (c), respectively] in order to obtain the collision integral in the diffusive limit. This is because of the exchange processes in the amplitudes (\ref{43a}) and (\ref{43c}) that exchange $k_{1'}$ with either $k_{2'}$ [channels (a) and (c)] or $k_{3'}$ [channel (c)]. In these processes, the momentum difference that is small in the diffusive limit is $k_{2'}-k_1$ or $k_{3'}-k_1$, but not $k_{1'}-k_1$ the characteristic value of which is much larger than $1/a$. The contribution to the collision integral of the processes with small $k_{2'}-k_1$ [channels (a) and (c)] or $k_{3'}-k_1$ [channel (a)] is, however, the same as that of the processes with small $k_{1'}-k_1$. Therefore, the current $J^{(3)}(k)$ is obtained in the diffusive limit by expanding in $q$ in Eqs.~(\ref{53})-(\ref{55}) {\it and} multiplying the result by a factor of 3 in channel (a) and a factor of 2 in channel (c). More specifically, expanding the products $P_{a,b,c}(k',k'+q)g_-(k')$ in the integrands to first order in $k'-k$, taking $\bar{P}_{a,b,c}(k',k'+q)$ out from under the integral sign at the point $k'=k$, and using the property of the kernel $P_{a,b,c}(k,k+q)$
\begin{equation}
\int_{-\infty}^\infty\!dq\,\int_{k-q}^k\!dk'\,P_{a,b,c}(k',k'+q)=0~,
\label{60}
\end{equation}
which follows from the vanishing of the collision integral in Eqs.~(\ref{46})-(\ref{48}) at $g_-(k)={\rm const}(k)$ and the condition that the current in momentum space is zero at $|k|\to\infty$, we have
\begin{equation}
J^{(3)}(k)\simeq D^{(3)}(k){\partial g_-(k)\over\partial k}
-\bar{C}^{(3)}(k)~,
\label{61}
\end{equation}
where
\begin{eqnarray}
&&D^{(3)}(k)={1\over 2}\int\!{dq\over 2\pi}\,q^2[\,3P_a(k,k+q)+P_b(k,k+q)+2P_c(k,k+q)\,]~,\label{62}\\
&&\bar{C}^{(3)}(k)=\int\!{dq\over 2\pi}\,q\,[\,3\bar{P}_a(k,k+q)-\bar{P}_b(k,k+q)+2\bar{P}_c(k,k+q)\,]~.\label{63}
\end{eqnarray}
The collision integral ${\rm St}^{(3)}\{f\}=-\partial J^{(3)}(k)/\partial k$ with $J^{(3)}(k)$ from Eq.~(\ref{61}) conserves total momentum and total energy.

Note the absence of a drift component [proportional to $g_-(k)$ itself, not its derivative $\partial g_-(k)/\partial k$] in the contribution to $J^{(3)}(k)$ that comes from the terms proportional to $P_{a,b,c}(k',k'+q)$. This is a direct consequence of the exact condition (\ref{60}), which can be represented in the limit of small momentum transfer (characteristic $q\to 0$) as
\begin{equation}
\int\!dq\,qP_{a,b,c}(k,k+q)={1\over 2}{\partial\over\partial k}\int\!dq\,q^2P_{a,b,c}(k,k+q)~.
\label{64}
\end{equation}
It is also worth noting that, while expanding $P_{a,b,c}(k',k'+q)g_-(k')$ in $k'-k$ in Eqs.~(\ref{53})-(\ref{55}) yields the first term in Eq.~(\ref{61}), expanding $\bar{P}_{a,b,c}(k',k'+q)$ would be beyond the accuracy of the diffusive approximation. Indeed, doing so would produce the term $\partial\bar{D}^{(3)}(k)/\partial k$ in $J^{(3)}(k)$, where
\begin{equation}
\bar{D}^{(3)}(k)={1\over 2}\int\!{dq\over 2\pi}\,q^2[\,3\bar{P}_a(k,k+q)-\bar{P}_b(k,k+q)+2\bar{P}_c(k,k+q)\,]~.
\label{65}
\end{equation}
\end{widetext}
The integrands of $\bar{P}_{a,b,c}(k,k+q)$ [Eqs.~(\ref{57})-(\ref{59})] for scattering $k\to k+q$ of electron 1 contain as factors the linear combinations of the differences of the distribution functions before and after scattering for electrons 2 and 3. For $q\to 0$, the conservation of momentum and energy gives two solutions for the pair $k_{2'}-k_2$ and $k_{3'}-k_3$: either $k_{2'}=k_2$ and $k_{3'}=k_3$ or $k_{2'}=k_3$ and $k_{3'}=k_2$. In the former case, the expansion of the differences $g(k_{2'})-g(k_2)$ and $g(k_{3'})-g(k_3)$ around the $q=0$ solution gives higher powers of $q$ compared to $q^2$ already present in Eq.~(\ref{65}). In the latter case, the expansion yields \cite{remark21} the factor $k_3-k_2$, whose characteristic value in the integrand at $q\to 0$ is of the order of the typical momentum transfer $1/a$. In either case, adding $\partial\bar{D}^{(3)}(k)/\partial k$ to $J^{(3)}(k)$ for the characteristic transferred energy $qk/m$ much smaller than $T$ only gives rise to small corrections to the diffusive approximation (\ref{61}). Similarly, in Eq.~(\ref{63}) for $\bar{C}^{(3)}(k)$, it suffices to expand $g(k_{2'})-g(k_2)$ and $g(k_{3'})-g(k_3)$ in the integrands of $\bar{P}_{a,b,c}(k,k+q)$ to first order in $k_{2'}-k_2$ and $k_{3'}-k_3$, provided that the characteristic transferred momenta are much smaller than $T/v_F$.

In the derivation of the Fokker-Planck equation (\ref{61}), we assumed that the characteristic change in energy of the diffusing electron with momentum $k$ in a single scattering event is much smaller than $T$. This allowed us to treat $g_-(k)$, $P_{a,b,c}(k,k+q)$, and $\bar{P}_{a,b,c}(k,k+q)$ as smooth functions of $k$ on the characteristic scale of $q$. One model in which this condition is satisfied for {\it arbitrary} $k$ is that of $V_{11}(q)$ and $V_{12}(q)$ falling off sufficiently rapidly as $|q|$ increases beyond the same characteristic scale $1/a\ll T/v_F$. Below, we employ this model for
estimating the relative weight of various scattering processes in Sec.~\ref{s3b2} and solving the Fokker-Planck equation analytically in Sec.~\ref{s3b}.
Recall, however, that for the case of Coulomb interaction, as can be seen from Appendix \ref{aA}, the functions $V_{11}(q)$ and $V_{12}(q)$ behave essentially differently with increasing $|q|$. Namely, $V_{12}(q)$ falls off exponentially for $|q|a\gg 1$, whereas $V_{11}(q)$ falls off only logarithmically (for $|q|d\gg 1$). In the Coulomb case, the Fokker-Planck expansion in Eqs.~(\ref{53})-(\ref{55}) is justified for $Ta/v_F\gg 1$ for scattering processes that involve the interwire interaction, but is not justified for channel (a). Importantly, however, the diffusive character of the current $J^{(3a)}(k)$ remains intact even in the case of Coulomb interaction for $|k|\ll k_F$, as will be seen in Sec.~\ref{s3b2}. Electron scattering at the bottom of the spectrum (which bottlenecks the right-left equilibration) can thus be treated in the Coulomb case within the Fokker-Planck approach also for channel (a). The gradient expansion of the intrawire contribution to the integral term $\bar{C}^{(3)}(k)$, on the other hand, would not be justified in the case of Coulomb interaction. We will return to the Coulomb case at the end of Sec.~\ref{s3b}.

In contrast to the differential Fokker-Planck equation for the case of two-particle scattering [Eq.~(\ref{14})], $J^{(3)}(k)$ from Eq.~(\ref{61}) gives an integro-differential equation---because the term $\partial\bar{C}^{(3)}(k)/\partial k$ in ${\rm St}^{(3)}(k)$ is an integral of the distribution function. With the integral kernel from $\bar{C}^{(3)}(k)$, the equation is not exactly soluble, even in the diffusive limit. \cite{remark20}
To proceed, we make two approximations, one of which is parametrically accurate in the drag problem for $T/\epsilon_F\ll 1$, the other---for a particular relation between the strength of the inter- and intrawire interaction potentials (the exact condition will be formulated in Sec.~\ref{s3b2}).

\subsection{Identifying relevant scattering processes}
\label{s3b2}

We now simplify the Fokker-Planck equation in the limits mentioned in the last paragraph of Sec.~\ref{s3b1}. The first step is to realize that, similar to the case of pair collisions, the right-left equilibration due to triple collisions is bottlenecked by the slowing down of diffusive motion in energy space around the point $k=0$. That is, when the equilibration rate limits the drag rate $1/\tau_{\rm D}$ from Eq.~(\ref{1.3a}), it is sufficient to calculate $J^{(3)}(k)$ for $|k|\ll k_F$ and the momenta $k_2$ and $k_3$ of two other electrons close to the Fermi surface. This separation of scales in momentum space is justified in the limit $T/\epsilon_F\ll 1$.

The second step is to compare the contribution to the equilibration rate, induced by three-particle scattering, of the region in momentum space in which $k_2$ and $k_3$ belong to the same chiral branch ($k_2\simeq k_3\simeq\pm k_F$) and the contribution of the region in which $k_2$ and $k_3$ are on the opposite sides of the Fermi surface ($k_2\simeq -k_3\simeq\pm k_F$), see Fig.~\ref{triple}. Specifically, let us estimate the contributions to the diffusion coefficient of a hole with $k\to 0$, $D_{h,++}^{(3)}$ and $D_{h,+-}^{(3)}$, coming from interactions with electrons on the Fermi surface with the same $(++)$ or opposite $(+-)$ chirality. The corresponding terms in the electron diffusion coefficient $D^{(3)}(k\to 0)$ are smaller by a factor of $\exp (-\epsilon_F/T)$.

The conservation of momentum and energy for $|k|\ll |k_2|,|k_3|$ gives
\begin{equation}
q_3\simeq q_2~,\quad q\simeq -2q_2
\label{66}
\end{equation}
for $k_2\simeq -k_3$ and
\begin{equation}
q_3\simeq -q_2~,\quad q\simeq q_2{k_{2'}-k_3\over k_3}
\label{67}
\end{equation}
for $k_2\simeq k_3$, where $q$ is the transferred momentum for electron 1, $q_2=k_{2'}-k_2$ and $q_3=k_{3'}-k_3$. Importantly, while in the former case all three transferred momenta $q,q_2,q_3$ are of the same order of magnitude, in the latter case $|q|\ll |q_2|,|q_3|$. Specifically, the characteristic value of $|q|$ is of the order of the characteristic value of $|q_2|,|q_3|\sim\min\{T/v_F,1/a\}$ in the former case and is smaller by a factor of $T/\epsilon_F$ in the latter. This means that, for $Ta/v_F\gg 1$, the typical length of an elementary step in the diffusion process near $k=0$ is of order $1/a$ for $k_2\simeq -k_3$ and of order $T/a\epsilon_F$ for $k_2\simeq k_3$.

To estimate the scattering rate near $k=0$, calculate first the density of final states
\begin{equation}
\rho(q;k,k_2,k_3)=\int\!{dq_2\over 2\pi}\!\int\!dq_3\,\delta(\ldots)\delta(q+q_2+q_3)~,
\label{67a}
\end{equation}
where $\delta(\ldots)$ is the delta-function that describes the conservation of energy in Eqs.~(\ref{40})-(\ref{42}), (\ref{46})-(\ref{48}), (\ref{56})-(\ref{59}), for scattering $2\to 2'$ on the surface in momentum space on which the conservation of both total energy and total momentum is satisfied for given $q,k,k_2,k_3$. For $|k|\ll |k_2|,|k_3|$ and $k_2\simeq -k_3\simeq \pm k_F$ [Eq.~(\ref{66})], $\rho(q;k,k_2,k_3)\simeq 1/4\pi v_F$, whereas the
characteristic value of $\rho(q;k,k_2,k_3)$ for momenta from Eq.~(\ref{67}) (and $|k_{2'}-k_3|\sim T/v_F$) can be seen to be a factor of $\epsilon_F/T$ larger. Next, observe that the characteristic width of the integration regions in two remaining integrals in $P_{a,b,c}(k,k+q)$, over $k_2$ and $k_3$, is $T/v_F$ for both $k_2\simeq -k_3$ and $k_2\simeq k_3$. Now, compare the characteristic values of the kernel $W_{a,b,c}(1',2',3'|1,2,3)$ for momenta given by Eqs.~(\ref{66}) and (\ref{67}). To do so in an efficient manner, it is convenient to use the following properties of some of the fractions that appear in $A_{a,b,c}^{\rm irr}(1',2',3'|1,2,3)$ from Eqs.~(\ref{43a})-(\ref{43c}) on the three-particle mass-shell:
\begin{widetext}
\begin{eqnarray}
&&{1\over (k_{3'}-k_3)(k_{3'}-k_2)}+{1\over (k_{1'}-k_2)(k_{1'}-k_1)}=-{1\over (k_{3'}-k_2)(k_{1'}-k_2)}~,\nonumber\\
&&{1\over (k_{1'}-k_3)(k_{1'}-k_1)}+{1\over (k_{2'}-k_2)(k_{2'}-k_3)}=-{1\over (k_{2'}-k_3)(k_{1'}-k_3)}~,\nonumber\\
&&{1\over (k_{2'}-k_2)(k_{2'}-k_1)}+{1\over (k_{3'}-k_3)(k_{3'}-k_1)}=-{1\over (k_{3'}-k_1)(k_{2'}-k_1)}~.
\label{68}
\end{eqnarray}
In particular, Eqs.~(\ref{68}) show that, for the momenta from both Eq.~(\ref{66}) and Eq.~(\ref{67}), two terms in each of the sums almost exactly cancel each other for the case of $k_Fa\gg 1$ (exponentially suppressed direct backscattering at the Fermi level): the characteristic value of the sum is a factor of $k_Fa$ smaller than the characteristic value of one of the terms. The use of the characteristic values of the momenta in the estimates is justified by the regularization rule (\ref{50}).

\begin{figure}
\centerline{\includegraphics[width=0.75\columnwidth]{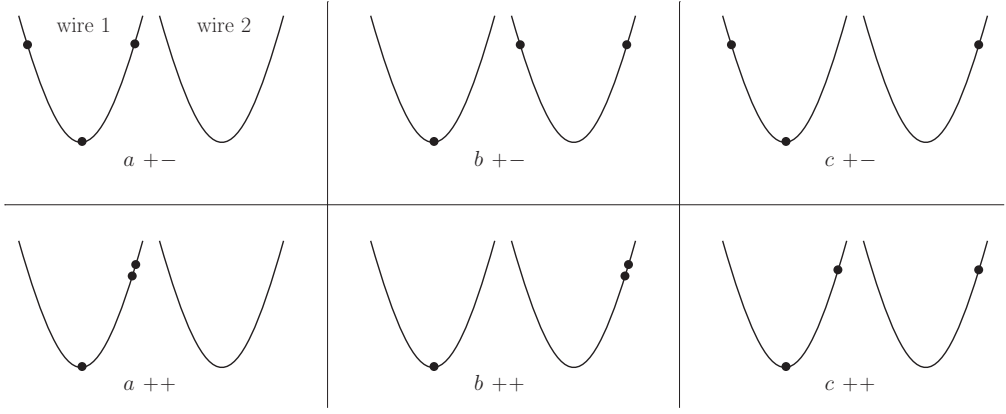}}
\caption{Momentum configurations for three-particle scattering in channels (a), (b), and  (c) (first, second, and third column, respectively) with two electrons having energies close to the Fermi level and a third electron having energy close to the bottom of the spectrum. The electrons on the Fermi surface can have the opposite ($+-$) or same (++) chiralities (first and second row, respectively).
The relative contributions of the different configurations to the diffusion coefficient (in energy space) of the cold electron are discussed around Eqs.~(\ref{73})-(\ref{74b}).
}
\label{triple}
\end{figure}

Let us denote $(A_{a,b,c}^{\rm irr})_{\rm dir}$ the amplitudes of direct scattering in Eqs.~(\ref{43a})-(\ref{43c}). These are associated with the terms in Eqs.~(\ref{43a})-(\ref{43c}) inside the square brackets. For $|k|\ll |k_2|\simeq |k_3|\simeq k_F$ and $k_2\simeq -k_3$ [Eq.~(\ref{66})], they simplify significantly:
\begin{eqnarray}
&&\left(A_a^{\rm irr}\right)_{\rm dir}\simeq {m\over L^2}\,{V_{11}(q/2)\left[\,V_{11}(q)-V_{11}(q/2)\,\right]\over k_F^2}~,\label{69a}\\
\hspace{-2cm}
k_2\simeq -k_3\,:\qquad\qquad&&\left(A_b^{\rm irr}\right)_{\rm dir}\simeq {m\over L^2}\,{V_{11}(q/2)V_{12}(q)-V_{12}^2(q/2)\over k_F^2}~,
\label{69b}\\
&&\left(A_c^{\rm irr}\right)_{\rm dir}\simeq {m\over L^2}\,{V_{12}(q/2)\left[\,V_{11}(q)+V_{12}(q)-2V_{11}(q/2)\,\right]\over 2k_F^2}~.
\label{69c}
\end{eqnarray}
The main simplification is that the only variable on which the amplitudes in Eqs.~(\ref{69a})-(\ref{69c}) depend is the momentum transfer $q$ (with the characteristic value of $q$ of order $1/a$, according to the above). For $|k|\ll |k_2|\simeq |k_3|\simeq k_F$ and $k_2\simeq k_3$ [Eq.~(\ref{67})], the amplitudes are written as
\begin{eqnarray}
\left(A_a^{\rm irr}\right)_{\rm dir}\!\!&\simeq&\!\!{m\over L^2}\,{V_{11}(q_3)\left[\,V_{11}(q_3)-V_{11}(q)\,\right]+q_3V'_{11}(q_3)V_{11}(q)\over k_F^2}~,\label{70a}\\
\hspace{-0.75cm}
k_2\simeq k_3\,:\qquad\qquad\quad\left(A_b^{\rm irr}\right)_{\rm dir}\!\!&\simeq&\!\! {m\over L^2}\,{V_{12}^2(q_3)-V_{11}(q_3)V_{12}(q)+q_3V'_{11}(q_3)V_{12}(q)\over k_F^2}~,
\label{70b}\\
\left(A_c^{\rm irr}\right)_{\rm dir}\!\!&\simeq&\!\! {m\over L^2}\,{V_{12}(q_3)\left[\,V_{11}(q_3)-V_{12}(q_3)\,\right]+q_3V'_{12}(q_3)V_{11}(q)\over k_F^2}\nonumber\\
&-&\!\!{m\over L^2}\,{q_3\over q}\,{V_{12}(q_3)\left[\,V_{11}(q)-V_{12}(q)\,\right]\over k_F^2}~,
\label{70c}
\end{eqnarray}
\end{widetext}
where $V'_{\sigma\sigma'}(q_3)=dV_{\sigma\sigma'}(q_3)/dq_3$. It is worth mentioning that while the product $q_3V'_{\sigma\sigma'}(q_3)$ might seem to imply that $|q_3|$ is assumed to be small compared to the characteristic scale on which $V_{\sigma\sigma'}(q_3)$ changes, taking the derivative $V'_{\sigma\sigma'}(q_3)$ in Eqs.~(\ref{70a})-(\ref{70c}), in fact, only assumes that the transferred momentum that is small in this sense is $q$, which is a much weaker (in the parameter $T/\epsilon_F\ll 1$) condition in view of Eq.~(\ref{67}).

Note that there are two essentially different types of strong cancellations between various terms in the derivation of Eqs.~(\ref{70a})-(\ref{70c}). One, controlled by the parameter $1/k_Fa\ll 1$, is described by Eqs.~(\ref{68}). The other, controlled by the parameter $T/\epsilon_F\ll 1$, is related to the destructive interference between two terms given, in case (a), by the first and second lines in Eq.~(\ref{43a}) and, in case (b), by the first and second lines in Eq.~(\ref{43b}), respectively. Importantly, the latter type of cancellation does not occur in case (c) [Eq.~(\ref{43c})]---because of the difference between the inter- and intrawire interaction potentials---which gives rise to the large factor $q_3/q\simeq k_3/(k_3-k_{2'})$ in the term of $\left(A_c^{\rm irr}\right)_{\rm dir}$ on the second line of Eq.~(\ref{70c}). The singularity of the amplitude at $k_3\to k_{2'}$ is of the type discussed in Sec.~\ref{s3a} and should be regularized in the kernel of the collision integral according to Eq.~(\ref{50}). As follows from this regularization rule, the contribution of $\left(A_c^{\rm irr}\right)_{\rm dir}$ to the collision integral can be estimated by substituting $T/v_F$ as a characteristic value of the difference $k_3-k_{2'}$. We see, then, that the characteristic value of the amplitude in channel (c) contains an additional factor of $\epsilon_F/T\gg 1$ compared to the amplitudes in channels (a) or (b), so that---unless $V_{11}(q)$ and $V_{12}(q)$ are very close to each other---channel (c) gives the main contribution to the collision kernel for the case of electrons 2 and 3 having the same chirality.

Let us now compare the terms in the diffusion coefficient of a hole with $k\to 0$, $D^{(3)}_{h,++}$ and $D^{(3)}_{h,+-}$, that come from interactions with electrons with $k_2\simeq k_3$ and $k_2\simeq -k_3$, respectively. Each of the two terms is a sum of the contributions of channels (a), (b), and (c) [Fig.~\ref{triple}]. It is instructive to estimate the relative weight of the six contributions to the total diffusion coefficient by splitting each of them into a product $\left<q^2\right>\!R/2$, where $\left<q^2\right>$ is the average of $q^2$ in the diffusion process whose elementary step is momentum transfer $q$ and $R$ is the characteristic scattering rate for these elementary steps. The scattering rates include the density of states (\ref{67a}) whose characteristic value was estimated below Eq.~(\ref{67a}) to be a factor of $\epsilon_F/T$ larger for the case of $k_2\simeq k_3$ compared to the case of $k_2\simeq -k_3$. The characteristic values of $q$ for $|k|\ll k_F$ were discussed below Eq.~(\ref{67}) and the characteristic values of momentum differences for electrons on the Fermi surface---below Eq.~(\ref{67a}). Piecing everything together, we estimate the scattering rates for $k_2\simeq -k_3$ in channels (a),(b),(c) as
\begin{eqnarray}
&&\hspace{-1.25cm}R_{+-}^{(3a)}\sim{v_F\over a}\left[{V_{11}(1/a)\over v_F}\right]^4\!\left({T\over\epsilon_F}\right)^2,\label{71a}\\
&&\hspace{-1.25cm}R_{+-}^{(3b),(3c)}\!\sim{v_F\over a}\left[{V_{11}(1/a)\over v_F}\right]^2\left[{V_{12}(1/a)\over v_F}\right]^2\!\left({T\over\epsilon_F}\right)^2.
\label{71b}
\end{eqnarray}
Similarly, for $k_2\simeq k_3$:
\begin{eqnarray}
&&\hspace{-0.3cm}R_{++}^{(3a)}\!\sim {v_F\over a}\left[{V_{11}(1/a)\over v_F}\right]^2\left[{V_{11}(T/a\epsilon_F)\over v_F}\right]^2\!\left({T\over\epsilon_F}\right)^2,\label{72a}\\
&&\hspace{-0.3cm}R_{++}^{(3b)}\!\sim {v_F\over a}\left[{V_{11}(1/a)\over v_F}\right]^2\left[{V_{12}(T/a\epsilon_F)\over v_F}\right]^2\!\left({T\over\epsilon_F}\right)^2,\label{72b}\\
&&\hspace{-0.3cm}R_{++}^{(3c)}\!\sim {v_F\over a}\left[{V_{12}(1/a)\over v_F}\right]^2\left[{V_{11}(T/a\epsilon_F)-V_{12}(T/a\epsilon_F)\over v_F}\right]^2.\nonumber\\
\label{72c}
\end{eqnarray}
In these estimates, we assume that $Ta/v_F\agt 1$. We also assume the most common behavior of the intra- and interwire potentials, namely (as sufficient conditions) that $|V_{11}(1/a)|\agt |V_{12}(1/a)|$ and $|V_{11}(T/a\epsilon_F)|\agt |V_{11}(1/a)|$. Here and in the estimates, the arguments of $V_{11}$ and $V_{12}$ are understood as characteristic scales of transferred momentum. The estimate for $R_{++}^{(3c)}$ [Eq.~(\ref{72c})] is written under the assumption that $|V_{11}(q)-V_{12}(q)|/|V_{11}(q)|\agt T/\epsilon_F$ for $|q|\sim T/a\epsilon_F$. Note that $R_{++}^{(3c)}$ differs from all other scattering rates in Eqs.~(\ref{71a})-(\ref{72c}) in that it does not contain the small factor $(T/\epsilon_F)^2$.

The hole diffusion coefficient $D_{h,+-}^{(3)}$, which results from triple collisions with electrons of opposite chirality, is thus estimated, by substituting $\left<q^2\right>\sim 1/a^2$, as
\begin{equation}
D_{h,+-}^{(3)}\sim {V_{11}^4(1/a)\over (v_Fa)^3}\left({T\over\epsilon_F}\right)^2.
\label{73}
\end{equation}
Provided $V_{11}(1/a)\gg V_{12}(1/a)$, the main contribution to $D_{h,+-}^{(3)}$ comes from scattering in channel (a) [Eq.~(\ref{71a})]. The diffusion coefficient $D_{h,++}^{(3)}$, associated with interactions with electrons of the same chirality and characterized by $\left<q^2\right>\sim (T/a\epsilon_F)^2$, is determined by two competing terms from channels (a) and (c):
\begin{eqnarray}
&&\hspace{-0.5cm}D_{h,++}^{(3a)}\!\sim {V_{11}^4(1/a)\over (v_Fa)^3}\left({T\over\epsilon_F}\right)^4,\label{74a}\\
&&\hspace{-0.5cm}D_{h,++}^{(3c)}\!\sim {V_{12}^2(1/a)\left[V_{11}(T/a\epsilon_F)-V_{12}(T/a\epsilon_F)\right]^2\over (v_Fa)^3}\left({T\over\epsilon_F}\right)^2.\nonumber\\
\label{74b}
\end{eqnarray}

Note that, in channel (a), interactions with electrons of opposite chirality are much more effective than with electrons of the same chirality, because $D_{h,++}^{(3a)}$ in Eq.~(\ref{74a}) has two more powers of the small parameter $T/\epsilon_F$ compared to $D_{h,+-}^{(3)}$ in Eq.~(\ref{73}). Therefore, the total diffusion coefficient can be estimated as a sum of only two terms, $D_{h,+-}^{(3)}$ [Eq.~(\ref{73})] and $D_{h,++}^{(3c)}$ [Eq.~(\ref{74b})]. Now, we observe that the term $D_{h,++}^{(3c)}$ is small in comparison to $D_{h,+-}^{(3)}$ in two limiting cases: if the interaction between the wires is much weaker than inside the wires, or if the two interaction potentials are very close to each other. That is, in both limits of a large and small distance between the wires, interactions of a hole at the bottom of the spectrum with electrons of the same chirality on the Fermi surface can be neglected. Moreover, in the crossover regime, when none of the conditions is satisfied, the contribution of $D_{h,++}^{(3c)}$ to the total diffusion coefficient is of the same order of magnitude as that of $D_{h,+-}^{(3)}$, thus not leading to any qualitatively new features, either.

In Sec.~\ref{s3b} below, we therefore focus on the contribution of three-particle scattering with two electrons having opposite chirality on the Fermi surface. In this case, the Fokker-Planck equation, upon substitution of Eqs.~(\ref{69a})-(\ref{69c}) for the scattering amplitudes, is exactly soluble for the right-left equilibration rate.

\subsection{Interplay of triple intrawire and pair interwire collisions}
\label{s3b}

Recalling the arguments of the very end of Sec.~\ref{s3b2}, we now consider the right-left equilibration due to triple collisions within the framework of the Fokker-Planck equation (\ref{61}) with the scattering amplitudes (\ref{69a})-(\ref{69c}). These amplitudes correspond to the momentum configuration in which two electrons of opposite chirality are close to the Fermi surface, while the third electron is close to the bottom of the spectrum. We treat three- and two-particle soft collisions on an equal footing by adding to the current in momentum space (\ref{61}), induced by three-particle scattering, the component induced by two-particle scattering [Eq.~(\ref{15})]. One important consequence of this is that the mechanisms of drag and right-left relaxation, rigidly connected to each other in the case of pair collisions in Sec.~\ref{s2c}, may now be disentangled. To describe the new physics that comes about from the interplay of triple and pair collisions, the most relevant example is that of drag mediated by pair collisions only, with triple collisions occurring between electrons all of which are from the same wire [channel (a) in the above]. Since drag is only possible in the presence of the processes of thermal equilibration between electrons of opposite chirality (Sec.~\ref{s2b}), drag may (as already noted at the beginning of Sec.~\ref{s3}) be strongly enhanced by {\it intrawire} triple collisions. These do not lead to any drag effect directly---but do affect drag indirectly by providing an additional channel for the thermalization processes which enhance friction induced by interwire pair collisions.

In the limit $Ta/v_F\gg 1$, substituting the amplitudes (\ref{69a})-(\ref{69c}) in the kernel of Eq.~(\ref{62}), we obtain the terms in the diffusion coefficient $D^{(3)}(k)$ [Eq.~(\ref{62})] at $|k|\ll k_F$ that result from interactions with electrons of opposite chirality in channels (a),(b),(c):
\begin{widetext}
\begin{eqnarray}
&&D^{(3a)}(k)\simeq{\zeta^2(k)\over 512 (\pi v_F)^3}\left({T\over\epsilon_F}\right)^2\!\int\!dq\,q^2\,V_{11}^2(q/2)[V_{11}(q)-V_{11}(q/2)]^2,\label{75a}\\
&&D^{(3b)}(k)\simeq{\zeta^2(k)\over 512 (\pi v_F)^3}\left({T\over\epsilon_F}\right)^2\!\int\!dq\,q^2\,[V_{11}(q/2)V_{12}(q)-V_{12}^2(q/2)]^2,\label{75b}\\
&&D^{(3c)}(k)\simeq{\zeta^2(k)\over 1024 (\pi v_F)^3}\left({T\over\epsilon_F}\right)^2\!\int\!dq\,q^2\,V_{12}^2(q/2)[V_{11}(q)+V_{12}(q)-2V_{11}(q/2)]^2.
\label{75c}
\end{eqnarray}
Without the factor $\zeta^2(k)/4$, Eqs.~(\ref{75a})-(\ref{75c}) give the diffusion coefficient for a hole with $k\to 0$, estimated in Eq.~(\ref{73}). Note that the terms corresponding to channels (a) and (b) have equal contributions of the modulus squared of the direct scattering amplitude [Eqs.~(\ref{70a}),(\ref{70b})], in which $|q_2|,|q_3|\ll k_F$ and $|k_2-k_3|\simeq 2k_F$, and of the modulus squared of the exchange amplitude with $|k_{2'}-k_3|,|k_{3'}-k_2|\ll k_F$ and $|k_2-k_3|\simeq 2k_F$. Altogether, taking into account the factors of 3, 1, 2 in front of $P_{a,b,c}(k,k+q)$ in Eq.~(\ref{62}), the exchange processes thus lead to multiplication of the contributions of the processes in which all three transferred momenta $q,q_2,q_3$ are small compared to $k_F$ by factors of 6, 2, 2 in channels (a), (b), (c), respectively. In the limit $Ta/v_F\gg 1$, for the relation between the transferred momenta from Eq.~(\ref{66}), the integral term $\bar{C}^{(3)}(k)$ [Eq.~(\ref{63})] reduces to a sum of two terms coming from channels (a) and (b), while the contribution of channel (c) can be neglected:
\begin{equation}
\bar{C}^{(3)}(k)\simeq D_{\rm i}(k)\,\left<\partial g_-/\partial k\right>~,
\label{76}
\end{equation}
where
\begin{eqnarray}
&&D_{\rm i}(k)=D^{(3a)}(k)-D^{(3b)}(k)~,\label{77a}\\
&&\left<\partial g_-/\partial k\right>=\int\!{dk\over k_T}\,\zeta^2(k){\partial g_-(k)\over\partial k}~,
\label{77}
\end{eqnarray}
and $k_T=\int\! dk\,\zeta^2=8\pi T\partial n/\partial\mu\simeq 8T/v_F$. In Eq.~(\ref{76}), the function $g_-(k)$ only enters $\bar{C}^{(3)}(k)$ through the $k$-independent average (\ref{77}), which greatly simplifies the solution of the Fokker-Planck equation.

Let us rewrite the Fokker-Planck equation
\begin{equation}
{e(E_1-E_2)\zeta^2(k)k\over 8mT}=-{\partial\over\partial k}\left[\,D_{\rm t}(k){\partial g_-(k)\over \partial k}-\bar{C}^{(3)}(k)\,\right]
\label{78a}
\end{equation}
in an integral form:
\begin{equation}
g_-(k)={1\over 4}\int_0^k\!dp\,{1\over D_{\rm t}(p)}\left\{e(E_1-E_2)\left[\,1-\tanh \left({p^2-k_F^2\over 4mT}\right)\,\right]+4\bar{C}^{(3)}(p)\right\}~,
\label{78}
\end{equation}
where the total diffusion coefficient
\begin{equation}
D_{\rm t}(k)=D^{(2)}(k)+D^{(3)}(k)
\label{79}
\end{equation}
describes both pair and triple collisions. The term describing pair collisions in Eq.~(\ref{79}) is related to the diffusion coefficient $D(k)$ from Eq.~(\ref{16}) by $D^{(2)}(k)=\zeta^2(k)D(k)/4$. Unlike Eq.~(\ref{22}) for the case of pair collisions [$D^{(3)}\to 0$, $\bar{C}^{(3)}\to 0$ in Eq.~(\ref{78})], this is not a solution but an integral equation for $g_-(k)$. Substituting Eq.~(\ref{76}) in Eq.~(\ref{78}) and integrating Eq.~(\ref{78}) with a factor $\zeta^2(k)\partial/\partial k$, we have an algebraic equation for $\left<\partial g_-/\partial k\right>$, the solution of which gives
\begin{equation}
\left<\partial g_-/\partial k\right>={e\over 4}(E_1-E_2)\int\!{dk\over k_T}\,\zeta^2(k){1\over D_{\rm t}(k)}\left[\,1-\tanh \left({k^2-k_F^2\over 4mT}\right)\,\right] \bigg/ \left[\,1-\int\!{dk\over k_T}\,\zeta^2(k){D_{\rm i}(k)\over D_{\rm t}(k)}\,\right]~.
\label{80}
\end{equation}
Using Eq.~(\ref{80}) in Eq.~(\ref{76}) and substituting the thus obtained $\bar{C}^{(3)}(k)$ back in Eq.~(\ref{78}) yields the solution of the Fokker-Planck equation in terms of two $k$-dependent combinations of the diffusion coefficients, $D_{\rm t}(k)$ and $D_{\rm i}(k)$.

To characterize the relative strength of two- and three particle scattering, we now introduce two constants (independent of $k$) ${\cal D}_2$ and ${\cal D}_3$ according to $D^{(2)}(k)={\cal D}_2\zeta^4(k)$ and $D^{(3)}(k)={\cal D}_3\zeta^2(k)$,
so that
\begin{equation}
D_{\rm t}(k)=\zeta^2(k)[\,{\cal D}_2\zeta^2(k)+{\cal D}_3\,]~.
\label{82}
\end{equation}
The constant ${\cal D}_2$ is related to the constant $c$ [Eqs.~(\ref{13}),(\ref{13a})] by ${\cal D}_2=mc/16=3k_F^2/4\tau_{\rm D}^\infty$. Note that ${\cal D}_2$ does not depend on $T$, whereas ${\cal D}_3$ is proportional to $(T/\epsilon_F)^2$ [Eqs.~(\ref{75a})-(\ref{75c})]. Similarly, we introduce the constant ${\cal D}_{\rm i}$:
\begin{equation}
D_{\rm i}(k)={\cal D}_{\rm i}\zeta^2(k)~.
\label{83}
\end{equation}
The shape of the function $g_-(k)$ in Eq.~(\ref{78}) and the resulting resistivity $\rho_{\rm D}$ will now be parametrized by the ``diffusion constants" ${\cal D}_2$, ${\cal D}_3$, ${\cal D}_{\rm i}$, and the ratio $T/\epsilon_F\ll 1$. Note that the relation between ${\cal D}_{\rm 2}$ and ${\cal D}_{\rm 3}$ can be arbitrary, while ${\cal D}_{\rm i}< {\cal D}_{\rm 3}$ [and, depending on the relative strength of channels (a) and (b),  ${\cal D}_{\rm i}$ can, in general, be of either sign]. In the limit of large separation between the wires, the main contribution to both ${\cal D}_{\rm 3}$ and ${\cal D}_{\rm i}$ comes from channel (a) and ${\cal D}_{\rm i}\simeq {\cal D}_{\rm 3}$; in contrast, in the limit of small separation, ${\cal D}_{\rm i}\ll {\cal D}_3$.

Let us first calculate the average (\ref{80}). The integral in the denominator of Eq.~(\ref{80}) can be neglected compared to unity in the limit ${\cal D}_3\ll {\cal D}_2$, while in the opposite limit it is close to unity, which makes the denominator small, if ${\cal D}_{\rm i}\simeq {\cal D}_3$:
\begin{equation}
\int\!{dk\over k_T}\,{{\cal D}_{\rm i}\zeta^2\over {\cal D}_2\zeta^2+{\cal D}_3}\simeq {{\cal D}_{\rm i}\over {\cal D}_3}\left(1-{2\over 3}{{\cal D}_2\over {\cal D}_3}\right)~,\quad {\cal D}_2\ll {\cal D}_3~.
\label{84}
\end{equation}
The integral in the numerator of Eq.~(\ref{80}) behaves differently depending on the parameter ${\cal D}_3e^{\epsilon_F/T}/{\cal D}_2$ for ${\cal D}_3\ll {\cal D}_2$, which gives rise to three different types of behavior for $\left<\partial g_-/\partial k\right>$:
\begin{eqnarray}
\hspace{-0.45cm}\left<\partial g_-/\partial k\right>\simeq {e\over 4}(E_1-E_2)\left({\epsilon_F\over T}\right)^{1/2}\!\!\times\left\{
\begin{array}{ll}
{(1/4\cal D}_2)\left[\,\ln^{1/2}({\cal D}_2/{\cal D}_3)+(\pi^{1/2}/ 2)e^{\epsilon_F/T}\,\right]~,\quad & {\cal D}_3\ll {\cal D}_2e^{-\epsilon_F/T}~, \\ \\
(1/{\cal D}_3)\ln^{1/2}\!\left({\cal D}_3e^{\epsilon_F/T}/{\cal D}_2\right)~,\quad & {\cal D}_2e^{-\epsilon_F/T}\ll{\cal D}_3\ll {\cal D}_2~, \\ \\
\left(\epsilon_F/ T\right)^{1/2}\!/({\cal D}_3-{\cal D}_{\rm i}+2{\cal D}_2{\cal D}_{\rm i}/3{\cal D}_3)~,\quad & {\cal D}_2\ll {\cal D}_3~.
\end{array}\right.
\label{85}
\end{eqnarray}
The range of $k$ that gives the main contribution to the integral in the numerator of Eq.~(\ref{80}) and thus determines Eq.~(\ref{85}) in the three regimes is: $|k|<(2mT)^{1/2}\ln^{1/2}({\cal D}_2/{\cal D}_3)$ for the first term and $|k|\alt (mT)^{1/2}$ for the second term in the first line of Eq.~(\ref{85}), $|k|<(2mT)^{1/2}\ln^{1/2}({\cal D}_3e^{\epsilon_F/T}/{\cal D}_2)$ in the second line, and $|k|<k_F$ in the third. The logarithmic divergency  as ${\cal D}_3\to 0$ in the first line of Eq.~(\ref{85}) only occurs within the diffusive approximation and is cut off \cite{remark11} when the characteristic $|k|\sim (mT)^{1/2}\ln^{1/2}({\cal D}_2/{\cal D}_3)$ becomes of the order of $mTa$.

We notice from Eq.~(\ref{85}) that $\left<\partial g_-/\partial k\right>$ contains the large factor $\epsilon_F/T$ to a certain power (different depending on the relation between ${\cal D}_3$ and ${\cal D}_2$). This means that, for thermally excited electrons in the vicinity of the Fermi surface, the derivative $\partial g_-(k)/\partial k$ from Eq.~(\ref{78}) is mainly given by the integral term proportional to $\bar{C}^{(3)}(k)$---unless ${\cal D}_{\rm i}/{\cal D}_3$ is small in the parameter $T/\epsilon_F$ or, if ${\cal D}_{\rm i}/{\cal D}_3\sim 1$, the ratio ${\cal D}_3/{\cal D}_2\ll (T/\epsilon_F)^{1/2}e^{-\epsilon_F/T}$. That is, unless the above conditions are satisfied, two terms in the total current in momentum space, $D_{\rm t}(k)\partial g_-(k)/\partial k$ and $-\bar{C}^{(3)}(k)$, almost compensate each other near the Fermi surface. The current in real space, however, is much less sensitive to the presence of the integral term in the diffusion equation for three-particle scattering, as will be seen below.

Substituting Eq.~(\ref{78}) in Eq.~(\ref{17}) for the electric current $j_-$ [using the relation (\ref{17}) between $g_-$ and $f_-$], we obtain $j_-$ as a double integral which is reducible, by integration by parts, to a single one. Using further Eq.~(\ref{29b}), we thus have
\begin{equation}
\rho_{\rm D}^{-1}={e^2T\over 16\pi}\int\!dk\,{1+e^{-(\epsilon-\epsilon_F)/T}\over {\cal D}_2+{\cal D}_3\cosh^2[(\epsilon-\epsilon_F)/2T]}\left[\,1+e^{-(\epsilon-\epsilon_F)/T}+8{\cal D}_{\rm i}\,{\left<\partial g_-/\partial k\right>\over e(E_1-E_2)}\,\right]~,
\label{86}
\end{equation}
where $\left<\partial g_-/\partial k\right>$ is given by Eqs.~(\ref{80}),(\ref{85}). From Eqs.~(\ref{80}),(\ref{86}), $\rho_{\rm D}$ is represented as $\rho_{\rm D}^{-1}=\rho_{\rm D1}^{-1}+\rho_{\rm D2}^{-1}$, where
\begin{equation}
\rho_{\rm D1}^{-1}={e^2T\over 16\pi}\int\!dk\,{\left[\,1+e^{-(\epsilon-\epsilon_F)/T}\,\right]^2\over {\cal D}_2+{\cal D}_3\cosh^2[(\epsilon-\epsilon_F)/2T]}\label{87}
\end{equation}
is a direct generalization of the result for pair collisions and
\begin{equation}
\rho_{\rm D2}^{-1}=32\,T^2\,{\cal D}_{\rm i}\,{\partial n\over\partial\mu}\left({\left<\partial g_-/\partial k\right>\over E_1-E_2}\right)^2\left\{1-{\cal D}_i\!\int\!{dk\over k_T}\,{1\over {\cal D}_2+{\cal D}_3\cosh^2[(\epsilon-\epsilon_F)/2T]}\,\right\}\label{88}
\end{equation}
comes from the integral term of the Fokker-Planck equation. The integral in the curly brackets in Eq.~(\ref{88}) is discussed above [Eq.~(\ref{84})].
\end{widetext}

In the limit ${\cal D}_3\to 0$ and ${\cal D}_{\rm i}\to 0$, we reproduce Eq.~(\ref{32}) for $\rho_{\rm D}$ induced by pair collisions [note that
the integration in Eqs.~(\ref{86}),(\ref{87}) is understood \cite{remark11} as limited by $|k|\ll mTa$]. Now, we observe that the integral (\ref{86}) is determined by $\epsilon\alt T$ in a wide range of the ratio ${\cal D}_3/{\cal D}_2$ (the exact condition is specified below)---because of the exponential functions $e^{-(\epsilon-\epsilon_F)/T}$ in the numerator that rapidly decay away from the bottom of the spectrum. It follows that triple collisions become essentially important already for ${\cal D}_3\gg {\cal D}_2e^{-\epsilon_F/T}$, when the second term in the denominator of the integrand of Eq.~(\ref{86}) at $\epsilon=0$ becomes much larger than the first one, i.e., when the scattering rate for an electron at the bottom of the spectrum is strongly enhanced by triple collisions.

Inspection of Eqs.~(\ref{86})-(\ref{88}) shows that $\rho_{\rm D}$ is given by $\rho_{\rm D1}$ for all ${\cal D}_3\ll {\cal D}_2e^{\epsilon_F/T}\left(T/\epsilon_F\right)^{3/2}$, and for ${\cal D}_3\gg {\cal D}_2e^{-\epsilon_F/T}$ reads:
\begin{eqnarray}
&&\hspace{-1cm}\rho_{\rm D}\simeq {4\pi {\cal D}_3\over e^2k_F}\left({\epsilon_F\over \pi T^3}\right)^{1/2}e^{-\epsilon_F/T}~,\nonumber\\
&&\hspace{-1cm}{\cal D}_2e^{-\epsilon_F/T}\ll {\cal D}_3\ll {\cal D}_2e^{\epsilon_F/T}\left(T/\epsilon_F\right)^{3/2}~.
\label{89}
\end{eqnarray}
Recall that ${\cal D}_3\propto T^2$, so that the pre-exponential factor in Eq.~(\ref{89}) scales with $T$ as $T^{1/2}$. Notice that both conditions on the ratio ${\cal D}_3/{\cal D}_2$ in Eq.~(\ref{89}) are very weak for $T/\epsilon_F\ll 1$. The one that limits ${\cal D}_3/{\cal D}_2$ from above comes from a comparison of $\rho_{\rm D1}$ and $\rho_{\rm D2}$: for larger ${\cal D}_3/{\cal D}_2$, $\rho_{\rm D}$ is mainly given by $\rho_{\rm D2}$. To see this, let us write $\rho_{\rm D2}$ for ${\cal D}_2\ll {\cal D}_3$, by substituting the last line in Eq.~(\ref{85}) together with Eq.~(\ref{84}) in Eq.~(\ref{88}). The result is
\begin{equation}
\rho_{\rm D2}\simeq {\pi\over e^2k_F}{1\over\epsilon_F}{{\cal D}_3\over {\cal D}_{\rm i}}\left({\cal D}_3-{\cal D}_{\rm i}+{2{\cal D}_2{\cal D}_{\rm i}\over 3{\cal D}_3}\right)~,\qquad {\cal D}_2\ll {\cal D}_3~.
\label{90}
\end{equation}
Equation (\ref{90}) shows that $\rho_{\rm D2}$ can only be smaller than $\rho_{\rm D1}$ in Eq.~(\ref{89}) if the wires are sufficiently far away from each other, so that ${\cal D}_3-{\cal D}_{\rm i}\ll {\cal D}_2\ll {\cal D}_3$. In this limit, $({\cal D}_3/{\cal D}_{\rm i})({\cal D}_3-{\cal D}_{\rm i}+2{\cal D}_2{\cal D}_{\rm i}/3{\cal D}_3)\to 2{\cal D}_2/3$ and friction from $\rho_{\rm D2}$ becomes much larger than that from $\rho_{\rm D1}$ if ${\cal D}_2e^{\epsilon_F/T}\left(T/\epsilon_F\right)^{3/2}\ll {\cal D}_3$. That is,
\begin{equation}
\rho_{\rm D}\simeq {2\pi\over 3e^2k_F}{{\cal D}_2\over\epsilon_F}~,\qquad {\cal D}_2e^{\epsilon_F/T}\left(T/\epsilon_F\right)^{3/2}\ll {\cal D}_3~.
\label{91}
\end{equation}

\begin{figure}
\centerline{\includegraphics[width=0.95\columnwidth]{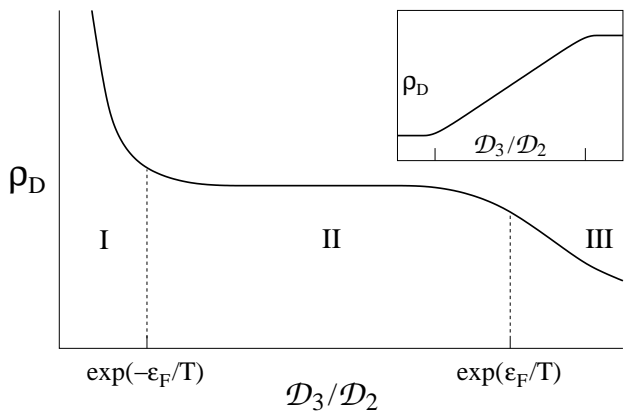}}
\caption{Schematic behavior of the drag resistivity $\rho_{\rm D}$ as a function of ${\cal D}_3/{\cal D}_2$ for fixed ${\cal D}_3$, where ${\cal D}_2$ and ${\cal D}_3$ characterize the strength of two- and three-particle scattering, respectively. Only the exponential factors are shown in the characteristic scales on the horizontal axis. Increasing the distance between the wires for the case of Coulomb interaction leads to a similar behavior of $\rho_{\rm D}$. In the plateau regime, $\rho_{\rm D}$ does not depend on the strength of interwire interactions. The dependence of $\rho_{\rm D}$ on $T$ in three regimes labeled in the figure for $Ta/v_F\gg 1$: (I) $T^{-3/2}e^{-2\epsilon_F/T}$ [Eq.~(\ref{32})], (II) $T^{1/2}e^{-\epsilon_F/T}$ [Eq.~(\ref{89})], (III) const$(T)$ [Eq.~(\ref{91})]. Inset: $\rho_{\rm D}$ as a function of ${\cal D}_3/{\cal D}_2$ for fixed ${\cal D}_2$, illustrates the growth of $\rho_{\rm D}$ with increasing strength of interactions inside the wires. The characteristic scales of ${\cal D}_3/{\cal D}_2$ in the inset are the same as in the main figure.}
\label{f1}
\end{figure}

The behavior of $\rho_{\rm D}$ as a function of ${\cal D}_3/{\cal D}_2$ with ${\cal D}_3$ held fixed is illustrated in Fig.~\ref{f1}. Note a highly nontrivial point: even if ${\cal D}_3$ is entirely due to interactions inside the wires and thus does not lead to any drag directly, $\rho_{\rm D}$ shows a plateau in the dependence on ${\cal D}_2$, i.e., the drag resistivity in this regime does not depend on the strength of interwire interactions [Eq.~(\ref{89})]. In particular, this means that varying the distance between the wires in this regime does not change $\rho_{\rm D}$. On the other hand, $\rho_{\rm D}$ in the plateau regime grows with increasing rate of three-particle scattering inside the wire, although this type of scattering by itself does not lead to any friction between electrons in different wires. Note also that the width of the plateau in the dependence on ${\cal D}_3/{\cal D}_2$ in Fig.~\ref{f1} is exponentially large in the parameter $\epsilon_F/T$. The inset in Fig.~\ref{f1} illustrates the behavior of $\rho_{\rm D}$ with increasing  ${\cal D}_3$ for fixed ${\cal D}_2$ [from Eq.~(\ref{32}) to Eq.~(\ref{89}) to Eq.~(\ref{91})].

\begin{figure}
\centerline{\includegraphics[width=0.95\columnwidth]{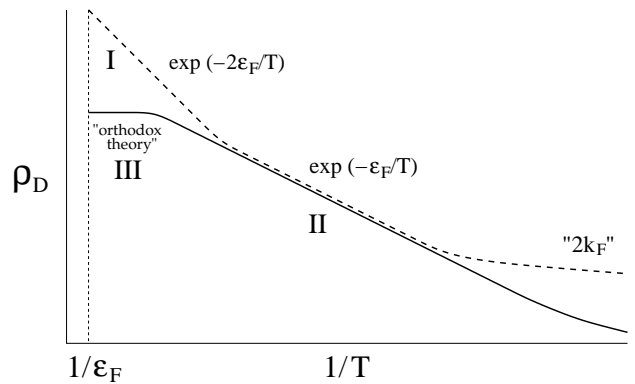}}
\caption{Schematic behavior of the drag resistivity $\rho_{\rm D}$ as a function of $1/T$ on the log-linear scale for two distances between the wires: larger (solid line) and smaller (dashed). Regimes I,II,III are labeled similar to Fig.~\ref{f1}. Drag is hindered by slow thermal equilibration between two electron subsystems with opposite chiralities, which results in the activation behavior of $\rho_{\rm D}$ (regime II: $\rho_{\rm D}\propto e^{-\epsilon_F/T}$, regime I: $\rho_{\rm D}\propto e^{-2\epsilon_F/T}$). In regime II, $\rho_{\rm D}$ does not depend on the distance between the wires. In the low-$T$ regime (labeled with ``$2k_F$"), drag is due to direct backscattering on the Fermi surface. For $T\alt\epsilon_F$, the conventional contribution to $\rho_{\rm D}$ (labeled with ``orthodox theory", regime III) is not suppressed only in the case of sufficiently strong interactions inside the wires in a narrow range of $T$ right below the Fermi energy $\epsilon_F$.}
\label{f2}
\end{figure}

The $T$ dependence of $\rho_{\rm D}$ is shown schematically in Fig.~\ref{f2}. In a wide range of $T$, the drag resistivity follows the Arrhenius plot with the activation energy equal to either $\epsilon_F$ or $2\epsilon_F$. If ${\cal D}_3\gg {\cal D}_2$ at $T\sim \epsilon_F$ (recall that ${\cal D}_3$ scales as $T^2$), there is a range of $T$ below $\epsilon_F$ within which the $T$-independent result of the orthodox theory [Eq.~(\ref{91})] is valid. As $T$ is lowered, $\rho_{\rm D}$ starts to behave as $e^{-\epsilon_F/T}$ [Eq.~(\ref{89})]. If ${\cal D}_3\ll {\cal D}_2$ at $T\sim\epsilon_F$, there is no room for the orthodox theory for $T$ below $\epsilon_F$. Instead, $\rho_{\rm D}$ behaves as $e^{-2\epsilon_F/T}$ [Eq.~(\ref{32})] in a range of $T$ right below $\epsilon_F$, before crossing over into the $e^{-\epsilon_F/T}$ regime. Eventually, drag crosses over into a low-$T$ regime in which it is associated with direct backscattering on the Fermi surface [Eq.~(\ref{32c})].

Another important point to note is that $\rho_{\rm D}$ in Eq.~(\ref{91}) coincides with the result of the orthodox theory \cite{pustilnik03} for $Ta/v_F\gg 1$, i.e., Eq.~(\ref{91}) can be represented as $\rho_{\rm D}\simeq\pi/e^2v_F\tau_{\rm D}^\infty$, where $\tau_{\rm D}^\infty$ is given by Eqs.~(\ref{11a}),(\ref{13a}), see also Eq.~(\ref{b9}). The reason for this is that Eq.~(\ref{91}) describes the limit in which three-particle scattering is strong enough to produce the right-left relaxation rate that is larger than the drag rate $1/\tau_{\rm D}$ [Eq.~(\ref{1.3a})]. In our formalism, $\rho_{\rm D}$ that results from the application of the drift ansatz is thus associated with $\rho_{\rm D2}$ [Eq.~(\ref{88})]. The condition requiring that the equilibration be sufficiently fast severely restricts the range of parameters in which the orthodox theory is valid: for given $T\ll\epsilon_F$, the orthodox theory is only justified if the distance $a$ between the wires is exponentially large in $\epsilon_F/T$, i.e., if drag is exponentially weak in this parameter [cf.\ the condition in Eq.~(\ref{91}), where ${\cal D}_2$ decreases with increasing $a$, whereas ${\cal D}_3$ in the limit of large $a$ is due to triple collisions inside the wires].

One more point worth discussing is the difference in the characteristic momenta $k$ that give the main contribution to the integral in Eq.~(\ref{86}) in two transport regimes, one described by Eqs.~(\ref{32}),(\ref{89}) and the other described by Eq.~(\ref{91}). These are momenta at the very bottom of the spectrum, $|k|\alt (mT)^{1/2}$, in the former case and all momenta below the Fermi surface, $|k|<k_F$, in the latter. We emphasize, however, that the distribution function $f_-(k)$ is sharply peaked at the Fermi surface and the integral over $k$ in Eq.~(\ref{17})---in contrast to Eq.~(\ref{86})---is determined by $|k-k_F|\sim T/v_F$ in both cases. What is different between the two regimes is the range of $k$ for the scattering processes that give the main contribution to the relaxation rate at $|k|\simeq k_F$. In the case of drag limited by the slow right-left relaxation in Eqs.~(\ref{32}),(\ref{89}), this range of $k$ is $|k|\alt (mT)^{1/2}$, as was already discussed in a similar context (for the case of pair collisions) in Sec.~\ref{s2c}. A subtle difference in the shape of the distribution function $f_-(k)=-g_-(k)/4\cosh^2[(\epsilon-\epsilon_F)/2T]$ in the two transport regimes is that $g_-(k)\simeq {\rm const}(k){\rm sgn}(k)$ for all $|k|\gg (mT)^{1/2}$ in the case of Eqs.~(\ref{32}),(\ref{89}), whereas $g_-(k)\propto k$ at $|k|\sim k_F$ in the case of Eq.~(\ref{91}). This means that, near the Fermi surface, electrons with the same chirality are at equilibrium in the {\it stationary} frame in the former case and in the {\it moving} frame in the latter.

\begin{figure}
\centerline{\includegraphics[width=0.95\columnwidth]{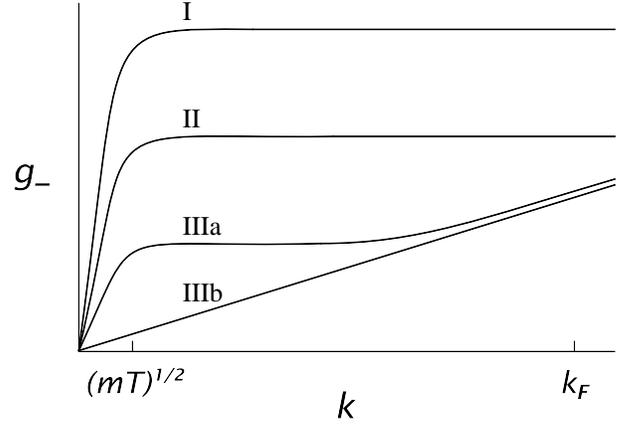}}
\caption{Schematic evolution of the distribution function $g_-(k)$ at and below the Fermi surface with increasing (I$\to$II$\to$IIIa$\to$IIIb) strength of three-particle scattering.
In regimes I and II, electrons with the same chirality are at equilibrium [except for $k\alt (mT)^{1/2}$ near the bottom of the spectrum] in the stationary frame. In regimes IIIa and IIIb, electrons near the Fermi surface are equilibrated in the moving frame. In regime IIIb, equilibrium in the moving frame extends down to $k=0$.
}
\label{g_-(k)}
\end{figure}

The reconstruction of $g_-(k)$ with increasing ${\cal D}_3$, with other parameters fixed, is illustrated in Fig.~\ref{g_-(k)}. Specifically, represent $g_-(k)$ as
\begin{equation}
g_-(k)\simeq {e\over 16}(E_1-E_2)[\,G_1(k)+G_2(k)\,]~
\end{equation}
where $G_1(k)$ and $G_2(k)$ describe the contributions to $g_-(k)$ of the first and second terms in the curly brackets in Eq.~(\ref{78}), respectively. For $G_1(k)$, we have
\begin{eqnarray}
G_1(k)&\simeq&\sqrt{\pi mT}\nonumber\\
&\times&\left\{\begin{array}{ll}\dfrac{e^{2\epsilon_F/T}}{4{\cal D}_2}\Phi\!\left(\dfrac{k}{\sqrt{mT}}\right)~, & {\cal D}_3\ll {\cal D}_2e^{-\epsilon_F/T}~,\\ \\
\dfrac{\sqrt{2}e^{\epsilon_F/T}}{{\cal D}_3}\Phi\!\left(\dfrac{k}{\sqrt{2mT}}\right)~, & {\cal D}_2e^{-\epsilon_F/T}\ll {\cal D}_3~,
\end{array}\right.\nonumber\\
\label{92a}
\end{eqnarray}
where $\Phi(x)$ is the error function [as defined below Eq.~(\ref{24})]. The term $G_1(k)$ determines $g_-(k)$ in regimes I and II in Fig.~\ref{g_-(k)}, with $G_2(k)\ll G_1(k)$ for all $k$. Regime I corresponds to the first line in Eq.~(\ref{92a}), regime II to the second.

The term $G_2(k)$ for ${\cal D}_2\ll {\cal D}_3$ is given by
\begin{equation}
G_2(k)\simeq {6\over {\cal D}_2}{\epsilon_F\over T}k~,\quad {\cal D}_2\ll {\cal D}_3~.
\label{92b}
\end{equation}
As ${\cal D}_3$ increases, the crossover from regime II to regime III occurs at $G_2(k_F)\sim G_1(k_F)$, i.e., at ${\cal D}_3\sim {\cal D}_2e^{\epsilon_F/T}\left(T/\epsilon_F\right)^{3/2}$. In regime IIIa in Fig.~\ref{g_-(k)}, the function $g_-(k)$ below the Fermi surface is equilibrated in the moving frame for $k\gg k_*=k_F({\cal D}_2/{\cal D}_3)(T/\epsilon_F)^{3/2}e^{\epsilon_F/T}$, while for $k\ll k_*$ it is still equilibrated in the stationary frame. Equilibrium in the moving frame extends down to $k=0$ (regime IIIb in Fig.~\ref{g_-(k)}) at larger ${\cal D}_3$, namely ${\cal D}_3\gg {\cal D}_2e^{\epsilon_F/T}T/\epsilon_F$.

The behavior of $\rho_{\rm D}$ with varying $T$ [${\cal D}_3(T)$ can be represented as ${\cal D}_3(\epsilon_F)(T/\epsilon_F)^2$] and strength of inter- and intrawire interactions (${\cal D}_2$ parametrizes the strength of interwire interactions and decreases as the distance $a$ between the wires is increased, whereas ${\cal D}_3$ in the limit of large $a$ is determined by interactions inside the wires) is conveniently summarized in the following form:
\begin{eqnarray}
\rho_{\rm D}\simeq {2\over 3}{\pi\over e^2k_F}{{\cal D}_2\over\epsilon_F}\times\left\{
\begin{array}{ll}
\dfrac{48}{\sqrt{2\pi}}\left(\dfrac{\epsilon_F}{T}\right)^{3/2}e^{-2\epsilon_F/T}~, & {\rm I}~,\\ \\
\dfrac{6}{\sqrt{\pi}}\dfrac{{\cal D}_3}{{\cal D}_2}\left(\dfrac{\epsilon_F}{T}\right)^{3/2}e^{-\epsilon_F/T}~, & {\rm II}~,\\ \\
1~, & {\rm III}~,
\end{array}\right.\nonumber\\
\label{92}
\end{eqnarray}
where regimes I, II, III correspond to those in Figs.~\ref{f1}-\ref{f2}. Note that regime III, when present, is always separated by regime II from the ``$2k_F$"-regime (Fig.~\ref{f2}) in the limit of $k_Fa\gg 1$, since in this limit $(1/k_Fa)\ln[{\cal D}_3(\epsilon_F)/{\cal D}_2]$ is small compared to unity. This is because ${\cal D}_3(\epsilon_F)/{\cal D}_2\sim \beta_{\rm f}^2k_F^2a^2$, where $\beta_{\rm f}$ parametrizes the strength of forward scattering, cf.\ Eq.~(\ref{1.1}). As a result, for $k_Fa\gg 1$, the drift-ansatz regime for the case of drag dominated by forward scattering can only be realized if $Ta/v_F\gg 1$, when the orthodox theory yields $T$-independent drag [Eq.~(\ref{92}), regime III; Eq.~(14) and the plateau regime in Fig.~2 in Ref.~\onlinecite{pustilnik03}]. In the opposite limit of $k_Fa\ll 1$, drag is determined by backward scattering on the Fermi surface (regime ``$2k_F$" in Fig.~\ref{f2}) for all $T\alt\epsilon_F$ considered in this paper. This implies, in turn, that there is no room for the $T^2$ drag resistivity [Eq.~(\ref{1.1})] even if the thermal equilibration is strong enough to establish the drift-ansatz regime.

We are now in a position to return to the case of Coulomb interaction (recall the discussion at the end of Sec.~\ref{s3b1}). As we see from the calculation for $Ta/v_F\gg 1$ [where $1/a$ was assumed to be a single scale characterizing both functions $V_{11}(q)$ and $V_{12}(q)$, beyond which they fall off fast enough to neglect momentum transfer with $|q|a\gg 1$], the main contribution to $\rho_{\rm D}^{-1}$ in the whole range of ${\cal D}_3/{\cal D}_2\ll e^{\epsilon_F/T}(T/\epsilon_F)^{3/2}$ comes from $\rho_{\rm D1}^{-1}$, with the integral term in the current in momentum space producing only a small correction to $\rho_{\rm D}$. The resulting drag resistivity is determined by scattering of cold electrons with $|k|\ll k_F$.
The characteristic energy transfer for these electrons is much smaller than $T$ even if that for electrons on the Fermi surface is of the order of $T$, as is the case for the intrawire three-particle scattering due to Coulomb interaction. It follows that the Fokker-Planck description of drag for ${\cal D}_3/{\cal D}_2\ll e^{\epsilon_F/T}(T/\epsilon_F)^{3/2}$ is also accurate for the Coulomb case. However, the diffusion coefficient in channel (a), $D^{(3a)}(k)$ [Eq.~(\ref{62})], should be calculated in the Coulomb case [$V_{11}(q)$ from Eq.~(\ref{a2})] without treating the thermal factors in Eq.~(\ref{49}) as smoothly changing functions of $q$ compared to the matrix elements, in contrast to Eq.~(\ref{75a}). Assuming that ${\cal D}_3$ is mainly due to three-particle scattering inside the wires, we obtain
\begin{equation}
{\cal D}_3={\pi(\ln 2)^2\over 15}{T^5\over \epsilon_F^2v_F^2}\left({e^2\over v_F}\right)^4\ln^2\!\left({v_F\over Td_0}\right)
\end{equation}
for $Td/v_F\gg 1$, while for smaller $T$ the diffusion coefficient in channel (a) acquires four more powers of $T$. In the opposite limit of fast equilibration in the frame moving with the drift velocity, i.e., for ${\cal D}_3/{\cal D}_2\gg e^{\epsilon_F/T}(T/\epsilon_F)^{3/2}$, the three-particle rate drops out from the expression for $\rho_{\rm D}$, independently of the character of three-particle scattering. Therefore, Eq.~(\ref{91}) describes the Coulomb case as well.

\section{Summary}
\label{s4}

We have presented a theory of Coulomb drag between clean (no disorder) quantum wires based on the kinetic equation approach. One conceptually important aspect of Coulomb drag that we have highlighted in this paper is an inherent link between this phenomenon and the processes of thermal equilibration. We have demonstrated that the dc drag resistivity $\rho_{\rm D}$ is exactly zero in the absence of equilibration between right- and left-moving electrons. Another way to state this is that forward scattering near the Fermi surface with small momentum transfer is not sufficient to produce a nonzero drag resistivity.

We have given a detailed discussion of the equilibration processes in quantum wires. Crucially, in one-dimensional geometry, the right-left equilibration requires backscattering---either directly in the vicinity of the Fermi surface or via diffusion in energy space with small energy transfer in one scattering event. The latter type of backscattering is favored if the wires are not too close to each other. We have shown that the slow diffusion in energy space is bottlenecked by scattering of cold electrons at the bottom of the spectrum, as a result of which $\rho_{\rm D}$ shows an activation behavior---in contrast to the conventional for the drag effect power-law dependence on the temperature---with the activation energy equal to the Fermi energy $\epsilon_F$ or $2\epsilon_F$, for the cases of three- or two-particle scattering, respectively. We have demonstrated a nontrivial interplay between the pair and triple collisions; in particular, $\rho_{\rm D}$ in a wide range of the parameters of the problem does not depend on the strength of interwire interactions, while depending strongly on the strength of interactions inside the wires.

\acknowledgments

We thank D.~Aristov, D.~Bagrets, L.~Glazman, V.~Gurevich, A.~Ioselevich, V.~Kachorovskii, M.~Muradov, B.~Narozhny, P.~Ostrovsky, and M.~Pustilnik for interesting discussions. The work was supported by the DFG/CFN, the DFG-RFBR project within the framework of the DFG/SPP ``Halbleiter Spintronik", GIF Grant No.\ 965, the RFBR, Rosnauka Grant No.\ 02.740.11.5072, and RF President Grant No.\ NSh-5442.2012.2.

\appendix

\section{Interwire interaction potential}
\label{aA}
\renewcommand{\theequation}{A.\arabic{equation}}
\setcounter{equation}{0}

Let us denote as ${\rm v}_{11}(x)$ and ${\rm v}_{12}(x)$ the potentials of interaction between electrons residing in one wire and in different wires, respectively. They are expressed in terms of the potential ${\rm v}(r)$ created by a charge in the plane of the wires (where $r$ is the distance in this plane) as ${\rm v}_{11}(x)={\rm v}(|x|)$ and ${\rm v}_{12}(x)={\rm v}(\sqrt{x^2+a^2})$, where $a$ is the distance between the wires. The potential ${\rm v}(r)$ is determined by the polarization properties of the medium around the wires; in particular, by the position and dimensions of a nearby metallic gate. In general, the relation between the characteristic spatial scales of ${\rm v}(r)$ and ${\rm v}_{12}(x)$ depends in an essential way on the shape of ${\rm v}(r)$. For instance, if ${\rm v}(r)$ is a monotonically decaying function characterized by a single spatial scale $d$, the characteristic radius of ${\rm v}_{12}(x)$ is given by $d$ for $d\gg a$, while for $d\ll a$ it may be either larger or smaller than $d$, depending on whether ${\rm v}(r)$ decreases slower or faster than the Gaussian function. Note also that ${\rm v}_{12}(x)$ is not necessarily characterized by a single scale even if ${\rm v}(r)$ is a single-scaled function.

For definiteness, let us consider ${\rm v}(r)$ in the presence of a perfectly screening metallic plate located at a distance $d$ from the wires (parallel to them). Then
\begin{equation}
{\rm v}_{11}(x)=e^2\left({1\over \sqrt{x^2+d_0^2}}-{1\over \sqrt{x^2+d_0^2+4d^2}}\right)~, \label{a1}
\end{equation}
where $d_0$ is the ``radius of the wire" (which is supposed to be the smallest spatial scale in the problem) and the dielectric constant of the medium in which the wires are imbedded is set to be equal to 1. The Fourier-component of ${\rm v}_{11}(x)$ from Eq.~(\ref{a1}) is given by
\begin{eqnarray}
V_{11}(q)&=&2e^2\left[\,K_0(qd_0)-K_0\left(q\sqrt{d_0^2+4d^2}\right)\,\right]\nonumber\\
&\simeq& 2e^2\ln\left({1\over\max\{|q|,d^{-1}\}d_0}\right)~, \label{a2}
\end{eqnarray}
where $K_0(x)$ is the Macdonald function. The Fourier-component $V_{12}(q)$ of the potential ${\rm v}_{12}(x)$ is given by Eq.~(\ref{a2}) with the change $d_0^2\to d_0^2+a^2\simeq a^2$ in the argument of the Macdonald functions and shows the following behavior depending on whether the distance to the gate is larger or smaller than the distance between the wires:
\begin{eqnarray}
\hspace{-0.45cm}V_{12}(q)\simeq \left\{
\begin{array}{ll}
2e^2\ln (d/a)~, & \!\!\!|q|\ll 1/d\ll 1/a~,\\
2e^2\ln (1/|q|a)~, & \!\!\!1/d\ll |q|\ll 1/a~,\\
4e^2d^2/a^2~, & \!\!\!|q|\ll 1/a\ll 1/d~,\\
e^2(2\pi/|q|a)^{1/2}e^{-|q|a} \\
\times\left(1-e^{-2|q|d^2/a}\right), & \,\,\,\,\,1/a\ll |q|~.
\end{array}\right.
\label{a3}
\end{eqnarray}
Note the emergence of the characteristic scale $|q|\sim a/d^2$ in the factor in the last line of Eq.~(\ref{a3}). One can see, however, that the characteristic scale of $|q|$ on which $V_{12}(q)$ starts to decay exponentially with increasing $|q|$ is the inverse distance between the wires $1/a$, {\it independently} of the ratio $a/d$.

\section{Relation between Eqs.~(\ref{10}),(\ref{11}) and the orthodox theory}
\label{aB}
\renewcommand{\theequation}{B.\arabic{equation}}
\setcounter{equation}{0}

The drift ansatz of Ref.~\onlinecite{pustilnik03} [see the discussion in Sec.~\ref{s1} below Eq.~({\ref{1.3})] is the result of an extension of the orthodox theory of drag \cite{jauho93,zheng93,kamenev95,flensberg95} to one dimension. As shown in Secs.~\ref{s2b} and \ref{s2c}, this approach fails totally in one dimension for the description of bulk drag due to forward scattering. In this appendix, we rewrite the kinetic equation (\ref{10}) in the form that allows one to explicitly identify the approximation that is made in the orthodox theory but contradicts the solution of the kinetic equation. To this end, let us represent the collision integral (\ref{11}) in terms of the equilibrium polarization operators
\begin{equation}
\Pi(\omega,q)=\int\!{dk\over 2\pi}\,{f_T(k+q)-f_T(k)\over\omega+i0-[(k+q)^2-k^2]/2m}
\label{b1}
\end{equation}
for two wires [cf.\ Eq.~(\ref{1.3})], whose imaginary parts are given by
\begin{equation}
{\rm Im}\,\Pi(\omega,q)=-\frac{m}{|q|}\left[ f_T\left( \frac{m\omega}{q}+\frac{q}{2}\right)-f_T\left(\frac{m\omega}{q}-\frac{q}{2}\right) \right]~.
\label{b2}
\end{equation}
Using the identity
\begin{equation}
f_T(k')[\,1-f_T(k)\,]=\frac{f_T(k')-f_T(k)}{1-\exp[\,(\epsilon'-\epsilon)/T\,]}~,
\label{b3}
\end{equation}
Eq.~(\ref{11}) is rewritten as
\begin{widetext}
\begin{equation}
{\rm st}_-\{g\}=\frac{2 m}{\zeta^2(k)} \int\!{dk'\over 2\pi}\,{|V(k'-k)|^2\over |k'-k|}
\,\frac{[\,f_T(k')-f_T(k)\,]^2}{\sinh^2[\,(\epsilon'-\epsilon)/2T\,]}\,[\,g_-(k')-g_-(k)\,]~.
\label{b4}
\end{equation}
Combining Eqs.~(\ref{b2}),(\ref{b4}) and changing variables to $q=k-k'$ and $\omega=\epsilon-\epsilon'$, we get
\begin{equation}
{\rm st}_-\{g\}=\frac{2}{m \zeta^2(k)}\int\!d\omega\!\int\!{dq\over 2\pi}\,{|q| |V(q)|^2
\,\left[\,{\rm Im}\,\Pi(\omega,q)\,\right]^2\over \sinh^2(\omega/2T)}\,\left[\,g_-(k-q)-g_-(k)\,\right]\, \delta\!\left(\omega-{k q\over m}+{q^2\over 2 m}\right)~.
\label{b5}
\end{equation}
Substituting Eq.~(\ref{b5}) into Eq.~(\ref{10}), multiplying the kinetic equation by $ek\zeta^2/4 m$, and integrating over $k$, we have the equation for $j_-$ of the form
\begin{equation}
-i\omega j_--{e^2(E_1-E_2)n\over 2m}={e\over 4m}\int\!{d\omega'\over 2\pi} \int\!{dq\over 2\pi}\,{q|V(q)|^2
\left[\,{\rm Im}\Pi(\omega',q)\,\right]^2\over\sinh^2(\omega'/2T)}\,\left[\,g_-\left(\frac{m\omega'}{q}-\frac{q}{2}\right)-g_-\left(\frac{m\omega'}{q}+\frac{q}{2}\right)\,\right]~.
\label{b6}
\end{equation}
The result for the dc drag resistivity obtained in Ref.~\onlinecite{pustilnik03} corresponds to the drift-ansatz replacement
\begin{equation}
g_-\left(\frac{m\omega'}{q}+\frac{q}{2}\right)-g_-\left(\frac{m\omega'}{q}-\frac{q}{2}\right) \to \frac{q}{e n T} j_-
\label{b7}
\end{equation}
in Eq.~(\ref{b6}) [i.e., $g_-(k)\to kj_-/enT$] at $\omega\to 0$. If one employs the drift ansatz (\ref{b7}) for finite $\omega$ as well, this leads to
\begin{equation}
j_- = \frac{e^2(E_1-E_2)n}{2m\left(-i\omega + 2/\tau_{\rm D}^\infty\right)}
\label{b8}
\end{equation}
with
\begin{equation}
\frac{1}{\tau_{\rm D}^\infty}={1\over 8 nmT}
\int\!\frac{d\omega}{2\pi}\int\!{dq\over 2\pi}\,\frac{q^2 |V(q)|^2\,
\left[\,{\rm Im}\,\Pi(\omega,q)\,\right]^2}{\sinh^2(\omega/2T)}
\label{b9}
\end{equation}
[Eq.~(\ref{b9}) coincides with Eq.~(\ref{11a})].
The Lorentzian shape of the $\omega$ dispersion for $j_-$ in Eq.~(\ref{b8}), with the $\omega$ independent damping rate (\ref{b9}), was posited in Ref.~\onlinecite{aristov07}. In fact, however, as discussed in Sec.~\ref{s2b}, the damping rate shows a strong dependence on $\omega$, vanishing  in the dc limit within the model of Refs.~\onlinecite{pustilnik03,aristov07,rozhkov08}.

It is also instructive to note that Eq.~(\ref{b5}) clearly demonstrates that the contact drag resistance, discussed in Sec.~\ref{s1.5}, depends on the setup. Indeed, in the limit of short wires (in which the distribution function is only slightly modified by drag), one can substitute in the collision integral the ``unperturbed" distribution function incident from the leads. The result depends in an essential way on whether the leads supply the drift-ansatz distribution function [Eq.~(\ref{b7})], corresponding to equilibrium in the moving frame, or the distribution function that is equilibrium in the stationary frame. In the latter case (Fermi leads), $g_-(k)\propto {\rm sgn}(k)$ and thus drops out of Eq.~(\ref{b5}) for all $q$ such that $k$ and $k-q$ in Eq.~(\ref{b5}) belong to the same chiral branch (forward-scattering drag). This leads to a strong suppression of drag compared to the orthodox theory.

\section{Three-particle scattering amplitude}
\label{aE}
\renewcommand{\theequation}{C.\arabic{equation}}
\setcounter{equation}{0}

Explicitly, the normalized determinants in Eqs.~(\ref{34a})-(\ref{34c}) read
\begin{eqnarray}
&&{\rm D}_a(k_1,k_2,k_3)={1\over (6L^3)^{1/2}}\begin{vmatrix} e^{ik_1x_1} & e^{ik_1x_2} & e^{ik_1x_3} \\ e^{ik_2x_1} & e^{ik_2x_2} & e^{ik_2x_3} \\ e^{ik_3x_1} & e^{ik_3x_2} & e^{ik_3x_3} \end{vmatrix}~,\nonumber\\ &&{\rm D}_b(k_2,k_3)={1\over (2L^2)^{1/2}}\begin{vmatrix} e^{ik_2x_2} & e^{ik_2x_3} \\ e^{ik_3x_2} & e^{k_3x_3} \end{vmatrix}~,
\quad {\rm D}_c(k_1,k_2)={1\over (2L^2)^{1/2}}\begin{vmatrix} e^{ik_1x_1} & e^{ik_1x_2} \\ e^{ik_2x_1} & e^{k_2x_2} \end{vmatrix}~.\label{e1}
\end{eqnarray}
The matrix elements (\ref{38}) are written as
\begin{eqnarray}
&&A^{(1)}_a(1,2,3|4,5,6)={1\over L}\left\{\left[\delta_{k_3,k_6}\delta_{k_1+k_2,k_4+k_5}V_{11}(k_1-k_4)+\delta_{k_1,k_4}
\delta_{k_2+k_3,k_5+k_6}V_{11}(k_2-k_5)\right.\right.\nonumber\\ &&\hspace{3.25cm}+\left.\left.\delta_{k_2,k_5}
\delta_{k_1+k_3,k_4+k_6}V_{11}(k_3-k_6)\right]-(k_2\leftrightarrow k_3)\right\}\nonumber\\ &&\hspace{3.25cm}-(k_1\leftrightarrow k_2)-(k_1\leftrightarrow k_3)~,\\
&&A^{(1)}_b(1,2,3|4,5,6)={1\over L}\left[\delta_{k_3,k_6}\delta_{k_1+k_2,k_4+k_5}V_{12}(k_1-k_4)+\delta_{k_1,k_4}\delta_{k_2+k_3,k_5+k_6}V_{11}(k_2-k_5)\right.\nonumber\\
&&\hspace{3.25cm}+\left.\delta_{k_2,k_5} \delta_{k_1+k_3,k_4+k_6}V_{12}(k_3-k_6)\right]-(k_2\leftrightarrow k_3)~,\\ &&A^{(1)}_c(1,2,3|4,5,6)={1\over
L}\left[\delta_{k_3,k_6}\delta_{k_1+k_2,k_4+k_5}V_{11}(k_1-k_4)+\delta_{k_1,k_4}\delta_{k_2+k_3,k_5+k_6}V_{12}(k_2-k_5)\right.\nonumber\\
&&\hspace{3.25cm}+\left.\delta_{k_2,k_5} \delta_{k_1+k_3,k_4+k_6}V_{12}(k_3-k_6)\right]-(k_1\leftrightarrow k_2)
\end{eqnarray}
(the terms in the third line for $A^{(1)}_a$ are understood to exchange momenta in the whole expression within the curly brackets, i.e., for 3 ``direct" terms there are 15 exchange terms). The irreducible parts of the amplitudes $A_{a,b,c}$ [Eq.~(\ref{37})] are given by
\begin{eqnarray}
A_a^{\rm irr}(1',2',3'|1,2,3)&=&{1\over L^2}\delta_{k_1+k_2+k_3,k_{1'}+k_{2'}+k_{3'}}\nonumber\\ &\times&\left\{\left[\,\,V_{11}(k_3-k_{3'})V_{11}(k_1-k_{1'})\left({1\over
\Delta_{233'}}+{1\over \Delta_{211'}}\right)\right.\right.\nonumber\\ &&\left.+\,V_{11}(k_1-k_{1'})V_{11}(k_2-k_{2'})\left({1\over \Delta_{311'}}+{1\over
\Delta_{322'}}\right)\right.\nonumber\\ &&\left.\left.\!+\,V_{11}(k_2-k_{2'})V_{11}(k_3-k_{3'})\left({1\over \Delta_{122'}}+{1\over
\Delta_{133'}}\right)\,\right]-(k_{2'}\leftrightarrow k_{3'})\right\}~\nonumber\\
&&\,-\,(k_{1'}\leftrightarrow k_{2'})-(k_{1'}\leftrightarrow k_{3'})~, \label{43a}\\
A_b^{\rm irr}(1',2',3'|1,2,3)&=&{1\over L^2}\delta_{k_1+k_2+k_3,k_{1'}+k_{2'}+k_{3'}}\nonumber\\ &\times&\,\left[\,\,V_{11}(k_3-k_{3'})V_{12}(k_1-k_{1'})\left({1\over
\Delta_{233'}}+{1\over \Delta_{211'}}\right)\right.\nonumber\\ &&\left.+\,V_{12}(k_1-k_{1'})V_{11}(k_2-k_{2'})\left({1\over \Delta_{311'}}+{1\over
\Delta_{322'}}\right)\right.\nonumber\\ &&\left.+\,V_{12}(k_2-k_{2'})V_{12}(k_3-k_{3'})\left({1\over \Delta_{122'}}+{1\over
\Delta_{133'}}\right)\,\right]-(k_{2'}\leftrightarrow k_{3'})~, \label{43b}\\
A_c^{\rm irr}(1',2',3'|1,2,3)&=&{1\over L^2}\delta_{k_1+k_2+k_3,k_{1'}+k_{2'}+k_{3'}}\nonumber\\
&\times&\,\left[\,\,V_{12}(k_3-k_{3'})V_{11}(k_1-k_{1'})\left({1\over \Delta_{233'}}+{1\over \Delta_{211'}}\right)\right.\nonumber\\
&&\left.+\,V_{12}(k_1-k_{1'})V_{12}(k_2-k_{2'})\left({1\over \Delta_{311'}}+{1\over \Delta_{322'}}\right)\right.\nonumber\\
&&\left.+\,V_{11}(k_2-k_{2'})V_{12}(k_3-k_{3'})\left({1\over \Delta_{122'}}+{1\over \Delta_{133'}}\right)\,\right]-(k_{1'}\leftrightarrow k_{2'})~,
\label{43c}
\end{eqnarray}
where
\begin{equation}
\Delta_{233'}=\epsilon_2+\epsilon_3-\epsilon_{3'}-\epsilon_{2+3-3'}=-{1\over m}(k_{3'}-k_3)(k_{3'}-k_2)~,\quad {\rm etc.}~,
\label{44}
\end{equation}
\end{widetext}
$\epsilon_{2+3-3'}=(k_2+k_3-k_{3'})^2/2m$. The sign $(k_2\leftrightarrow k_3)$ means that only the momenta $k_2$ and $k_3$ are transposed (but $k_{2'}$ and $k_{3'}$ are not). The amplitudes $A_{a,b,c}^{\rm irr}(1',2',3'|1,2,3)$ in Eqs.~(\ref{43a})-(\ref{43c}) coincide with those derived in Ref.~\onlinecite{lunde07} (see also Refs.~\onlinecite{levchenko11,levchenko11a}). It is worth mentioning once more, however, that while the amplitude of three-particle scattering is the same in our work and in Refs.~\onlinecite{lunde07,levchenko11,levchenko11a}, the corresponding contributions to the collision integral are not. This is because the combinatorial factors in Eqs.~(\ref{40})-(\ref{42}), necessary to prevent double counting of the initial and final states in the collision integral, are missing in Refs.~\onlinecite{lunde07,levchenko11,levchenko11a}.

Note that if it were not for the difference between ${\rm v}_{11}(x)$ and ${\rm v}_{12}(x)$, the ``direct" terms (as opposed to the exchange terms) in the amplitudes $A_b^{\rm irr}(1',2',3'|1,2,3)$ and $A_c^{\rm irr}(1',2',3'|1,2,3)$ [those shown in Eqs.~(\ref{43b}),(\ref{43c}) with the positive sign] would be expressible as series resulting from the cyclic permutations ($k_1k_{1'}\to k_2k_{2'}\to k_3k_{3'}\to k_1k_{1'}$). Moreover, the direct terms would then become the same in $A_b^{\rm irr}(1',2',3'|1,2,3)$ and $A_c^{\rm irr}(1',2',3'|1,2,3)$. In fact, the whole kinetic problem for three-particle scattering of spinless electrons in a double wire would then become identical to that for three-particle scattering of spinful electrons in a single wire with spin-independent interaction. Our drag problem, in which generically ${\rm v}_{11}(x)\neq {\rm v}_{12}(x)$ and the structure of $A_b^{\rm irr}(1',2',3'|1,2,3)$ and $A_c^{\rm irr}(1',2',3'|1,2,3)$ is therefore less symmetric, can be viewed as a generalization of the spinful problem in a single wire to the case of Ising-type anisotropy of the interaction potential in spin space.

\section{Cancellation of three-particle singularities in one dimension}
\label{aC}
\renewcommand{\theequation}{D.\arabic{equation}}
\setcounter{equation}{0}

As discussed in Sec.~\ref{s3a1}, one of the important differences between two- and three-particle scattering is the occurrence of nonintegrable singularities in the differential cross-section in the three-particle case. These occur if the cross-section is written as the modulus squared of (the connected part of) the three-particle $T$-matrix---this would be a straightforward extension of the conventional formalism for the two-particle case. In fact, the finite collision integral that describes triple collisions in the kinetic equation contains a counterterm [Eq.~(\ref{51})] that cancels the contribution of the singularities. The purpose of this appendix is to provide technical details that explicitly demonstrate the cancellation between the essential singularities in the cross-sections of many-particle scattering. Specifically, we focus here on the singular behavior of three-particle scattering in the case of one-dimensional electrons.

The amplitude of three-particle scattering in Eqs.~(\ref{43a})-(\ref{43c}) shows a pole-type singularity as a function of momenta each time the energy $\Delta$, endowed with indices according to the definition in Eq.~(\ref{44}), transferred in the virtual transition into the intermediate state is equal to zero. In the case of scattering of type (a), when all colliding electrons are in the same wire, the residue of each of the poles can be shown to vanish linearly in $\Delta$---i.e., the singularity is, in fact, absent---provided the initial and final momenta of the three-particle amplitude conserve total momentum and total energy as $\Delta$ varies. Importantly, the regular behavior of the amplitude in channel (a) at $\Delta=0$ results from a compensation of the direct and exchange processes in the residue (for a calculation of the total scattering rate in the spinless case see Ref.~\onlinecite{khodas07}). For the amplitude of three-particle scattering that involves electrons from different wires, the compensation is not complete---because the exchange interaction in the absence of tunneling between the wires is only allowed within the same wire---and the residue does not vanish (a similar situation occurs for spinful electrons in a single wire). Thus triple collisions between electrons belonging to different wires yield a nonintegrable singularity in the modulus squared of the three-particle $T$-matrix: at second order in the interaction potential for the amplitude, the singularity in the differential cross-section is of the type $1/\Delta^2$.

There is one more important aspect of the divergency of the thus defined triple-collision rate that is specific to one dimension. The divergency does not rely on a particular form of the dispersion law; in particular, the singularity is present---and remains nonintegrable---in the limit $1/m\to 0$. The divergent triple-collision rate for electrons with a linear dispersion relation in one dimension raises the question as to how the kinetic equation approach relates to the Dzyaloshinskii-Larkin theorem \cite{dzyaloshinskii74,giamarchi04} which says that, at thermal equilibrium, the one-dimensional electron system with a linear dispersion relation is exactly described in terms of the random-phase approximation. This approximation includes pair collisions only. That is, according to the theorem (and the whole bosonization approach \cite{giamarchi04} for that matter), triple collisions are ``exactly absent" at equilibrium. The condition of equilibrium is important; however, the divergency occurs at the level of the structure of the kernel of the collision integral, so that, e.g., the out-scattering rate from Eqs.~(\ref{43a})-(\ref{43c}) diverges in the linear-response limit as well, similar to the inverse lifetime of a particle due to triple collisions at equilibrium. Below, we resolve the apparent conflict between the Dzyaloshinskii-Larkin theorem and the divergency in the three-particle scattering channel by calculating the scattering rate at order $V_{12}^4$ ``by brute force" diagrammatically for an arbitrary dispersion relation $\xi_k$.

The singularity at zero $\Delta$ in the matrix elements in either channel (b) or (c) is not related to the difference between $V_{11}(q)$ and $V_{12}(q)$ (the singularity survives when the difference is neglected) but is only due to the ``lack" of exchange processes in these channels compared to channel (a). Since the singularity is entirely associated with scattering of electrons belonging to different wires, we neglect intrawire interactions throughout Appendix \ref{aC}. Moreover, since our purpose in this appendix is to illustrate the principle (discussed in Sec.~\ref{s3a}) on which the cancellation of the $1/\Delta^2$ divergencies is based, we do not calculate here the full set of out- and in-scattering nonequilibrium self-energies for two- and three-particle scattering but focus on the simplest quantity that exemplifies the problem. This is the inverse electron lifetime in an equilibrium electron bath, expanded in $V_{12}(q)$ to fourth order. In this calculation, the inverse lifetime will be seen to be a well-behaved scattering rate that experiences no infrared divergency from the vicinity of the point $\Delta=0$. The quantities of interest are thus the inverse lifetimes for an electron with momentum $k$ in channels (b) and (c),
\begin{equation}
1/\tau_{b,c}(k)=-2{\rm Im}\,\Sigma_{b,c}(i\epsilon_n\to\xi_k+i0,k)~,
\label{c1}
\end{equation}
where $\Sigma_{b,c}(i\epsilon_n,k)$ are the corresponding electron self-energies in the Matsubara representation, at order $V_{12}^4$ and zeroth order in $V_{11}$.

One can separate the contributions of direct ($H$) and exchange ($F$) processes in the self-energy in Eq.~(\ref{c1}),
\begin{equation}
\Sigma_{b,c}=\Sigma^H_{b,c}+\Sigma^F_{b,c}~.
\end{equation}
For the case of triple collisions, the $H$-term comes from the sum squared of the terms with sign $+$ in Eq.~(\ref{43b}) or (\ref{43c}) [for channel (b) and (c), respectively] plus the sum squared of the terms with sign $-$, while the $F$-term is given by twice the product of the two sums. In fact, the self-energy $\Sigma_{b,c}$ contains also a contribution of pair collisions at order $V_{12}^4$, for which one can similarly separate the direct and exchange processes. The role of pair collisions will be discussed below Eq.~(\ref{c23}). Since in channels (b) and (c) the $H$- and $F$-terms do not compensate each other, it suffices---for the purpose of describing the divergency of the triple-collision rate---to focus on one of the terms: below, we write down details of the calculation for the $H$-term only.

Let us begin with channel (c)---by calculating the scattering rate for an electron in wire 1 due to interaction with two other electrons, one of which is in wire 1 and the other is in wire 2. The self-energy $\Sigma_c^H$ of fourth order in interaction for the case $V_{11}=0$ is given by the diagram in Fig.~\ref{chan_c}a, where the thick wavy line is the effective interaction $V(i\Omega_m,q)$, shown in Fig.~\ref{chan_c}b and written as
\begin{eqnarray}
\phantom{a}\hspace{-1cm}V(i\Omega_m,q)&=&|V_{12}(q)|^2T\sum_{m'}\int\!{dq'\over 2\pi}\,|V_{12}(q')|^2\nonumber\\
&\times&A(i\Omega_m,q|i\Omega_{m'},q')\Pi(i\Omega_{m'},q')~.
\label{c2}
\end{eqnarray}
Here the polarization operator $\Pi(i\Omega_m,q)$ is the generalization of Eq.~(\ref{b1}) to arbitrary $\xi_k$,
\begin{equation}
\Pi(i\Omega_m,q)=\int\!{dk\over 2\pi}\,{f_T(k+q)-f_T(k)\over i\Omega_m-\xi_{k+q}+\xi_k}~,
\label{c3}
\end{equation}
and $A(i\Omega_m,q|i\Omega_{m'},q')$ is a sum of the four-leg loops in Fig.~\ref{four_leg} over all nonequivalent insertions of {\it one} of the legs:
\begin{widetext}
\begin{eqnarray}
A(i\Omega_m,q|i\Omega_{m'},q')=-T\sum_n\int\!{dk\over 2\pi}\!\!\!\!\!\!\!\!&&\,\,\,\left[\,{1\over (i\epsilon_n-\xi_k)(i\epsilon_n-i\Omega_m-\xi_{k-q})(i\epsilon_n-i\Omega_{m'}-\xi_{k-q'})(i\epsilon_n-i\Omega_{m-m'}-\xi_{k-q-q'})}\right.\nonumber\\
&&+\,\,{1\over (i\epsilon_n-\xi_k)^2 (i\epsilon_n-i\Omega_m-\xi_{k-q})(i\epsilon_n-i\Omega_{m'}-\xi_{k-q'})}\nonumber\\
&&+\left.{1\over (i\epsilon_n-\xi_k)(i\epsilon_n-i\Omega_m-\xi_{k-q})^2(i\epsilon_n-i\Omega_{m-m'}-\xi_{k-q-q'})}\,\right]~.
\label{c4}
\end{eqnarray}
For the case of a linear dispersion relation, $A$ vanishes (apart from the uncertainty at $i\Omega_m\to\xi_k+i0$ and similarly for $i\Omega_{m'}$) in agreement with the loop cancellation (Dzyaloshinskii-Larkin) theorem. \cite{dzyaloshinskii74,giamarchi04} However, as already mentioned above, the divergency of the integral of the modulus squared of the three-particle $T$-matrix elements over $\Delta$ around $\Delta=0$ survives the limit of the linearized dispersion law.
\begin{figure}
\centerline{\includegraphics[width=0.9\columnwidth]{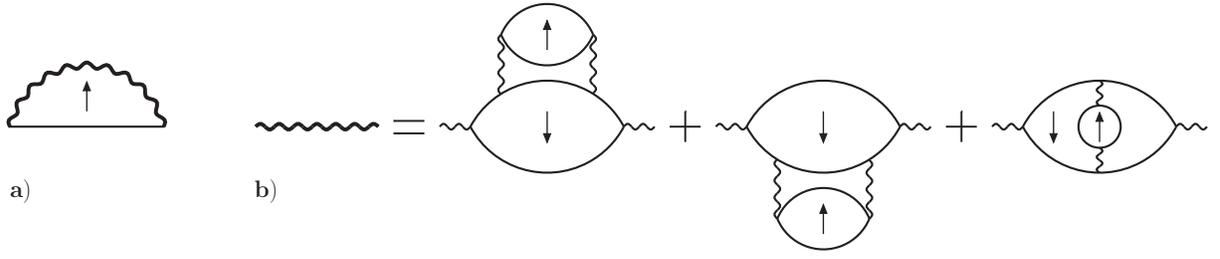}}
\caption{(a) Self-energy of the fourth order in the interwire interaction in channel (c) for direct scattering processes. (b) Effective interaction (thick wavy line) expressed in terms of the bare interwire interaction (thin wavy lines). The electron lines for different wires are labeled by the upward and downward arrows.}
\label{chan_c}
\end{figure}
\begin{figure}
\centerline{\includegraphics[width=0.6\columnwidth]{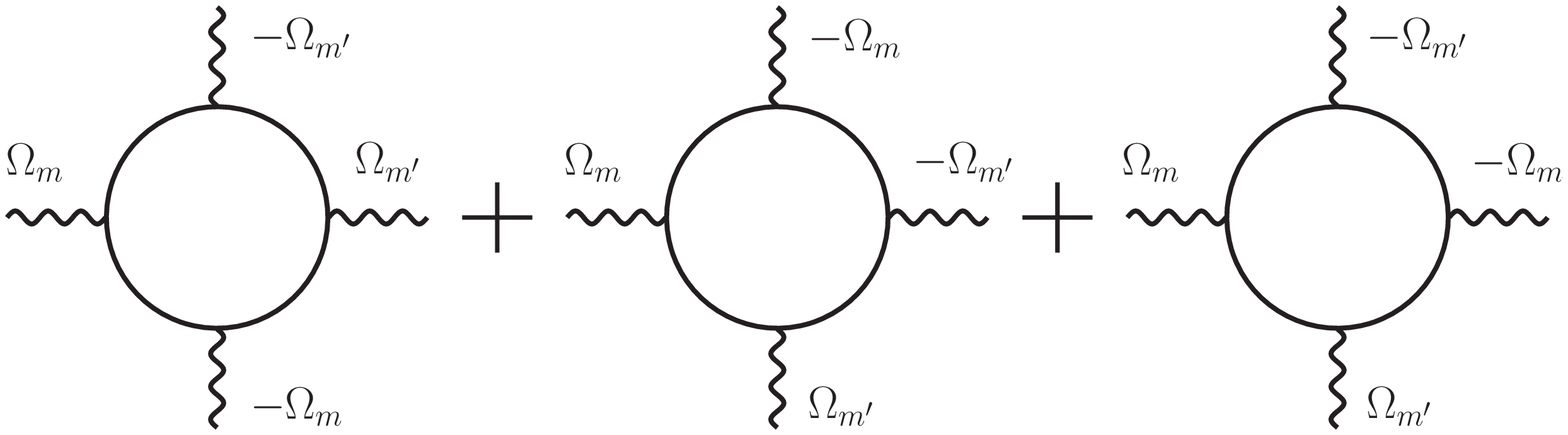}}
\caption{Sum of the four-leg loops from the effective interaction in Fig.~\ref{chan_c}b. The legs are labeled with the incoming frequencies.}
\label{four_leg}
\end{figure}

Summing over $\epsilon_n$ in Eq.~(\ref{c4}) and $\Omega_{m'}$ in Eq.~(\ref{c2}), $V(i\Omega_m,q)$ reads
\begin{eqnarray}
\phantom{a}\hspace{-10mm}V(i\Omega_m,q)&=&|V_{12}(q)|^2\int\!{dq'\over 2\pi}\,|V_{12}(q')|^2\int\!{dk'\over 2\pi}\,[f_T(k')-f_T(k'+q')]\int\!{dk\over 2\pi}\,{1\over i\Omega_m+\xi_{k-q}-\xi_k} \nonumber\\ \nonumber\\
&\times&\left\{\left[f_T(k-q)-f_T(k-q')\right]z_1+\left[f_T(k-q-q')-f_T(k-q)\right]z_2+\left[f_T(k)-f_T(k-q)\right]z_3\right.\nonumber\\ \nonumber\\
&+&\left.{\partial f_T(k)\over\partial\xi_k}\,{c(\xi_k-\xi_{k-q'})-c(\xi_{k'+q'}-\xi_{k'})\over\Delta_1}-{\partial f_T(k-q)\over\partial\xi_{k-q}}\,{c(\xi_{k-q}-\xi_{k-q-q'})-c(\xi_{k'+q'}-\xi_{k'})\over\Delta_2}\right\}~,
\label{c5}
\end{eqnarray}
where $c(\omega)=(1/2)\coth (\omega/2T)$,
\begin{equation}
\Delta_1=\xi_k+\xi_{k'}-\xi_{k-q'}-\xi_{k'+q'}~,\qquad \Delta_2=\xi_{k-q}+\xi_{k'}-\xi_{k-q-q'}-\xi_{k'+q'}~,
\label{c6}
\end{equation}
and the functions $z_{1,2,3}$ are given by
\begin{eqnarray}
\phantom{a}\hspace{-14mm}z_1\!\!&=&\!\!{1\over\Delta_1}\left[-{\partial c(\omega)\over\partial\omega}\Big\vert_{\omega=\xi_k-\xi_{k-q'}}+{c(\xi_k-\xi_{k-q'})-c(\xi_{k'+q'}-\xi_{k'})\over\Delta_1}-
{c(\xi_{k-q}-\xi_{k-q'})-c(\xi_{k'+q'}-\xi_{k'})\over i\Omega_m+\xi_{k-q}-\xi_{k-q'}-\xi_{k'+q'}+\xi_{k'}}\right]\nonumber\\
&+&\!\!{1\over\Delta_1}{c(\xi_{k-q}-\xi_{k-q'})-c(\xi_k-\xi_{k-q'})\over i\Omega_m+\xi_{k-q}-\xi_k}\nonumber\\
&+&\!\!{1\over i\Omega_m+\xi_{k-q-q'}-\xi_{k-q'}}
\left[{c(\xi_k-\xi_{k-q'})-c(\xi_{k'+q'}-\xi_{k'})\over\Delta_1}
-{c(\xi_{k-q}-\xi_{k-q'})-c(\xi_{k'+q'}-\xi_{k'})\over i\Omega_m+\xi_{k-q}-\xi_{k-q'}-\xi_{k'+q'}+\xi_{k'}}\right]~,\label{c7}\\
\nonumber\\
\phantom{a}\hspace{-14mm}z_2\!\!&=&\!\!{1\over\Delta_2}\left[-{\partial c(\omega)\over\partial\omega}\Big\vert_{\omega=\xi_{k-q}-\xi_{k-q-q'}}+{c(\xi_k-\xi_{k-q'})-c(\xi_{k'+q'}-\xi_{k'})\over\Delta_1}-
{c(\xi_k-\xi_{k-q-q'})-c(\xi_{k'+q'}-\xi_{k'})\over -i\Omega_m+\xi_k-\xi_{k-q-q'}-\xi_{k'+q'}+\xi_{k'}}\right]\nonumber\\
&-&\!\!{1\over\Delta_2}{c(\xi_k-\xi_{k-q-q'})-c(\xi_{k'+q'}-\xi_{k'})\over i\Omega_m+\xi_{k-q}-\xi_k}\nonumber\\
&-&\!\!{1\over i\Omega_m+\xi_{k-q-q'}-\xi_{k-q'}}
\left[{c(\xi_{k-q}-\xi_{k-q-q'})-c(\xi_{k'+q'}-\xi_{k'})\over\Delta_2}
-{c(\xi_{k-q}-\xi_{k-q'})-c(\xi_{k'+q'}-\xi_{k'})\over -i\Omega_m+\xi_k-\xi_{k-q-q'}-\xi_{k'+q'}+\xi_{k'}}\right]~,\label{c8}\\
\nonumber\\
\phantom{a}\hspace{-14mm}z_3\!\!&=&\!\!{1\over\Delta_1}\left[-{\partial c(\omega)\over\partial\omega}\Big\vert_{\omega=\xi_k-\xi_{k-q'}}+{c(\xi_k-\xi_{k-q'})-c(\xi_{k'+q'}-\xi_{k'})\over\Delta_1}\right]\nonumber\\
&-&\!\!{1\over i\Omega_m+\xi_{k-q}-\xi_k}\left[{c(\xi_k-\xi_{k-q'})-c(\xi_{k'+q'}-\xi_{k'})\over\Delta_1}-{c(\xi_k-\xi_{k-q-q'})-c(\xi_{k'+q'}-\xi_{k'})\over -i\Omega_m+\xi_k-\xi_{k-q-q'}-\xi_{k'+q'}+\xi_{k'}}\right]\nonumber\\
&-&\!\!{1\over i\Omega_m+\xi_{k-q-q'}-\xi_{k-q'}}\left[{c(\xi_{k-q}-\xi_{k-q-q'})-c(\xi_{k'+q'}-\xi_{k'})\over\Delta_2}
-{c(\xi_k-\xi_{k-q-q'})-c(\xi_{k'+q'}-\xi_{k'})\over -i\Omega_m+\xi_k-\xi_{k-q-q'}-\xi_{k'+q'}+\xi_{k'}}\right]~.
\label{c9}
\end{eqnarray}
The terms with $\partial f_T(k)/\partial\xi_k=-\zeta^2(k)/4T$ in Eq.~(\ref{c5}) arise from the double poles in the fermionic summation over $\epsilon_n$. The terms with $\partial c(\omega)/\partial\omega=-1/4T\sinh^2(\omega/2T)$ in Eqs.~(\ref{c7})-(\ref{c9}) arise from the double poles in the bosonic summation over $\Omega_{m'}$. The terms in the second lines in Eqs.~(\ref{c7})-(\ref{c9}) are proportional to $(i\Omega_m+\xi_{k-q}-\xi_k)^{-1}$ and produce, together with the same factor in the first line in Eq.~(\ref{c5}), double poles in $\Omega_m$.

The self-energy in Fig.~\ref{chan_c}a reads
\begin{equation}
\Sigma_c^H(i\epsilon_n,k_1)=-T\sum_m\int\!{dq\over 2\pi}\,{V(i\Omega_m,q)\over i\epsilon_n-i\Omega_m-\xi_{k_1-q}}~.
\label{c10}
\end{equation}
Doing the summation over $\Omega_m$ in the terms of $\Sigma_c$ that are proportional to $z_{1,2,3}$ gives
\begin{eqnarray}
&&T\sum_m{z_1\over (i\epsilon_n-i\Omega_m-\xi_{k_1-q})(i\Omega_m+\xi_{k-q}-\xi_k)}\nonumber\\
&&={1\over\Delta_1}\left[-{\partial c(\omega)\over\partial\omega}\Big\vert_{\omega=\xi_k-\xi_{k-q'}}+{c(\xi_k-\xi_{k-q'})-c(\xi_{k'+q'}-\xi_{k'})\over\Delta_1}\right]
I(i\epsilon_n,\xi_{k_1-q}|\xi_{k-q}-\xi_k)\nonumber\\
&&-{c(\xi_{k-q}-\xi_{k-q'})-c(\xi_{k'+q'}-\xi_{k'})\over\Delta_1^2}\left[I(i\epsilon_n,\xi_{k_1-q}|\xi_{k-q}-\xi_k)
-I(i\epsilon_n,\xi_{k_1-q}|\xi_{k-q}+\xi_{k'}-\xi_{k-q'}-\xi_{k'+q'})\right]\nonumber\\
&&-{c(\xi_{k-q}-\xi_{k-q'})-c(\xi_k-\xi_{k-q'})\over\Delta_1}{\partial I(i\epsilon_n,\xi_{k_1-q}|\omega)\over\partial\omega}\Big\vert_{\omega=\xi_{k-q}-\xi_k}\nonumber\\
&&+{c(\xi_k-\xi_{k-q'})-c(\xi_{k-q}-\xi_{k-q'})\over\Delta_1(\Delta_2-\Delta_1)}
\left[I(i\epsilon_n,\xi_{k_1-q}|\xi_{k-q-q'}-\xi_{k-q'})
-I(i\epsilon_n,\xi_{k_1-q}|\xi_{k-q}-\xi_k)\right]\nonumber\\
&&+{c(\xi_{k-q}-\xi_{k-q'})-c(\xi_{k'+q'}-\xi_{k'})\over\Delta_1\Delta_2}
\left[I(i\epsilon_n,\xi_{k_1-q}|\xi_{k-q-q'}-\xi_{k-q'})
-I(i\epsilon_n,\xi_{k_1-q}|\xi_{k-q}+\xi_{k'}-\xi_{k-q'}-\xi_{k'+q'})\right]~,\nonumber\\ \label{c11}\\
&&T\sum_m{z_2\over (i\epsilon_n-i\Omega_m-\xi_{k_1-q})(i\Omega_m+\xi_{k-q}-\xi_k)}\nonumber\\
&&={1\over\Delta_2}\left[-{\partial c(\omega)\over\partial\omega}\Big\vert_{\omega=\xi_{k-q}-\xi_{k-q-q'}}+{c(\xi_{k-q}-\xi_{k-q-q'})-c(\xi_{k'+q'}-\xi_{k'})\over\Delta_2}\right]
I(i\epsilon_n,\xi_{k_1-q}|\xi_{k-q}-\xi_k)\nonumber\\
&&-{c(\xi_k-\xi_{k-q-q'})-c(\xi_{k'+q'}-\xi_{k'})\over\Delta_2^2}\left[I(i\epsilon_n,\xi_{k_1-q}|\xi_{k-q}-\xi_k)
-I(i\epsilon_n,\xi_{k_1-q}|\xi_{k-q-q'}+\xi_{k'+q'}-\xi_k-\xi_{k'})\right]\nonumber\\
&&+{c(\xi_k-\xi_{k-q-q'})-c(\xi_{k-q}-\xi_{k-q-q'})\over\Delta_2}{\partial I(i\epsilon_n,\xi_{k_1-q}|\omega)\over\partial\omega}\Big\vert_{\omega=\xi_{k-q}-\xi_k}\nonumber\\
&&-{c(\xi_{k-q}-\xi_{k-q-q'})-c(\xi_{k'+q'}-\xi_{k'})\over\Delta_2(\Delta_2-\Delta_1)}
\left[I(i\epsilon_n,\xi_{k_1-q}|\xi_{k-q-q'}-\xi_{k-q'})
-I(i\epsilon_n,\xi_{k_1-q}|\xi_{k-q}-\xi_k)\right]\nonumber\\
&&+{c(\xi_k-\xi_{k-q-q'})-c(\xi_{k'+q'}-\xi_{k'})\over\Delta_1(\Delta_2-\Delta_1)}
\left[I(i\epsilon_n,\xi_{k_1-q}|\xi_{k-q-q'}-\xi_{k-q'})-I(i\epsilon_n,\xi_{k_1-q}|\xi_{k-q}-\xi_k)\right]\nonumber\\
&&+{c(\xi_k-\xi_{k-q-q'})-c(\xi_{k'+q'}-\xi_{k'})\over \Delta_1\Delta_2}
\left[I(i\epsilon_n,\xi_{k_1-q}|\xi_{k-q}-\xi_k)
-I(i\epsilon_n,\xi_{k_1-q}|\xi_{k-q-q'}+\xi_{k'+q'}-\xi_k-\xi_{k'})\right]~,\nonumber\\ \label{c12}\\
&&T\sum_m{z_3\over (i\epsilon_n-i\Omega_m-\xi_{k_1-q})(i\Omega_m+\xi_{k-q}-\xi_k)}\nonumber\\
&&={1\over\Delta_1}\left[-{\partial c(\omega)\over\partial\omega}\Big\vert_{\omega=\xi_k-\xi_{k-q'}}+{c(\xi_k-\xi_{k-q'})-c(\xi_{k'+q'}-\xi_{k'})\over\Delta_1}\right]
I(i\epsilon_n,\xi_{k_1-q}|\xi_{k-q}-\xi_k)\nonumber\\
&&+{c(\xi_k-\xi_{k-q-q'})-c(\xi_{k'+q'}-\xi_{k'})\over\Delta_2^2}\left[I(i\epsilon_n,\xi_{k_1-q}|\xi_{k-q}-\xi_k)
-I(i\epsilon_n,\xi_{k_1-q}|\xi_{k-q-q'}+\xi_{k'+q'}-\xi_k-\xi_{k'})\right]\nonumber\\
&&+\left({1\over\Delta_1}-{1\over\Delta_2}\right)[c(\xi_k-\xi_{k-q-q'})-c(\xi_{k'+q'}-\xi_{k'})]{\partial I(i\epsilon_n,\xi_{k_1-q}|\omega)\over\partial\omega}\Big\vert_{\omega=\xi_{k-q}-\xi_k}\nonumber\\
&&-{c(\xi_k-\xi_{k-q'})-c(\xi_k-\xi_{k-q-q'})\over\Delta_1(\Delta_2-\Delta_1)}
\left[I(i\epsilon_n,\xi_{k_1-q}|\xi_{k-q}-\xi_k)
-I(i\epsilon_n,\xi_{k_1-q}|\xi_{k-q-q'}-\xi_{k-q'})\right]\nonumber\\
&&-{c(\xi_k-\xi_{k-q-q'})-c(\xi_{k'+q'}-\xi_{k'})\over\Delta_1\Delta_2}
\left[I(i\epsilon_n,\xi_{k_1-q}|\xi_{k-q}-\xi_k)
-I(i\epsilon_n,\xi_{k_1-q}|\xi_{k-q-q'}+\xi_{k'+q'}-\xi_k-\xi_{k'})\right]~,
\label{c13}
\end{eqnarray}
where
\begin{equation}
I(i\epsilon_n,\xi|\omega)={c(\omega)-t(\xi)\over i\epsilon_n-\xi+\omega}
\label{c14}
\end{equation}
and $t(\xi)=(1/2)\tanh(\xi/2T)$.

The imaginary part of the retarded self-energy $\Sigma_c^H$ comes from the functions $I$:
\begin{equation}
{\rm Im}\,I(i\epsilon_n\to\xi_{k_1}+i0,\xi_{k_1-q}|\omega)=-{\pi\over 2}\left[c(\xi_{k_1-q}-\xi_{k_1})-t(\xi_{k_1-q})\right]\delta(\xi_{k_1}-\xi_{k_1-q}+\omega)~.
\label{c15}
\end{equation}
Triple collisions are associated with the terms in Eqs.~(\ref{c11})-(\ref{c13}) that contain the functions $I$ with six electron energies in the denominator [i.e., six electron energies in the delta-function in Eq.~(\ref{c15})], namely $I(i\epsilon_n,\xi_{k_1-q}|\xi_{k-q}+\xi_{k'}-\xi_{k-q'}-\xi_{k'+q'})$ and $I(i\epsilon_n,\xi_{k_1-q}|\xi_{k-q-q'}+\xi_{k'+q'}-\xi_k-\xi_{k'})$. These are only present in the contributions to $\Sigma_c^H$ coming from the functions $z_{1,2,3}$. More specifically, they are absent in the combination $z_2+z_3$, so that regrouping the terms proportional to $z_{1,2,3}$ in Eq.~(\ref{c5}) as
\begin{equation}
[f_T(k-q)-f_T(k-q')]z_1+[f_T(k-q-q')-f_T(k)]z_2+[f_T(k)-f_T(k-q)](z_2+z_3)~,
\label{c16}
\end{equation}
only the first two differences $f_T(k-q)-f_T(k-q')$ and $f_T(k-q-q')-f_T(k)$ describe the rate of triple collisions. Thus we obtain the contribution $1/\tau_{c,3}^H(k_1)$ to the triple-collision rate in channel (c) from direct scattering:
\begin{eqnarray}
{1\over\tau_{c,3}^H(k_1)}&=&2\pi\int\!{dq\over 2\pi}\,|V_{12}(q)|^2\int\!{dq'\over 2\pi}\,|V_{12}(q')|^2
\int\!{dk'\over 2\pi}\left[\,c(\xi_{k_1-q}-\xi_{k_1})-t(\xi_{k_1-q})\,\right]\left[\,f_T(k'+q')-f_T(k')\,\right]\nonumber\\
&\times&\int\!{dk\over 2\pi}\left\{{1\over\Delta_1}\left({1\over\Delta_1}-{1\over\Delta_2}\right)[\,c(\xi_{k-q}-\xi_{k-q'})-c(\xi_{k'+q'}-\xi_{k'})\,]
\,[\,f_T(k-q)-f_T(k-q')\,]\right.\nonumber\\
&&\hspace{38.5mm}\times\,\,\delta(\xi_{k_1}+\xi_{k-q}+\xi_{k'}-\xi_{k_1-q}-\xi_{k-q'}-\xi_{k'+q'})\nonumber\\
&&\hspace{7.5mm}+\,\,{1\over\Delta_2}\left({1\over\Delta_2}-{1\over\Delta_1}\right)[\,c(\xi_k-\xi_{k-q-q'})-c(\xi_{k'+q'}-\xi_{k'})\,]
\,[\,f_T(k-q-q')-f_T(k)\,]\nonumber\\
&&\left.\hspace{38mm}\times\,\,\delta(\xi_{k_1}+\xi_{k-q-q'}+\xi_{k'+q'}-\xi_{k_1-q}-\xi_k-\xi_{k'})\right\}~.
\label{c17}
\end{eqnarray}
Two different delta-functions in Eq.~(\ref{c17}) correspond to two different amplitudes: the one proportional to $\Delta_1^{-1}(\Delta_1^{-1}-\Delta_2^{-1})$ comes from the process shown in Fig.~\ref{two_pro}a, the other---from the process shown in Fig.~\ref{two_pro}b.
The energy denominators $\Delta_1$ and $\Delta_2$ for the processes in Figs.~\ref{two_pro}a and \ref{two_pro}b are then identified in terms of the energies in Eq.~(\ref{44}) as follows:
\begin{eqnarray}
&&{\rm Fig.\,3a}\!:\,\,\,\,\,\,\,\Delta_1=-\Delta_{311'}~,\,\,\,\,\,\Delta_2=\Delta_{322'}~,\\
&&{\rm Fig.\,3b}\!:\,\,\,\,\,\,\,\Delta_1=\Delta_{311'}~,\,\,\,\,\,\Delta_2=-\Delta_{322'}~.
\label{c18}
\end{eqnarray}
The sum of the contributions of the two processes to Eq.~(\ref{c17}) gives the cross-section proportional to
\begin{equation}
\left({1\over\Delta_{311'}}+{1\over\Delta_{322'}}\right)^2
\end{equation}
[cf.\ Eq.~(\ref{43c})]. This is because the factor
\begin{equation}
[c(\xi_{k_1-q}-\xi_{k_1})-t(\xi_{k_1-q})][f(k'+q')-f(k')][f(k-q)-f(k-q')][c(\xi_{k-q}-\xi_{k-q'})-c(\xi_{k'+q'}-\xi_{k'})]
\label{c19}
\end{equation}
from the process in Fig.~\ref{two_pro}a and its counterpart from the process in Fig.~\ref{two_pro}b,
\begin{equation}
[c(\xi_{k_1-q}-\xi_{k_1})-t(\xi_{k_1-q})][f(k'+q')-f(k')][f(k-q-q')-f(k)][c(\xi_k-\xi_{k-q-q'})-c(\xi_{k'+q'}-\xi_{k'})]~,
\label{c20}
\end{equation}
become identically equal to each other when written in terms of the energies $\epsilon_{1,2,3,1',2',3'}$ in Fig.~\ref{two_pro}. Specifically, each of them is written as
\begin{equation}
{1\over 16}[\coth(1-1')+\tanh(1')][\tanh(2)-\tanh(2')][\tanh(3)-\tanh(3')][\coth(3-3')+\coth(2-2')]~,
\label{c21}
\end{equation}
where $\coth(1-1')=\coth[(\epsilon_1-\epsilon_{1'})/2T]$, etc. Further, for $\epsilon_1+\epsilon_2+\epsilon_3=\epsilon_{1'}+\epsilon_{2'}+\epsilon_{3'}$, Eq.~(\ref{c21}) can be reduced to
\begin{equation}
{\cosh(1)\over 16\cosh(2)\cosh(3)\cosh(1')\cosh(2')\cosh(3')}~,
\label{c22}
\end{equation}
which is recognized as the factor that appears in the collision integral (\ref{33}) for the scattering rate $1\to 1'$, namely
\begin{equation}
f(k_2)f(k_3)[1-f(k_{1'})][1-f(k_{2'})][1-f(k_{3'})]+f(k_{1'})f(k_{2'})f(k_{3'})[1-f(k_2)][1-f(k_3)]~,
\label{c23}
\end{equation}
taken at thermal equilibrium. We have thus reproduced the inverse lifetime due to triple collisions diagrammatically.

\begin{figure}
\centerline{\includegraphics[width=0.6\columnwidth]{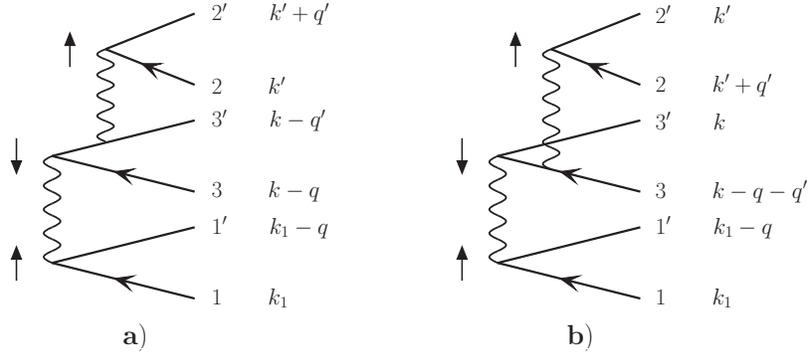}}
\caption{Three-particle scattering amplitudes at second order in the bare interwire interaction (wavy lines). They contribute to the first (a) and second (b) terms in the kernel of the triple-collision rate in Eq.~(\ref{c17}). The electron lines for different wires are labeled by the upward and downward arrows.}
\label{two_pro}
\end{figure}
\begin{figure}
\centerline{\includegraphics[width=0.2\columnwidth]{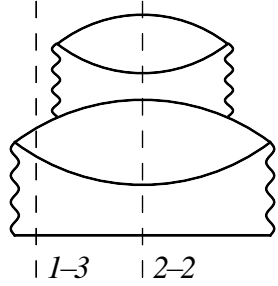}}
\caption{Cuts labeled 2-2 and 1-3 in this particular term in the three-particle self-energy contribute to the total three-particle scattering rate and to a reduction of it that comes from two consecutive two-particle scattering events, respectively. Each of the contributions is diverging (with opposite signs), their sum is finite.}
\label{cross-sec}
\end{figure}

The scattering rate in Eq.~(\ref{c17}) diverges because of the factors $1/\Delta_1^2$ and $1/\Delta_2^2$. But, as can be seen from Eqs.~(\ref{c11})-(\ref{c13}), the factors $1/\Delta_1^2$ and $1/\Delta_2^2$ are present in the self-energy (\ref{c1}) not only in the part associated with triple collisions [six fermionic energies in the argument of the delta-function in Eq.~(\ref{c15})], but also in the part that contains the delta-function of a sum of four fermionic energies and is therefore identified with the contribution of {\it pair} collisions at order $V_{12}^4$. These scattering processes are associated with the product of two amplitudes in which one is of order $V_{12}$, the other---of order $V_{12}^3$. For example, if one considers the diagram for the three-particle self-energy in Fig.~\ref{cross-sec}, the two-particle processes correspond to the cut labeled 1-3 (one amplitude is of the first order in interaction, the other---of the third order), in contrast to the cut labeled 2-2 which contributes to $1/\tau_{c,3}^H(k_1)$ in Eq.~(\ref{c17}). Specifically, we have for the sum $R_{c,2}(k_1)$ of the terms in $1/\tau_c(k_1)$ that originate from direct scattering in pair collisions and are proportional to either $1/\Delta_1^2$ or $1/\Delta_2^2$:
\begin{eqnarray}
R_{c,2}(k_1)&=&2\pi\int\!{dq\over 2\pi}\,|V_{12}(q)|^2\int\!{dq'\over 2\pi}\,|V_{12}(q')|^2\int\!{dk'\over 2\pi}\left[\,c(\xi_{k_1-q}-\xi_{k_1})-t(\xi_{k_1-q})\,\right]\left[\,f_T(k'+q')-f_T(k')\,\right]\nonumber\\
&\times&\int\!{dk\over 2\pi}\left\{{1\over\Delta_1^2}[\,c(\xi_k-\xi_{k-q'})-c(\xi_{k-q}-\xi_{k-q'})\,]
\,[\,f_T(k-q)-f_T(k-q')\,]\right.\nonumber\\
&&\hspace{7.5mm}+\,\,{1\over\Delta_1^2}[\,c(\xi_k-\xi_{k-q'})-c(\xi_{k'+q'}-\xi_{k'})\,]
\,[\,f_T(k)-f_T(k-q)\,]\nonumber\\
&&\hspace{7.5mm}+\,\,{1\over\Delta_2^2}[\,c(\xi_{k-q}-\xi_{k-q-q'})-c(\xi_k-\xi_{k-q-q'})\,]
\,[\,f_T(k-q-q')-f_T(k-q)\,]\nonumber\\
&&\hspace{7.2mm}\left.+\,\,{1\over\Delta_2^2}[\,c(\xi_k-\xi_{k-q-q'})-c(\xi_{k'+q'}-\xi_{k'})\,]
\,[\,f_T(k)-f_T(k-q)\,]\right\}\nonumber\\
&&\hspace{7.5mm}\times\,\,\delta(\xi_{k_1}+\xi_{k-q}-\xi_{k_1-q}-\xi_k)~,
\label{c24}
\end{eqnarray}
which diverges in the same manner as $1/\tau_{c,3}^H(k_1)$. To see how the sum of the two contributions to $1/\tau_c$ behaves, compare the factors in front of $1/\Delta_1^2$ in Eqs.~(\ref{c17}) and (\ref{c24}). Importantly, the delta-functions $\delta(\xi_{k_1}+\xi_{k-q}+\xi_{k'}-\xi_{k_1-q}-\xi_{k-q'}-\xi_{k'+q'})$ in Eq.~(\ref{c17}) and $\delta(\xi_{k_1}+\xi_{k-q}-\xi_{k_1-q}-\xi_k)$ in Eq.~(\ref{c24}) become identical at $\Delta_1=0$. Further, the difference $c(\xi_k-\xi_{k-q'})-c(\xi_{k'+q'}-\xi_{k'})$ in the third line of Eq.~(\ref{c24}) vanishes at $\Delta_1=0$, while the remaining factor in front of $1/\Delta_1^2$ in Eq.~(\ref{c24}) and its counterpart in Eq.~(\ref{c17}) exactly cancel each other. A similar cancellation occurs with the terms proportional to $1/\Delta_2^2$. We thus arrive at the conclusion that, in channel (c), the divergency in the total triple-collision rate, defined via the integral of the modulus squared of the matrix elements of the three-particle $T$-matrix, is exactly canceled by the divergency associated with pair collisions. How this happens diagrammatically in the above calculation of the inverse lifetime at equilibrium exemplifies the subtraction of the counterterm (\ref{51}) in the three-particle collision integral.

\begin{figure}
\centerline{\includegraphics[width=0.5\columnwidth]{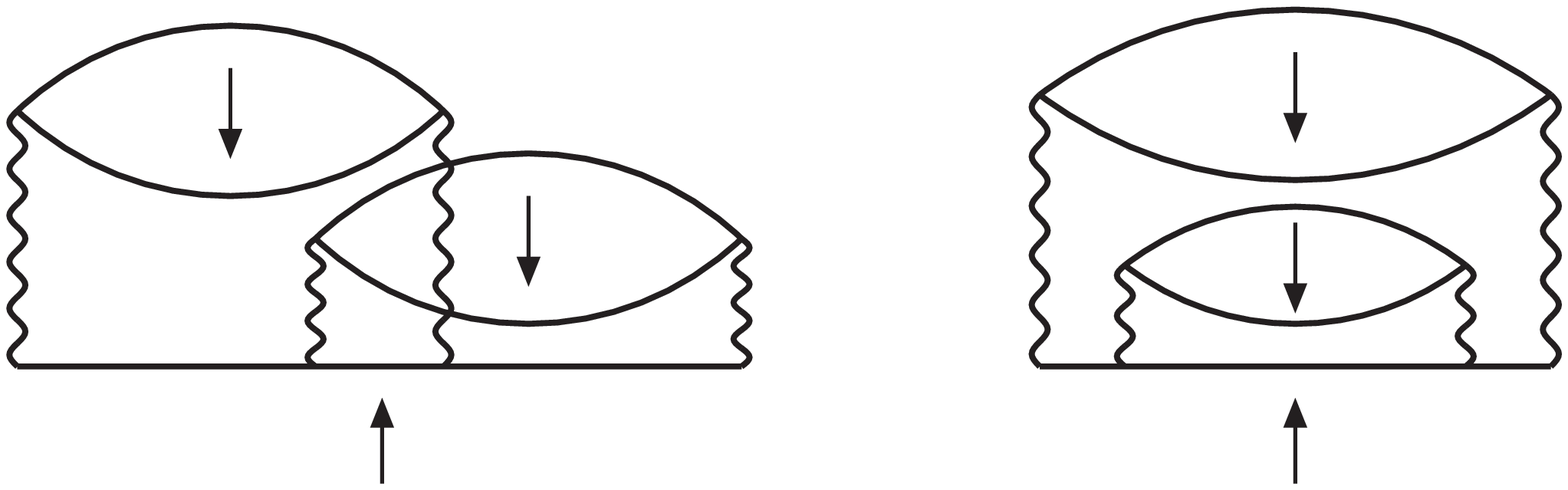}}
\caption{Two contributions to the self-energy of the fourth order in the interwire interaction (wavy lines) in channel (b) for direct scattering processes. The electron lines for different wires are labeled by the upward and downward arrows.}
\label{chan_b}
\end{figure}

Let us now turn to channel (b), where the self-energy $\Sigma_b^H(i\epsilon_n,k_1)$ is given (at order $V_{12}^4$ and zeroth order in $V_{11}$, similar to the above) by the sum of two diagrams in Fig.~\ref{chan_b}:
\begin{equation}
\Sigma_b^H(i\epsilon_n,k_1)=T^2\sum_{mm'}\int\!{dq\over 2\pi}\,|V_{12}(q)|^2\int\!{dq'\over 2\pi}\,|V_{12}(q')|^2B(i\epsilon_n,k_1|i\Omega_m,q;i\Omega_{m'},q')\Pi(i\Omega_m,q)\Pi(i\Omega_{m'},q')~,
\label{c25}
\end{equation}
where
\begin{eqnarray}
B(i\epsilon_n,k_1|i\Omega_m,q;i\Omega_{m'},q')&=&{1\over (i\epsilon_n-i\Omega_m-\xi_{k_1-q})(i\epsilon_n-i\Omega_{m+m'}-\xi_{k_1-q-q'})}\nonumber\\
&\times&\left({1\over i\epsilon_n-i\Omega_{m'}-\xi_{k_1-q'}}+{1\over i\epsilon_n-i\Omega_m-\xi_{k_1-q}}\right)~.
\label{c26}
\end{eqnarray}
As far as the singularities are concerned, the structure of $\Sigma_b^H$ is much simpler than that of $\Sigma_c^H$, because the nonintegrable singularities of the type $1/\Delta_{1,2}^2$ encountered separately for triple and double collisions in Eq.~(\ref{c10}) are absent in Eq.~(\ref{c25}). Specifically, the only \cite{remark16} term in $\Sigma_b^H(i\epsilon_n,k_1)$ that contains a factor similar to $1/\Delta_{1,2}^2$ in channel (c) is
\begin{eqnarray}
&&\int\!{dq\over 2\pi}\,|V_{12}(q)|^2\int\!{dq'\over 2\pi}\,|V_{12}(q')|^2\int\!{dk\over 2\pi}\,[f_T(k)-f_T(k-q)]\int\!{dk'\over 2\pi}\,[f_T(k'+q')-f_T(k')]\nonumber\\
&&\times{1\over\Delta_3^2}[c(\xi_{k_1-q}-\xi_{k_1-q-q'})-c(\xi_{k'+q'}-\xi_{k'})]
[I(i\epsilon_n,\xi_{k_1-q-q'}|\xi_{k-q}+\xi_{k'}-\xi_k-\xi_{k'+q'})-I(i\epsilon_n,\xi_{k_1-q}|\xi_{k-q}+\xi_k)]~,\nonumber\\
\label{c27}
\end{eqnarray}
where $\Delta_3=\xi_{k_1-q-q'}+\xi_{k'+q'}-\xi_{k_1-q}-\xi_{k'}$. However, in contrast to channel (c), the difference of the bosonic distribution functions $c(\xi_{k_1-q}-\xi_{k_1-q-q'})-c(\xi_{k'+q'}-\xi_{k'})$ in Eq.~(\ref{c27}) vanishes at $\Delta_3=0$, so that the singularity reduces to $1/\Delta_3$ and the integral (taken in the principal value sense for the scattering rate) is finite.

Note that, at order $V_{12}^4$ and zeroth order in $V_{11}$, the electron self-energy contains more terms than the diagrams with two electron loops, shown in Figs.~\ref{chan_c} and \ref{chan_b}, and their exchange counterparts (with one loop less). These come from further renormalizations of the two-particle $T$-matrix (with one loop and three loops) not included in the exchange counterparts of Figs.~\ref{chan_c} and \ref{chan_b}; in particular, from the diagram with a chain of three loops, which corresponds to the effective interaction in the random-phase approximation. Summing up contributions of two-particle cuts of higher-order self-energy diagrams will give the modulus squared of the matrix elements of the exact (in $V_{12}$, in this particular case) two-particle $T$-matrix. For the precise meaning of an $M$-particle cut, see Ref.~\onlinecite{bezzerides68}.

As seen from the above calculation, the divergencies in the triple- and pair-collision contributions to the scattering rate and their cancellation occur for an arbitrary form of the dispersion relation of colliding particles. In particular, this means that the nonintegrable singularity in the differential cross-section for triple collisions that comes from the modulus squared of the three-particle $T$-matrix is present in the Luttinger liquid model (linear dispersion) as well. The three-particle singularity is, however, canceled by the other one that comes, in the kinetic-equation formalism, from those two-particle collisions in which a given particle participates twice. This resolves the apparent conflict, mentioned at the beginning of this appendix, between the divergency of the triple-collision rate and the Dzyloshinskii-Larkin theorem for the linear dispersion law.

\end{widetext}


\begin{thebibliography}{99}
\bibitem{remark34} Reference \onlinecite{pustilnik03} calculated a dimensionless ``drag resistivity" $r_{\rm D}$ defined as the ``drag resistance per unit length of the interacting region" according to $r_{\rm D}=(e^2/2\pi)R_{\rm D}(L)/L$ with $R_{\rm D}(L)=-\lim_{j_1\to 0}(dV_2/dj_1)\vert_{j_2=0}$. Here $L$ is the length of the region in which the interwire interaction is present and $V_2$ is the bias voltage applied to the passive wire to maintain $j_2=0$. An implicit condition restricting $L$ from below in the calculation of Ref.~\onlinecite{pustilnik03} follows from the form of Eq.~(5) in Ref.~\onlinecite{pustilnik03}. Indeed, at the lowest order in the interwire interaction, the friction-force density experienced by electrons in the passive wire can be represented as a single integral over the transferred momentum only if $L\gg\max\{a,v_F/T\}$. It should be emphasized that the friction-force density was found in Ref.~\onlinecite{pustilnik03} to be $L$-independent---implying that the drag resistance $R_{\rm D}(L)$ grows linearly with $L$---while no explicit condition restricting $L$ from above was given. Moreover, the only explicit condition for using the lowest-order perturbative expression for $V_2$ was specified in Ref.~\onlinecite{pustilnik03} as the condition that $T$ is large enough to prevent the formation of the zig-zag ordered state (see Sec.~\ref{s1.2})---this condition does not depend on $L$. This suggests that $r_{\rm D}$ calculated in Ref.~\onlinecite{pustilnik03} describes bulk drag, i.e., $\rho_{\rm D}\stackrel{?}{=}(2\pi/e^2)r_{\rm D}$, where $\rho_{\rm D}$ is the drag resistivity \cite{remark24} of infinite wires.
\bibitem{remark24} The dc resistivity is defined (conventionally and throughout this paper) as describing the response of a macroscopically homogeneous infinite system to a perturbation at frequency $\omega\to 0$ and wavevector $q\to 0$ with the following order of taking the limits: $\omega\to 0$ after $q\to 0$. The calculations of the dc drag resistivity and/or conductivity within the orthodox theory in Refs.~\onlinecite{jauho93,zheng93,kamenev95,flensberg95} are also based on this definition.
\bibitem{remark7} In Ref.~\onlinecite{pustilnik03}, the $T^5$ suppression of $\rho_{\rm D}$ in the limit of low $T$ was obtained as well, but only for the case of quantum wires with different Fermi velocities $v_{F1}\neq v_{F2}$ in the presence of intrawire interactions. For nonidentical wires, Ref.~\onlinecite{aristov07} gives the $T^5$ scaling of $\rho_{\rm D}$ with a prefactor depending on the difference $\Delta v_F=v_{F1}-v_{F2}$ as $(\Delta v_F)^{-2}$, which is parametrically larger, for small $\Delta v_F$, than the prefactor proportional to $(\Delta v_F)^{-1}$ in Ref.~\onlinecite{pustilnik03}.
\bibitem{remark8a} It is worth noting, however, that the result of the orthodox theory does not necessarily imply the equilibrium state in the moving frame.
\bibitem{remark8} This condition, being particularly restrictive in one dimension, is general: in the case of Coulomb drag between generic two-dimensional electron systems, it is also crucially important to compare the characteristic inelastic interlayer scattering rates and the equilibration rates inside the layers. Full equilibration in the moving frame yields \cite{remark8a} the result of the orthodox theory for $\rho_{\rm D}$ independently of the dimensionality of the system.
\bibitem{remark27} Equilibration in the moving frame in the double-wire system is only achieved, as shown in Sec.~\ref{s3b}, if intrawire three-particle scattering is sufficiently strong compared to interwire two-particle scattering.
\bibitem{remark30} In the setup of Ref.~\onlinecite{remark25}, electrons incident on the ``drag region" interact with electrons of opposite chirality that have passed through it. By this mechanism, the distribution function of {\it incident} electrons (for which the boundary condition fixes it in the form of the drift ansatz for distances from the drag region larger than $l_{\rm b}$) is modified by drag already in the leads. This feedback effect can, however, be neglected in the calculation of $R_{\rm D}$ if the correction to the distribution function induced by interwire interactions is small.
\bibitem{remark29} It is worth noting that equilibration in the stationary frame between particles of the same chirality, induced by pair collisions in the double-wire system, is characterized by a relaxation length $l_0$ that is much shorter than $l_{\rm f}$ for $T\ll\epsilon_F$ [the spatial scales $l_{\rm f}$ and $l_0$ are associated, respectively, with the time scales $\tau_{\rm D}^\infty$ from Eq.~(\ref{11a}) and  $1/M(0,k_F)$ from Eq.~(\ref{i9})]. Specifically, $l_0\sim l_{\rm f} (T/\epsilon_F)^2$, independently of the parameter $Ta/v_F$. In particular, $l_0\sim 1/\beta^2_{\rm f}k_F$ for $T\alt v_F/a$ does not depend on $T$ [assuming, similar to Eq.~(\ref{1.1}), that $\beta_{\rm f}\epsilon_F\ll T$]. More precisely, the scale $l_0$ describes equilibration of the difference of the distribution functions in two wires for particles of the same chirality whose momenta are within a distance of the order of $T/v_F$ from $\pm k_F$. Spreading of the equilibrated part of the distribution function down to the bottom of the spectrum is characterized by the much larger scale $l_{\rm b}$, with the scale $l_{\rm f}$ in between ($l_0\ll l_{\rm f}\ll l_{\rm b}$) in the limit of small momentum transfer.
\bibitem{remark17} Note that, unless the wires are identical [as is the case in Eq.~(\ref{11a})], the phenomenon of drag, in general, is not characterized at given $\omega$ by a single scattering rate (``drag rate") for electrons in both wires (the asymmetry of the drag rates in nonidentical wires is a direct consequence of the symmetry property $\sigma_{12}=\sigma_{21}$ of the drag components of the conductivity matrix, which holds independently of the possible difference $\sigma_{11}\neq\sigma_{22}$ of the diagonal components). However, the conceptual framework developed here---highlighting the close link between Coulomb drag and thermal equilibration---remains intact in the case of drag between nonidentical conductors. In particular, $\rho_{\rm D}=0$ in the absence of right-left equilbration independently of whether the wires are identical or not.
\bibitem{remark9} For the Born matrix element, the presence of the $\theta$-function in Eq.~(\ref{Vtheta}) would not conform to the condition of spatial homogeneity in Eq.~(\ref{1}), because homogeneity implies that the Fourier component of the interaction potential depends only on the difference $k'-k$. However, at the model level, Eqs.~(\ref{1}) and (\ref{4}) with the collision kernel proportional to $\theta (kk')$ can be viewed as describing homogeneous wires at higher orders in interaction, where the scattering matrix depends on $k$ and $k'$ separately.
\bibitem{remark15} Note that $f_-$ is zero at thermal equilibrium, so that the condition $J^{(2)}=0$ at equilibrium is satisfied trivially---without invoking the Einstein relation. The nonzero solution $f_-\propto\zeta^2$, which also nullifies $J^{(2)}$, describes nonidentical wires each of which is at equilibrium within itself, and corresponds to a shift of the chemical potentials in the wires with respect to each other.
\bibitem{remark11} The $1/k$ dependence of $f_-$ in the limit of large $k$ is an artefact of the Fokker-Planck approximation, i.e., only valid for $|k|\ll mTa$, where this approximation is accurate. For $a\gg v_F/T$ (which is the condition of the calculation in Sec.~\ref{s2c}), the range of momenta $|k|\ll mTa$ includes $|k|\alt k_F$, i.e., the Fokker-Planck description only breaks down for $|k|$ much larger than $k_F$. Note that in the limit of large $a$ the main contribution to the current (\ref{17}) does come from the $1/k$ tail of $f_-$ at $|k|\gg k_F$; however, this requires that $a$ be exponentially large in $\epsilon_F/T$, namely $a\agt (v_F/T)(T/\epsilon_F)^{1/2}\!\exp (2\epsilon_F/T)$, which we exclude from the consideration.
\bibitem{remark12} Note, however, that if $f_-$ were given exactly by Eq.~(\ref{28}) for arbitrary $k$, it would lead to $J^{(2)}(k)\propto \delta(k)$ [cf.\
Eq.~(\ref{15})], vanishing for any $k\neq 0$. That is, the finite stationary value of $f_-$ in Eq.~(\ref{28}) is only due to the fact that $f_-$, given exactly by Eq.~(\ref{23}), is {\it not} exactly proportional to $\zeta^2$.
\bibitem{remark33} The expression $\rho_{\rm D}\propto 1/j_-$ emphasizes the Galilean invariance of $\rho_{\rm D}$ in our problem. In particular, it gives $\rho_{\rm D}$ in the case of $j_2\neq 0$. Note also that if the drag problem is formulated for given $j_1$ and $j_2$, Eq.~(\ref{29a}) yields $E_1=-E_2$ for arbitrary $j_{1,2}$.
\bibitem{remark23} For the case of $V(q)$ that falls off with increasing $|q|$ faster than an exponential function of $|q|$, the asymmetry factor is exponentially large in the parameter $|k|/mTa\gg 1$.
\bibitem{remark13} More accurately, in view of the exponential behavior of $\rho_{\rm D}$ in both Eqs.~(\ref{32d}),(\ref{32}), the crossover between two regimes for $k_Fa\gg 1$ is sharp; moreover, in the case of $V(q)\propto e^{-|q|a}$ [Eq.~(\ref{a3})], the crossover at $T\simeq T_c=v_F/2a$ has the form of a steep jump of width in $T$ of the order of $T_c [\ln (k_Fa)/k_Fa]^{1/2}$. The appearance of the jump is related to the suppression of the diffusion in energy space through the bottom of the spectrum at $T<T_c$, induced by the strong asymmetry between hopping up and hopping down in energy in the range $2mTa <|k|<k_F$ (see the discussion at the end of Sec.~\ref{s2c}). However, for $k_Fa\gg 1$, the jump is masked by the contribution of three-particle scattering if the forward-scattering coupling constant $\beta_{\rm f}$ is not exponentially small in the parameter $k_Fa$.
\bibitem{remark31} In the case of three-particle scattering (or scattering of a larger number of particles), $g_+(k)$ does not relax for special types of nonequilibrium only [this is in contrast to the case of two-particle scattering, where the collision integral for $g_+(k)$ vanishes for any type of nonequilibrium)]. These are characterized by the inhomogeneous (driving) term in the linearized kinetic equation for $g_+(k)$ that is a polynomial of $k$ of degree two or less, with powers of zero, one, and two corresponding to the conservation of particle number, total momentum, and total energy, respectively (cf.\ Ref.~\onlinecite{micklitz11}). In our problem, the inhomogeneous term, describing a spatially homogeneous driving force, is linear in $k$.
\bibitem{remark22} This kind of divergency was encountered in Ref.~\onlinecite{karzig10}, where the energy relaxation of hot electrons in a single wire was studied in the spinful case, for the total scattering rate and was argued not to affect the solution of the kinetic equation if the ``electron distribution has a finite spectral width". In fact, however, the singularity is absent both in the collision integral, as follows from Sec.~\ref{s3a1}, and in the total scattering rate, as follows from Appendix \ref{aC}.
\bibitem{remark20} In Ref.~\onlinecite{micklitz10}, which deals with triple collisions in a short (not equilibrated) single wire [channel (a) in our nomenclature], the validity of the Fokker-Planck approximation for Coulomb interaction was argued to be controlled by the parameters $T/\epsilon_F\ll 1$ and $|k|\ll k_F$---since the characteristic transferred momentum is then given by $T/v_F$ for each of three colliding electrons and the characteristic change of energy for the diffusing electron is $T|k|/k_F\ll T$. However, the integral term $\bar{C}^{(3)}(k)$ was missing in the Fokker-Planck equation written in Ref.~\onlinecite{micklitz10} [because the nonequlibrium corrections to the coefficients $A$ and $B$ in Eqs.~(B.5),(B.6) in Ref.~\onlinecite{micklitz10} were not included in the collision integral]. The omission of $\bar{C}^{(3)}(k)$ may strongly affect the
    relaxation rates and, moreover, violates the conservation of momentum. The condition for the Fokker-Planck expansion in the integral term $\bar{C}^{(3)}(k)$ is different from those specified in Ref.~\onlinecite{micklitz10}, namely---since the distribution function in $\bar{C}^{(3)}(k)$ is taken in the vicinity of the Fermi level---the characteristic transferred momentum for electrons with velocity $v_F$ should be smaller than $T/v_F$. This requires that $V_{11}(q)$ be a sharply peaked function on the scale of $T/v_F$, which means that $T$ is limited from below, in contrast to Ref.~\onlinecite{micklitz10}.
\bibitem{remark16} It is worth noting that $\Sigma^H_b(i\epsilon_n,k_1)$ contains other terms that have a transferred energy squared in the denominator; however, these have the structure of either $\partial I(i\epsilon_n,\xi|\omega)/\partial\omega$ or $\partial I(i\epsilon_n,\xi|\omega)/\partial\xi$, none of which leads to a divergency in ${\rm Im}\,\Sigma^H_b(i\epsilon_n\to \xi_{k_1}+i0,k_1)$.
\end{thebibliography}
\end{document}